\documentclass[12pt,aps,prd,preprint,nofootinbib,superscriptaddress,nobalancelastpage]{revtex4-1}
\pdfoutput=1
\usepackage{bbm}
\usepackage{amsfonts}
\usepackage{amssymb}
\usepackage{amsmath,amsthm,slashed,mathtools,mathrsfs,multirow}
\usepackage{graphicx}
\usepackage{array}
\usepackage{color}
\usepackage{ulem}
\usepackage{xspace}
\usepackage{verbatim}
\usepackage[usenames,dvipsnames]{xcolor}
\usepackage{subfigure}
\usepackage{verbatim}
\usepackage{dsfont}
\usepackage{braket}
\usepackage{mathtools}
\usepackage{adjustbox}
\usepackage{mleftright}
\usepackage{scrextend}
\usepackage{hyperref}\hypersetup{colorlinks,bookmarksopen,bookmarksnumbered,citecolor=blue,linkcolor=blue,pdfstartview=FitH,urlcolor=blue}

\newcommand{\D}{D\hspace{-8pt}\slash}
\newcommand{\be}{\begin{equation}}
\newcommand{\ee}{\end{equation}}
\newcommand{\bea}{\begin{eqnarray}}
\newcommand{\eea}{\end{eqnarray}}

\allowdisplaybreaks

\def\INFNg{Dipartimento di Fisica, Universit\`a di Genova and INFN - Sezione di Genova, via Dodecaneso 33, 16146 Genova, Italy}

\def\ift{Instituto de F\'isica Te\'orica UAM/CSIC, Calle Nicol\'as Cabrera 13-15, Cantoblanco E-28049 Madrid, Spain}

\def\uam{Departamento  de  F\'{\i}sica Te\'{o}rica,  Universidad  Aut\'{o}noma  de  Madrid, Cantoblanco  E-28049  Madrid,  Spain}

\def\elte{Institute for Theoretical Physics, ELTE E\"{o}tv\"{o}s Lor\'{a}nd University, P\'azm\'any P\'eter s\'et\'any 1/A, H-1117 Budapest, Hungary}

\def\IFIC{Instituto de F\'{\i}sica Corpuscular, Universidad de Valencia and CSIC, Edificio Institutos Investigaci\'on,
Catedr\'atico Jos\'e Beltr\'an 2, 46980, Spain}

\begin{document}

\preprint{FTUV-19-1118.7550}
\preprint{IFIC/19-50}
\preprint{FTUAM-19-23}
\preprint{IFT-UAM/CSIC-19-158}

\title{Global Bounds on the Type-III Seesaw}
\vspace{1.0cm}
\author{Carla Biggio}
\email{carla.biggio@ge.infn.it}
\affiliation{\INFNg}

\author{Enrique Fernandez-Martinez}
\email{enrique.fernandez-martinez@uam.es}
\affiliation{\ift}
\affiliation{\uam}

\author{Manuele Filaci}
\email{manuele.filaci@ge.infn.it}
\affiliation{\INFNg}

\author{Josu Hernandez-Garcia}
\email{josu.hernandez@ttk.elte.hu}
\affiliation{\elte}

\author{Jacobo Lopez-Pavon}
\email{jacobo.lopez@uv.es}
\affiliation{\IFIC \vspace*{1cm}}

\begin{abstract}
\vspace{0.5cm}

We derive general bounds on the Type-III Seesaw parameters from a
global fit to flavor and electroweak precision data. We explore
and compare three Type-III Seesaw realizations: a general scenario,
where an arbitrary number of heavy triplets is integrated out without
any further assumption, and the more constrained cases in which only 3 
or 2 (minimal scenario) additional heavy states are included. The latter assumption implies rather non-trivial
correlations in the Yukawa flavor structure of the model so as to reproduce the neutrino
masses and mixings as measured in neutrino oscillations experiments and thus
qualitative differences can be found with the more general
scenario. In particular, we find that, while the bounds on most elements of the dimension 6 operator coefficients are of order $10^{-4}$ for the general and
3-triplet cases, the 2-triplet scenario is more strongly constrained
with bounds between $10^{-5}$ and $10^{-7}$ for the different
flavours. We also discuss how these correlations affect the present
CMS constraints on the Type-III Seesaw in the minimal 2-triplet
scenario. 
\end{abstract}

\maketitle


\section{Introduction}
\label{s:introduction}

The presence of a mass for at least two neutrinos is nowadays a
well-established fact required by the observation of the neutrino oscillation phenomenon. However, the nature and the origin of this
mass is far from being understood. One possibility is to simply add to the
Standard Model (SM) particle content at least two
right-handed neutrinos, in order to give a Dirac mass to them, as for the
charged fermions. However, unless lepton number (LN) conservation is
promoted to a new fundamental symmetry of the theory, Majorana masses
for the right-handed neutrinos,
inducing LN violation, would also be allowed at the Lagrangian level. 

A rather appealing explanation for the origin of neutrino masses,
accounting not only for their presence, but also for their smallness,
are the Seesaw mechanisms, in which the new scale at which LN is
broken is assumed to be much larger than the electroweak (EW)
scale. Originally conceived with right-handed neutrinos (Type-I
Seesaw)~\cite{Minkowski:1977sc,Mohapatra:1979ia,Yanagida:1979as,GellMann:1980vs},
they have later been considered in extensions of
the SM involving scalar $SU(2)$ triplets (Type-II)~\cite{Magg:1980ut,Schechter:1980gr,Wetterich:1981bx,Lazarides:1980nt,Mohapatra:1980yp} and
fermionic $SU(2)$ triplets (Type-III)~\cite{Foot:1988aq}. Besides these minimal tree level Seesaw models, a plethora of radiative models, also linked to the breaking of lepton number, can generate neutrino masses at the loop-level~\cite{Zee:1980ai,Cheng:1980qt,Zee:1985rj,Zee:1985id,Babu:1988ki,Cai:2017jrq}.

In all these cases a new, heavy scale at which LN is violated is introduced and the smallness
of neutrino masses, which in these scenarios are Majorana
particles, is inversely proportional to this scale. Indeed, if the heavy degrees of freedom present at this scale are integrated out, the corresponding effective theory contains the
unique $d=5$ operator that can be constructed with SM
fields~\cite{PhysRevLett.43.1566}
\begin{equation}
\delta{\cal L}^{d=5} =\frac{1}{2}\, c_{\alpha \beta}^{d=5} \,
\left( \overline{\ell_L^c}_{\alpha} \tilde \phi^* \right) \left(
\tilde \phi^\dagger \, {\ell_L}_{ \beta} \right) + {\rm h.c}.\,  .
\label{eq:Weinberg}
\end{equation}
The coefficient in front of this operator depends on the high-energy
completion of the SM but it
is generically proportional to the
Yukawa coupling of the new particles with the lepton doublets $Y$ and inversely
proportional to the mass of the heavy particles being integrated out $M$. In the minimal Seesaw models, light neutrino masses are typically
obtained with $O(1)$ Yukawas and very heavy scales, $M\sim M_{GUT}\sim 10^{14}~\textrm{GeV}$.

This picture, even if very appealing as a rationale for the smallness of neutrino masses, has two drawbacks: the new large scale introduced induces a hierarchy problem for the Higgs mass~\cite{Vissani:1997ys} and it is also impossible to test for all practical purposes. Indeed, all the other phenomenological consequences of the Seesaw models
are driven by higher-dimensional operators. The leading new physics effects are expected to be generated by the $d=6$
operators, which are inversely proportional
to $M^2$, giving extremely small effects, of the order of
$m_\nu/M$. Moreover it is even more unlikely to directly produce and detect a particle of
such a huge mass.

In order to overcome these shortcomings while retaining a natural explanation of the smallness of neutrino masses, variants such as the inverse~\cite{Mohapatra:1986aw,Mohapatra:1986bd} or linear~\cite{Malinsky:2005bi} Seesaws were introduced. In these scenarios the new physics scale can be kept low and the new couplings can also be sizable since the
lightness of neutrino masses is instead guaranteed by a small parameter
breaking LN, which is otherwise a conserved symmetry. In this way the
new particles could be accessible at colliders and their effects in EW
physics, arising through the $d=6$ operator, can be sizable. These conditions can be realized in the context of the three minimal Seesaw models~\cite{Gavela:2009cd}.   

In Refs.~\cite{Kersten:2007vk, Abada:2007ux} it has been shown that, if one introduces only 3 or even 2
right-handed neutrinos, the approximate LN symmetry only allows
few textures for $Y$ and $M$, rendering the model quite constrained and predictive. This applies to the
Type-III Seesaw case as well, since the neutral component of the
triplet behaves exactly like a right-handed neutrino. In this paper we will focus on the Type-III Seesaw realizations 
which have better collider prospects since the triplets are charged and can be Drell-Yan produced, in contrast to the Type-I Seesaw singlets.
In particular, we will consider the scenarios with approximate LN conservation with only 2 or 3 triplets respectively, but also the most general scenario in which an arbitrary number of triplets is added without any further assumption. Our goal is
to update the low energy constraints on the general case, and to set bounds on the
restricted scenarios for the first time. Bounds on the general case
were firstly derived in Ref.~\cite{Abada:2007ux} and partially updated
in Ref.~\cite{He:2009tf},
while the other cases were considered mostly in relation to the LHC phenomenology~\cite{Franceschini:2008pz,Arhrib:2009mz,Li:2009mw,Eboli:2011ia,Agostinho:2017biv,Jana:2019tdm} so far. In Ref.~\cite{Kamenik:2009cb} bounds were derived for the general case with two triplets, imposing light neutrino mass constraints but without assuming approximate LN conservation.

 Both CMS~\cite{CMS:2012ra,Sirunyan:2017qkz} and ATLAS~\cite{ATLAS:2018ghc} have searched for fermionic triplets that decay like the ones introduced in the Type-III Seesaw model. In particular, CMS has set a lower bound on the heavy fermion triplet masses of 
$840$~GeV assuming flavor universality in the Yukawa couplings, while the bound
is relaxed to $390$~GeV if only couplings to the $\tau$ flavor are considered~\cite{Sirunyan:2017qkz}. However, even if a positive signal were found, this would not
be enough to claim that the Type-III Seesaw has been tested, because a
fingerprint of the relation with neutrino masses would still be
missing. On the other hand, in the concrete models with 2 and 3
triplets outlined above, imposing the generation of neutrino masses
implies stringent constraints on the parameters and
  consequently relations on
the Branching Ratios. Thus, in case of
a positive signal, it would be possible to test for this connection to the neutrino
mass generation mechanism at the LHC. Therefore, while the goal of this paper is
to set bounds on the considered models, its natural prosecution is the
analysis of LHC phenomenology, partially carried out in Ref.~\cite{Eboli:2011ia,Agostinho:2017biv}.

This paper is organized as follows. In Section~\ref{s:parametrization}, the parametrizations adopted for our analysis for the general and the two/three-triplet scenarios are introduced. In Section~\ref{s:observables}, the set of observables that have been used to probe for the new extra triplets is described. In Section~\ref{s:results} the results are presented and discussed. In Section~\ref{s:LHC} recasted LHC limits on the new physics scale of the two triplet scenario are presented, and finally, we conclude in Section~\ref{s:conclusions}.

\newpage

\section{Type-III Seesaw Scenarios and Parametrization}
\label{s:parametrization}

The Type-III Seesaw consists in the addition of $n$ extra $SU(2)$ fermion triplets with zero 
hypercharge $\Sigma_R=(\Sigma^1,\Sigma^2,\Sigma^3)$ to the Standard Model (SM) field content, leading to the following Lagrangian
\begin{equation}
\mathcal{L}=\mathcal{L}_\text{SM}+i\overline\Sigma_R\slashed{D}\Sigma_R-\dfrac{1}{2}\overline{\Sigma_R^i} \left(M_\Sigma\right)_{ij} \Sigma_R^{c\,j}-\left(Y_\Sigma\right)_{i \alpha}\overline{\Sigma_R^i}\tilde{\phi}^\dagger\tau \ell_L^\alpha+\text{h.c.}\,,
\label{eq:lagrangian}
\end{equation}
where $i=1,...,n$, $\tau = (\tau_1,\tau_2,\tau_3)$ are the Pauli matrices, $\tilde\phi=i\tau_2\phi^*$ with $\phi$ the SM Higgs field, $M_\Sigma$ is the allowed Majorana mass matrix for the triplets and $Y_\Sigma$ the Yukawa couplings between the 
triplets and the Higgs. After EW symmetry breaking, the vev of the Higgs, $v_\text{EW}$, 
generates Dirac neutrino masses $m_D=v_\text{EW} Y_\Sigma/\sqrt{2}$. The neutral 
eigenstates of the electric charge $\Sigma^0\equiv\Sigma^3$  mix with the active neutrinos playing
a similar role to the right-handed neutrinos in the Type-I Seesaw, while the remaining components 
$\Sigma^1, \Sigma^2$ combine into Dirac fermions that mix with the charged leptons.
In the usual Seesaw limit, for $M_\Sigma \gg m_D,v_\text{EW}$, the heavy triplets can be integrated out to obtain a series of non renormalizable effective operators suppressed
by the mass of the triplets. 
The least suppressed operator is the $d=5$ Weinberg operator \cite{PhysRevLett.43.1566} given by Eq.~(\ref{eq:Weinberg}) that generates Majorana masses for the active neutrinos after EW symmetry breaking
\begin{equation}
\label{eq:mhat}
\hat{m}  = - \dfrac{v_\text{EW}^2}{2}c_{\alpha\beta}^{d=5} \equiv -m_D^\top M_\Sigma^{-1} m_D = U^* m U^\dagger \,,
\end{equation}
where $m=\text{diag}\left(m_1,m_2,m_3\right)$ and $U=U_{23}\left(\theta_{23}\right)U_{13}\left(\theta_{13},\delta\right)U_{12}\left(\theta_{12}\right)\text{diag}\left(e^{-i\alpha_1/2},e^{-i\alpha_2/2},1\right)$ is 
the unitary mixing matrix that diagonalizes $\hat{m}$. 
At tree-level, only one $d=6$ operator is generated~\cite{Abada:2007ux}

\be\label{Ld6}
\delta{\cal L}^{d=6} = c^{d=6}_{\alpha \beta} \, \left( \overline{\ell_{L\alpha}} \vec\tau \tilde \phi
\right) i \D \left( \tilde \phi^\dagger \vec\tau \ell_{L \beta} \right),
\ee 
with 
\be\label{cLd6}
c^{d=6} = Y_\Sigma^\dagger \, \dfrac{1}{M_\Sigma^\dagger}\dfrac{1}{ M_\Sigma} \, Y_\Sigma \, .
\ee  
After symmetry breaking, the $d=6$ operator directly modifies the coupling between leptons and the $W$ and $Z$ bosons\footnote{Due to QED conservation, electromagnetic interactions are not affected.} and also induces non-canonical neutrino and charged lepton kinetic terms~\cite{Abada:2007ux}. Two unitary rotations are needed to diagonalize the $d=5$ and $d=6$ operators respectively. This, together with a rescaling to bring the neutrino and charged lepton kinetic terms to their canonical forms, finally induce non-unitary leptonic 
mixing, modifying not only the charged current (CC) lepton couplings but also the neutrino and charged lepton 
couplings to the $Z$. This leads to charged lepton Flavor Changing Neutral Currents (FCNC) 
already at tree-level, not present in the Type-I Seesaw case. In the basis in which the neutrino and 
lepton kinetic terms are canonical and their mass matrices diagonal and staying at leading order in the coefficient of the $d=6$ operator, their weak interactions read~\cite{Abada:2007ux}

\bea
\label{eq:Leff}
{\cal L}_{\rm eff}^{d\le 6}&=&{\cal L}_{\cal K}^l+{\cal L}_{\cal K}^\nu
+
\dfrac{g}{\sqrt{2}}\overline {l_L}{W\!\!\!\!\!/}\;^{-}\left(\mathbbm{1}+\eta\right)U\, \nu_L+
\nonumber\\
&+&
\dfrac{g}{c_\text{W}}\left\{-\dfrac{1}{2}
\overline{l_{L}}\,{Z\!\!\!\!/}\,(\mathbbm{1}+4\eta)\,l_L
+s^2_\text{W} \overline{l}\,{Z\!\!\!\!/}\;\;l
+\dfrac{1}{2}\overline{\nu_L} \,{Z\!\!\!\!/}
\left[U^\dagger(\mathbbm{1}-2\eta) \, U\right]\nu_L\,\right\} + \text{h.c.}
+... \,,
\eea
where $s_\text{W}\equiv \sin\theta_\text{W}$, $c_\text{W}\equiv \cos\theta_\text{W}$, and
$\eta$ is the coefficient of the $d=6$ operator given by the Hermitian matrix
\begin{equation}
\eta \equiv \dfrac{v_{EW}^2}{4}c^{d=6}=\dfrac{m_D^\dagger M_\Sigma^{-2} m_D}{2}\,.
\label{eq:dim6_eta} 
\end{equation}
Notice that the $W$ couplings to neutrinos and charged leptons are now characterized by the non-unitary mixing matrix $N\equiv \left(\mathbbm{1}+\eta\right)U$. 
In the decoupling limit of the triplets or when their couplings vanish $\eta \to 0$ and $N=U$ becomes the standard PMNS unitary matrix.

Non-unitarity is also often parameterized through an equivalent ``triangular'' 
parameterization~\cite{Xing:2007zj,Xing:2011ur,Escrihuela:2015wra}. This 
parameterization is more convenient in the context of neutrino oscillation analyses and can be easily mapped to 
the Hermitian parameterization that will be considered in this
work~\cite{Blennow:2016jkn}. Notice, however, that, apart from the
unitarity deviations in CC interactions, the Type-III Seesaw also
induces modifications in neutral current (NC) interactions both with neutrinos and charged leptons. Moreover, in the Type-I and Type-III Seesaw the deviations from unitarity induced due to the presence of heavy fermions have opposite sign. Indeed, in the Type-I Seesaw with heavy right-handed neutrinos $N$ of mass $M_N$ the mixing matrix $N=\left(\mathbbm{1}-\eta\right)U$ with
$\eta=\dfrac{m_D^\dagger M_N^{-2} m_D}{2}$, which is positive-definite by construction in both scenarios~. This difference stems from the fact that in the Type-III case the unitarity deviations not only arise from the mixing of neutrinos and charged leptons with the heavy fermions,
but also from a direct correction to the $W$ couplings via the operator in Eq.~(\ref{Ld6}), upon integrating out the heavy triplets which are in a non-trivial representation of $SU(2)$.
This also implies that the relation between CC and NC in the Type-III scenario is not simply a change of basis of the form $N^\dagger N$. Thus, NC matter effects in neutrino oscillations do not follow the usual pattern assumed in non-unitarity studies.

Below we will describe the three different Type-III Seesaw scenarios analyzed in the present work and its corresponding parameterizations.

\subsection{General scenario (G-SS)}

In this general scenario an arbitrary number of fermion triplets heavier than the EW scale is added to the field content 
without any further assumption. From now on we will refer to this unrestricted case as G-SS (general Seesaw). 
In this scenario, there are enough independent parameters for the coefficient of the $d=6$ operator $\eta$ to be completely independent from that of the 
$d=5$ that induces the light neutrino mass matrix $\hat{m}$. Therefore, $\eta$ cannot be constrained 
by the light neutrino masses and mixings~\cite{Broncano:2003fq, Antusch:2009gn}. 
We will thus treat all elements of $\eta$ as independent free parameters in this scenario. Nevertheless, Eq. (\ref{eq:dim6_eta}) 
shows that $\eta$ is a positive-definite matrix and the Schwartz inequality holds

\be
\label{eq:schwarz}
|\eta_{\alpha \beta}|\leq \sqrt{\eta_{\alpha \alpha}\eta_{\beta\beta}}~.
\ee

\subsection{Three triplets Seesaw scenario ($3\Sigma$-SS)}

We will also investigate the $ 3\Sigma$-SS case defined by the following requirements:

\begin{itemize}
\item only 3 fermion triplets are added to the SM;
\item the mass scale of the triplets is larger than the EW scale;
\item even though the neutrino masses are small, large, potentially observable $\eta$ is allowed;
\item the small neutrino masses are radiatively stable.
\end{itemize}

It turns out that the only way to simultaneously satisfy these requirements is through an underlying LN symmetry~\cite{Kersten:2007vk, Abada:2007ux,Eboli:2011ia,Agostinho:2017biv,Fernandez-Martinez:2015hxa}. In such a case we have
\begin{equation}
\label{eq:3TSSmasses}
m_D=\dfrac{v_{EW}}{\sqrt{2}}\left(\begin{matrix}
Y_{\Sigma_e} & Y_{\Sigma_\mu} & Y_{\Sigma_\tau} \\
\epsilon_1 Y_{\Sigma_e}'& \epsilon_1 Y_{\Sigma_\mu}' & \epsilon_1 Y_{\Sigma_\tau}' \\
\epsilon_2 Y_{\Sigma_e}''& \epsilon_2 Y_{\Sigma_\mu}'' & \epsilon_2 Y_{\Sigma_\tau}''
\end{matrix}\right)\qquad\text{and}\qquad 
M_\Sigma=\left(\begin{matrix}
\mu_1 & \Lambda & \mu_3 \\
\Lambda & \mu_2 & \mu_4 \\
\mu_3 & \mu_4 & \Lambda'
\end{matrix}\right),
\end{equation}
where $\epsilon_i$ and $\mu_i$ are small lepton number violating parameters. By setting all $\epsilon_i=0$ and $\mu_i=0$, LN symmetry is recovered if the 3 fermion triplets are assigned LN 1, $-1$ and 0 respectively. 
In this case, we find $m_i=0$ (3 massless neutrinos), $M_1=M_2=\sqrt{\Lambda^2+\sum_\alpha (|Y_{\Sigma_\alpha}|^2 v_{EW}^2/2)^2}$ 
(a heavy Dirac field) and $M_3=\Lambda'$ (a heavy decoupled Majorana fermion), where $m_i$ and $M_i$ are the mass eigenvalues of the full $6 \times 6$ mass matrix including $m_D$ and $M_\Sigma$. Substituting Eq.~(\ref{eq:3TSSmasses})
in Eq.~(\ref{eq:dim6_eta}), for the LN-conserving limit we find  

\begin{equation}
\label{eq:etaandthetas}
\eta=\dfrac{1}{2}\left(\begin{matrix}
\left|\theta_e\right|^2 & \theta_e\theta_\mu^* & \theta_e\theta_\tau^* \\
\theta_\mu\theta_e^* & \left|\theta_\mu\right|^2 & \theta_\mu\theta_\tau^* \\
\theta_\tau\theta_e^* & \theta_\tau\theta_\mu^* & \left|\theta_\tau\right|^2
\end{matrix}\right) \quad \text{with}\quad \theta_\alpha\equiv\dfrac{Y_{\Sigma_\alpha}v_{EW}}{\sqrt{2}\Lambda}\,.
\end{equation}
The parameters $\theta_\alpha$ represent the mixing of the active
neutrino $\nu_\alpha$ with the neutral component of the heavy Dirac
fermion triplet that is integrated out to obtain the $d=5$ and $d=6$ operators. As can be seen, given the underlying LN symmetry, this mixing, and hence the $d=6$ operator $\eta$, can be arbitrarily large while the $d=5$ operator is exactly zero and neutrinos are massless. In other words, the $d=5$ operator is protected by the LN symmetry while the $d=6$ is not and hence an approximate LN symmetry can alter the naive expectation of the $d=6$ operator being subdominant with respect to the $d=5$. 
When the small LN-violating parameters are not neglected so as to reproduce the correct pattern of masses and mixings, the following
relation has to be satisfied~\cite{Fernandez-Martinez:2015hxa}
\begin{eqnarray}
\theta_\tau&=&\dfrac{1}{\hat{m}_{e\mu}^2-\hat{m}_{ee}\hat{m}_{\mu\mu}}\Big(\theta_e\left(\hat{m}_{e\mu}\hat{m}_{\mu\tau}-\hat{m}_{e\tau}\hat{m}_{\mu\mu}\right)\nonumber\\
	&+&\theta_\mu\left(\hat{m}_{e\mu}\hat{m}_{e\tau}-\hat{m}_{ee}\hat{m}_{\mu\tau}\right)\pm\sqrt{\theta_e^2\hat{m}_{\mu\mu}-2\theta_e\theta_\mu\hat{m}_{e\mu}+\theta_\mu^2\hat{m}_{ee}} \label{eq:hellishRelation} \\
	&\times&\sqrt{\hat{m}_{e\tau}^2\hat{m}_{\mu\mu}-2\hat{m}_{e\mu}\hat{m}_{e\tau}\hat{m}_{\mu\tau}+\hat{m}_{ee}\hat{m}_{\mu\tau}^2+\hat{m}_{e\mu}^2\hat{m}_{\tau\tau}-\hat{m}_{ee}\hat{m}_{\mu\mu}\hat{m}_{\tau\tau}}\Big)\,,\nonumber
\end{eqnarray}
where only leading order terms in the Seesaw expansion have been considered and $\hat{m}$ contains the information on light neutrino masses and mixings through Eq.~(\ref{eq:mhat}).
This extra constraint leads to correlations among the $\eta$ matrix elements in Eq.~(\ref{eq:etaandthetas}) not present 
in the unrestricted scenario G-SS described above\footnote{Notice that
  the $\eta$ matrix also contains contributions driven by the small
  LN-violating parameters, $\epsilon_i$ and $\mu_i$, which are however subleading and thus neglected in our analysis.}. The value of the complex mixing $\theta_\tau$ is 
fixed by $\theta_e$ and $\theta_\mu$ through
Eq.~(\ref{eq:hellishRelation}), and by the SM neutrino masses and
mixings, that are encoded in the $d = 5$ operator
$\hat{m}$. Therefore, we will scan the allowed parameter space of the
model using as free parameters $\theta_e$ and $\theta_\mu$, as well as the unknown parameters that characterize $\hat m$, i.e. the Dirac phase $\delta$, the Majorana phases $\alpha_1 $ and $\alpha_2$, the smallest neutrino mass and the mass hierarchy (which can be normal or inverted). We will also consider in the fit the constraint on the sum of the light neutrino masses (from Planck) $\Sigma m_i < 0.12$ eV at 95\% CL \cite{Aghanim:2018eyx}, while we use the best fit values of the remaining oscillation parameters from Ref.~\cite{Esteban:2018azc} which are summarized in Table~\ref{tab:osc_params}.

\begin{table}[!t]

\begin{tabular}{|c|c|}
\hline
& Best fit $\pm 1\sigma$   
\\
 \hline
$\sin^2{\theta_{12}}$ & $0.310^{+0.013}_{-0.012}$ 
\\ 
\hline
$\sin^2{\theta_{23}}$ & $0.563^{+0.018}_{-0.024}$
\\ 
\hline           
$\sin^2{\theta_{13}}$ & $0.02237^{+0.00066}_{-0.00065}$ 
\\ \hline
$\Delta m_\text{sol}^2$& $7.39^{+0.21}_{-0.20}\cdot 10^{-5} \text{ eV}^2$ \\
\hline
$\vert \Delta m_\text{atm}^2\vert $ & $2.528^{+0.029}_{-0.031}\cdot 10^{-3} \text{ eV}^2$  \\ 
\hline
\end{tabular}
\caption{Present best fit values of the light neutrino mixing angles and two squared mass differences from~\cite{Esteban:2018azc}.}
\label{tab:osc_params}

\end{table}

\subsection{Two triplets Seesaw scenario ($2\Sigma$-SS)}

The $ 2\Sigma$-SS scenario  is a particular case of the $3\Sigma$-SS defined by the same conditions
but with the addition of only two fermion triplets instead of three. Notice that this is the most economic realization of the Type-III Seesaw model able to account for the two distinct mass splittings observed in neutrino oscillation experiments.

Analogously to the $3\Sigma$-SS scenario, for this case we have
\begin{equation}
\label{eq:2TSSmasses}
m_D=\dfrac{v_{EW}}{\sqrt{2}}\left(\begin{matrix}
Y_{\Sigma_e} & Y_{\Sigma_\mu} & Y_{\Sigma_\tau} \\
\epsilon_1 Y_{\Sigma_e}'& \epsilon_1 Y_{\Sigma_\mu}' & \epsilon_1 Y_{\Sigma_\tau}'
\end{matrix}\right)\qquad\text{and}\qquad 
M_\Sigma=\left(\begin{matrix}
\mu_1 & \Lambda \\
\Lambda & \mu_2 
\end{matrix}\right),
\end{equation}
where again $\epsilon_i$ and $\mu_i$ are small lepton number violating
parameters which, once set to zero, imply LN conservation if the two triplets have LN 1 and $-1$ respectively. In this limit the mass 
eigenvalues of the full $6\times6$ mass matrix are $M_1=M_2=\sqrt{\Lambda^2+\sum_\alpha (|Y_{\Sigma_\alpha}|^2 v_{EW}^2/2)^2}$ 
(which combine into a heavy Dirac pair) and the light neutrino masses vanish. Eq. (\ref{eq:etaandthetas}) still holds, 
showing that large $\eta$ entries are possible even in the LN-conserving limit with massless neutrinos. 
Analogously to the $3\Sigma$-SS scenario, upon switching on the LN-violating parameters in
Eq.~(\ref{eq:2TSSmasses}), neutrino masses and mixings $\hat{m}$ 
are generated. Given the reduced number of parameters in the Lagrangian, Eq. (\ref{eq:mhat}) now implies additional correlations 
\begin{eqnarray}
\theta_\mu &=& \dfrac{\theta_e}{\hat{m}_{ee}} \Big(\hat{m}_{e\mu}\pm i
               \sqrt{m_1
               m_2}\big(U^*_{13}U^*_{22}-U^*_{12}U^*_{23}\big)\Big)
               \quad \text{for normal hierarchy (NH)}\,, \nonumber \\
\theta_\mu &=& \dfrac{\theta_e}{\hat{m}_{ee}} \Big(\hat{m}_{e\mu}\pm i
               \sqrt{m_3
               m_2}\big(U^*_{12}U^*_{21}-U^*_{11}U^*_{22}\big)\Big)
               \quad \text{for inverted hierarchy (IH)}\,,
\label{eq:thetamu}
\end{eqnarray}
and
\begin{eqnarray}
\theta_\tau &=& \dfrac{\theta_e}{\hat{m}_{ee}} \Big(\hat{m}_{e\tau}\pm i \sqrt{m_1 m_2}\big(U^*_{13}U^*_{32}-U^*_{12}U^*_{33}\big)\Big) \quad \text{for} \quad \text{NH}\,, \nonumber \\
\theta_\tau &=& \dfrac{\theta_e}{\hat{m}_{ee}} \Big(\hat{m}_{e\tau}\pm i \sqrt{m_3 m_2}\big(U^*_{12}U^*_{31}-U^*_{11}U^*_{32}\big)\Big) \quad \text{for} \quad \text{IH}\,,
\label{eq:thetatau}
\end{eqnarray}
where both options for the sign in front of the square root are possible but the same choice has to be taken for $\theta_\mu$ and $\theta_\tau$.
 
Therefore, $\theta_\mu$ and $\theta_\tau$ are both proportional to $\theta_e$. In other words, once the known oscillation parameters are fixed to their best fit values, and the remaining unknown parameters\footnote{In this minimal scenario one of the light neutrino masses is zero and one of the Majorana phases is nonphysical ($\alpha_1$ can be set to zero).} characterizing $\hat{m}$ (the Dirac phase $\delta$, the Majorana phase $\alpha_2$ and the mass hierarchy) are specified, the $d=6$ operator is fixed up to an overall factor that we parametrize through $\theta_e$.\footnote{$\theta_e$ can thus be considered a real parameter since its associated phase becomes a global phase.} This same conclusion via a different parametrization was first derived in the context of the Type-I Seesaw in~\cite{Gavela:2009cd}, and applied to the Type-III Seesaw case in~\cite{Eboli:2011ia,Agostinho:2017biv}. 

The parameters characterizing the low energy new physics effects and correlations among them in each of the three cases described in this section are summarized in Table~\ref{tab:params}.

\begin{table}[!t]
\begin{center}
\begin{tabular}{|c|c|c|c|c|c|c|}
\hline
& $\eta_{ee}$  & $\eta_{\mu\mu}$ & $\eta_{\tau\tau}$ & $\eta_{e\mu}$ & $\eta_{e\tau}$ & $\eta_{\mu\tau}$ \\
 \hline
\multirow{2}{*}{G-SS} & $\eta_{ee}>0$ & $\eta_{\mu\mu}>0$ & $\eta_{\tau\tau}>0$ & $\left|\eta_{e\mu}\right|\leq\sqrt{\eta_{ee}\eta_{\mu\mu}}$ & $\left|\eta_{e\tau}\right|\leq\sqrt{\eta_{ee}\eta_{\tau\tau}}$ & $\left|\eta_{\mu\tau}\right|\leq\sqrt{\eta_{\mu\mu}\eta_{\tau\tau}}$ \\
& free & free & free & free & free & free \\ 
\hline
\multirow{2}{*}{3$\Sigma$-SS} & $\eta_{ee}=\dfrac{\left|\theta_e\right|^2}{2}$ & $\eta_{\mu\mu}=\dfrac{\left|\theta_\mu\right|^2}{2}$ & $\eta_{\tau\tau}=\dfrac{\left|\theta_\tau\right|^2}{2}$  & $\eta_{e\mu}=\dfrac{\theta_e\theta_\mu^*}{2}$ & $\eta_{e\tau}=\dfrac{\theta_e\theta_\tau^*}{2}$ & $\eta_{\mu\tau}=\dfrac{\theta_\mu\theta_\tau^*}{2}$ \\
& free & free & fixed by Eq.~(\ref{eq:hellishRelation}) & fixed by $\theta_e$, $\theta_\mu$ & fixed by $\theta_e$, $\theta_\tau$ & fixed by $\theta_\mu$, $\theta_\tau$ \\ 
\hline
\multirow{2}{*}{2$\Sigma$-SS} & $\eta_{ee}=\dfrac{\theta_e^2}{2}$ & $\eta_{\mu\mu}=\dfrac{\left|\theta_\mu\right|^2}{2}$ & $\eta_{\tau\tau}=\dfrac{\left|\theta_\tau\right|^2}{2}$  & $\eta_{e\mu}=\dfrac{\theta_e\theta_\mu^*}{2}$ & $\eta_{e\tau}=\dfrac{\theta_e\theta_\tau^*}{2}$ & $\eta_{\mu\tau}=\dfrac{\theta_\mu\theta_\tau^*}{2}$                  \\
& free & fixed by Eq.~(\ref{eq:thetamu}) & fixed by Eq.~(\ref{eq:thetatau}) & fixed by $\theta_e$, $\theta_\mu$ & fixed by $\theta_e$, $\theta_\tau$ & fixed by $\theta_\mu$, $\theta_\tau$ \\ \hline
\end{tabular}
\caption{Summary of the parameters that characterize the low energy new physics effects of
a totally general Type-III Seesaw model (G-SS), and the realizations with 3  and 
2 additional heavy triplets (3$\Sigma$-SS and 2$\Sigma$-SS
respectively). $\eta$ is the coefficient of the $d=6$ operator while
$\theta_\alpha$ corresponds to the mixing between $\nu_\alpha$ and the
neutral component of the heavy fermion triplets. In the 3$\Sigma$-SS case, $\theta_\tau$ is calculated via Eq.~(\ref{eq:hellishRelation}) as 
a function of $\theta_e$, $\theta_\mu$, $\delta$, $\alpha_1$, $\alpha_2$, the lightest neutrino mass and the
mass hierarchy. In the 2$\Sigma$-SS, $\theta_\mu$ and $\theta_\tau$ are computed via 
Eq.~(\ref{eq:thetamu}) and Eq.~(\ref{eq:thetatau}) respectively as functions of $\theta_e$, $\delta$, $\alpha_2$ 
and the mass hierarchy. The remaining oscillation parameters are fixed to their best fit values shown in Table~\ref{tab:osc_params}.}
\label{tab:params}
\end{center}
\end{table}

\section{Observables}
\label{s:observables}

In this section the leading order dependence on $\eta_{\alpha\beta}$ of the most constraining electroweak and flavor observables is presented and discussed. The SM loop corrections are relevant for these precision observables, and have therefore been taken into account~\cite{Tanabashi:2018oca} in the results presented in Section~\ref{s:results}. However, in order to simplify the discussion, they will not be included in the analytic expressions of the observables presented in this section as we are interested in highlighting the corrections stemming from $\eta_{\alpha\beta}$ instead. On the other hand, the 1-loop contributions of the new degrees of freedom are not expected to play an important role on the determination of the bounds on $\eta_{\alpha\beta}$, and therefore they will be neglected~\cite{Fernandez-Martinez:2015hxa}.

Finally, all the observables will be given in terms of $\alpha$, $M_Z$
and $G_F$ as measured in $\mu$ decay,
$G_\mu$~\cite{Tanabashi:2018oca}, making the SM predictions in terms
of this three parameters
\begin{eqnarray}
\alpha&=&\left(7.2973525698\pm0.0000000024\right)\cdot 10^{-3}\,, \nonumber\\
M_Z&=&\left(91.1876\pm0.0021\right) \text{ GeV}\,, \\
G_\mu&=&\left(1.1663787\pm0.0000006\right)\cdot 10^{-5} \text{ GeV}^{-2}\,. \nonumber
\end{eqnarray}

A summary of all the observables considered for the numerical fit is provided in Section~\ref{summary}, where we refer the reader not interested in the details of the $\eta_{\alpha\beta}$ dependence of each observable to be discussed in the following.

\subsection{Constraints from $\mu$-decay: $M_{W}$ and $\theta_{\text{W}}$}

\begin{figure}
\centering
\includegraphics[width=0.34\textwidth]{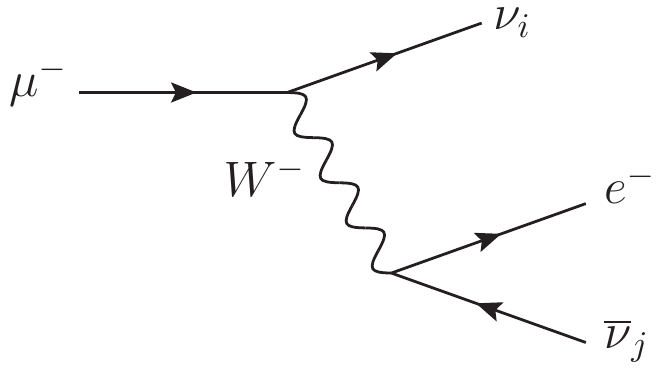} \quad \quad
\includegraphics[width=0.34\textwidth]{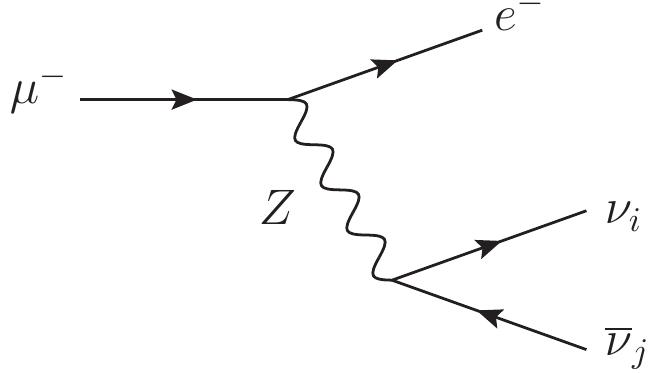}
\caption{Diagrams contributing to the $\mu \to e \nu_i \bar{\nu}_j$
  decay rate. At leading order in $\eta_{\alpha\beta}$, the
  contribution of the LFV decay mediated by the $Z$ boson is
  subleading and will be neglected.}
\label{fig:muon_decay}
\end{figure}

The presence of new degrees of freedom modifies not only the CC interactions, but also the NC interactions leading to charged Lepton Flavor Violation (LFV) already at tree level (see Eq.~(\ref{eq:Leff})). Therefore, the expected decay rate of $\mu \to e \nu_i \bar{\nu}_j$ will receive contributions mediated by both $W$ and $Z$ bosons, as shown in Figure~\ref{fig:muon_decay}. The contribution mediated by the $Z$ boson is however proportional to $\vert \eta_{\mu e}\vert^2$, and thus subleading with respect to the linear $\eta_{ee}$ and $\eta_{\mu \mu}$ corrections present in the $W$ exchange
\begin{equation}
\Gamma_{\mu}
= \dfrac{m_{\mu}^5 G_{F}^{2}}{192 \pi^3}\left(1+2\eta_{ee}+2\eta_{\mu\mu}\right)\equiv
\dfrac{m_{\mu}^5 G_{\mu}^{2}}{192 \pi^3}\, .
\end{equation}
Thus, the determination of $G_F$ itself through the muon decay acquires a correction from the $d=6$ operator $\eta_{\alpha\beta}$ that will affect all other electroweak observables
\begin{equation}
G_{F}=G_{\mu}\left(1-\eta_{ee}-\eta_{\mu\mu}\right)\,.
\label{eq:Gmu}
\end{equation}

In particular, the relation between $G_F$ and $M_W$ allows to
constrain the elements $\eta_{ee}$ and $\eta_{\mu \mu}$ through
kinematic measurements of $M_W$ together with $M_Z$ and $\alpha$,
unaffected by $\eta_{\alpha\beta}$
\begin{equation}
G_\mu = \dfrac{\alpha \pi M^2_Z\left(1+\eta_{ee}+\eta_{\mu\mu}\right)}{\sqrt{2}M^2_W\left(M^2_Z - M^2_W \right)}\,.
\label{eq:MW}
\end{equation}
\begin{table}[!t]
\centering
\begin{tabular}{|c|c|c|}
\hline
Observable & SM prediction & Experimental value \\
\hline
\hline
$M_{W}= M_{W}^{\text{SM}}\left(1-0.20\left(\eta_{ee}+\eta_{\mu\mu}\right)\right)$ & $\left(80.363\pm 0.006\right)$ GeV & $\left(80.379\pm0.012\right)$ GeV\\
\hline
\end{tabular}
\caption{The $W$ boson mass as input parameter to the fit: in the first column, the leading order corrections on the new physics parameters $\eta$; in the second, the SM prediction (including loop corrections) of $M_W$; in the third, the experimental value used in the fit.} 
\label{Tab:obs:MW:values}
\end{table}

In an analogous way, the weak mixing angle $s_{\text{W}}^2$ determined by $G_\mu$, $M_Z$ and $\alpha$ acquires a dependence on $\eta_{ee}$ and $\eta_{\mu \mu}$ 
\begin{equation}
s_{W}^{2}=\dfrac{1}{2}\left(1-\sqrt{1-\dfrac{2\sqrt{2}\alpha\pi}{G_{\mu}
M_{Z}^{2}}\left(1+\eta_{ee}+\eta_{\mu\mu}\right)}\right)\,.
\label{eq:sw}
\end{equation}
Thus, processes containing $Z$ boson couplings to quarks or charged leptons allow to further constrain these parameters.

\subsection{Constraints from $Z$ decays}

The different precision measurements performed by LEP and SLC at the $Z$ peak become a powerful tool to study the extra contributions to the $Z$ couplings under the presence of heavy fermion triplets. Among the possible observables containing $Z$ decays, we found that the invisible decay of the $Z$, 6 rates of $Z$ decays into different charged fermions, and 6 $Z$-pole asymmetries are the most relevant for the fit. Table~\ref{Tab:obs:Z:values} summarizes the expressions of these 13 observables at leading order in $\eta_{\alpha\beta}$, together with the SM predictions and their experimental values.

\subsubsection{Invisible $Z$ decay}

The invisible Z decay is corrected directly via the non-diagonal $Z$ coupling to neutrinos and indirectly via Eq.~(\ref{eq:Gmu})
\begin{equation}
\Gamma_{\text{inv}}
= \dfrac{G_{\mu}M_{Z}^{3}}{12\sqrt{2}\pi}\Big(3-7\left(\eta_{ee}+\eta_{\mu\mu}\right)-4\eta_{\tau\tau}\Big) \equiv \dfrac{G_{\mu}M_{Z}^{3} N_{\nu}}{12\sqrt{2}\pi}\,,
\end{equation}
where $N_\nu = 2.990 \pm 0.007$ is the number of active neutrinos determined through the invisible $Z$ decay measured by LEP~\cite{ALEPH:2005ab}.

\begin{table}[!ht]
\centering
\begin{tabular}{|c|c|c|}
\hline
Observable & SM prediction & Experimental value \\
\hline
\hline
$\Gamma_\text{inv}= \Gamma_\text{inv}^{\text{SM}}\left(1-2.33\left(\eta_{ee}+\eta_{\mu\mu}\right) -1.33\eta_{\tau\tau}\right)$ & $\left(0.50144\pm0.00004\right)$ GeV& $\left(0.4990\pm0.0015\right)$ GeV\\
\hline
$R_{e}= R_{e}^{\text{SM}}\left(1-8.83\eta_{ee}-0.26\eta_{\mu\mu}\right)$ & $20.737\pm 0.010$ & $20.804\pm0.050$\\
$R_{\mu}= R_{\mu}^{\text{SM}}\left(1-0.26\eta_{ee}-8.83\eta_{\mu\mu}\right)$ & $20.740\pm 0.010$ & $20.785\pm0.033$\\
$R_{\tau}= R_{\tau}^{\text{SM}}\left(1-0.26\left(\eta_{ee}+\eta_{\mu\mu}\right)-8.57\eta_{\tau\tau}\right)$ & $20.782\pm 0.010$ & $20.764\pm0.045$\\
$R_{c}= R_{c}^{\text{SM}}\left(1-0.12\left(\eta_{ee}+\eta_{\mu\mu}\right)\right)$ & $0.17221\pm 0.00003$ & $0.1721\pm0.0030$\\
$R_{b}= R_{b}^{\text{SM}}\left(1+0.06\left(\eta_{ee}+\eta_{\mu\mu}\right)\right)$ & $0.21582\pm0.00002$ & $0.21629\pm0.00066$\\
$\sigma^0_\text{had}=\sigma_\text{had}^{0\text{ SM}}\left(1+8.55\eta_{ee}-0.02\eta_{\mu\mu}-0.04\eta_{\tau\tau}\right)$ & $\left(41.481\pm 0.008\right)$ nb & $\left(41.541\pm0.037\right)$ nb\\
\hline
$A_e = A_e^\text{SM}\left(1 +30.6\eta_{ee} -16.5\eta_{\mu\mu} \right)$ & $0.1469 \pm0.0003$ & $0.1515\pm0.0019$ \\
$A_\mu = A_\mu^\text{SM}\left(1 -16.5\eta_{ee} +30.6\eta_{\mu\mu}\right)$ & $0.1469 \pm0.0003$ & $0.142\pm0.015$ \\
$A_\tau = A_\tau^\text{SM}\left(1 -16.5\eta_{ee} -16.5\eta_{\mu\mu} +47.1\eta_{\tau\tau} \right)$ & $0.1469 \pm0.0003$ & $0.143\pm0.004$ \\
$A_b = A_b^\text{SM}\left(1 -0.22\left(\eta_{ee} +\eta_{\mu\mu}\right) \right)$ & $0.9347 \pm 0.0001$ & $0.923\pm0.020$ \\
$A_c = A_c^\text{SM}\left(1 -1.66\left(\eta_{ee} +\eta_{\mu\mu}\right) \right)$ & $0.6677\pm0.0001$ & $0.670\pm0.027$ \\
$A_s = A_s^\text{SM}\left(1 -0.22\left(\eta_{ee} +\eta_{\mu\mu}\right) \right)$ & $0.9356 \pm 0.0001$ & $0.90\pm0.09$ \\
\hline
\end{tabular}
\caption{List of flavor conserving observables containing $Z$ couplings included in the global fit: in the first column, the leading order corrections on the parameters $\eta$; in the second, the SM predictions (including loopcorrections)~\cite{Tanabashi:2018oca}; in the third, the experimental values~\cite{Tanabashi:2018oca} used in the fit.}
\label{Tab:obs:Z:values}
\end{table}

\subsubsection{$Z$ decays into charged fermions}

The charged lepton NC interactions are also modified  with respect to the SM (see Eq.~(\ref{eq:Leff})) and thus the decay rate of $Z$ into charged leptons, $\Gamma\left(Z\rightarrow l_\alpha\bar{l}_\alpha\right)\equiv\Gamma_{l}$, is directly sensitive to $\eta_{\alpha\alpha}$
\be
\Gamma_{l}=
\dfrac{G_{\mu}M_{Z}^{3}}{3\sqrt{2}\pi}
\left\{ \left[s_\text{W}^4 +\left(s_\text{W}^2-\dfrac{1}{2}\right)^2\right]\left(1-\eta_{ee}-\eta_{\mu\mu}\right)
+ 4\left(\dfrac{1}{2}-s_\text{W}^2\right)\eta_{\alpha\alpha} \right\}\,,
\label{eq:Z:l_widths} 
\ee
where the indirect $\eta_{ee}$ and $\eta_{\mu\mu}$ corrections from
the determination of $G_F$ in muon decays have been explicitly added, while $s_\text{W}^2$ implicitly introduces extra corrections via Eq.~(\ref{eq:sw}).
 
On the other hand, even though the $Z$ boson couplings to quarks remain the same as in the SM, the decay rates of $Z$ into quarks $\Gamma\left(Z\rightarrow q\bar{q}\right)\equiv\Gamma_{q}$ present indirect corrections from $G_F$ and $s_\text{W}$

\begin{equation}
\Gamma_{q}=
\dfrac{3 G_{\mu}M_{Z}^{3}\left(\left(T_{q}-2Q_{q}s_{\text{W}}^{2}\right)^2+T_{q}^2\right)}{2\sqrt{2}\pi}\left(1-\eta_{ee}-\eta_{\mu\mu}\right)\,,
\label{eq:Z:q_widths}
\end{equation}
where $Q_q$ and $T_q$ are the electric charge and the third component of the weak isospin of the given quark $q$, respectively. Here, $\eta_{ee}$ and $\eta_{\mu \mu}$ corrections are also implicit in $s_{\text{W}}^{2}$.

These $Z$ decay rates into charged fermions are combined to construct the following observables
\begin{equation}
R_{q}=\dfrac{\Gamma_{q}}{\Gamma_{\text{had}}}\,,\quad 
R_{l}=\dfrac{\Gamma_{\text{had}}}{\Gamma_{l}}\,, \quad \text{and}\quad
\sigma_{\text{had}}^{0}=\dfrac{12 \pi\Gamma_{e}\Gamma_{\text{had}}}{M_{Z}^{2}\Gamma_{Z}^{2}}\,;
\end{equation} 
with $\Gamma_{\text{had}}\equiv\displaystyle\sum_{q\neq t}\Gamma_{q}$, and where $\Gamma_Z=\Gamma_{e}+\Gamma_{\mu}+\Gamma_{\tau}+\Gamma_{\text{inv}}+\Gamma_{\text{had}}$ is the total $Z$ width. 

\subsubsection{$Z$ asymmetry parameters}

Measurements of the $Z$-pole asymmetries, made by the LEP collaborations and by SLD at SLAC, are additional observables which include the polarization and the forward-backward asymmetry. These observables are ultimately
sensitive to the combination (see for instance~\cite{Tanabashi:2018oca})
\be
A_f= \dfrac{2 g_V^f g_A^f}{(g_V^f)^2+(g_A^f)^2}\, .
\label{Af}
\ee
In particular, including the Type-III Seesaw corrections at leading order, the corresponding expression for charged leptons with flavor $\alpha$ is given by
\be
A_\alpha= \dfrac{1-4s_{W}^{2}}{1-4s_{W}^{2}+8s_{W}^{4}}+\dfrac{64s_{W}^{4}(1-2 s_{W}^{2})}{(1-4s_{W}^{2}+8s_{W}^{4})^2}\,\eta_{\alpha\alpha},
\label{Al}
\ee
where $s_{W}^{2}$ implicitly introduces extra $\eta_{ee}$ and $\eta_{\mu \mu}$ corrections via Eq.~(\ref{eq:sw}). In the
quark case only these indirect corrections are present
\be
A_q= \dfrac{T_q (T_q -2Q_q s_{W}^{2})}{T_q^2 +2Q_q^2 s_{W}^{4}-2Q_qT_qs_{W}^{2}}.
\label{Aq}
\ee

\subsection{Constraints from weak interaction universality tests}

Lepton flavor universality of weak interactions can be probed for measuring ratios of decay rates of charged leptons, $W$, or mesons into charged leptons of different flavors. These decay rates mediated by the $W$ boson acquire corrections proportional to $(1 + 2\eta_{\alpha \alpha})$, where $\alpha$ is the flavor of the corresponding charged lepton. By doing the ratio between different flavors, 
the uncertainties of the common variables involved in the two decays cancel out. The weak interaction universality ratios, given by 
\begin{equation}
\dfrac{\Gamma^P_\alpha}{\Gamma^P_\beta} \equiv \dfrac{\Gamma^P_\alpha\vert_\text{SM}}{\Gamma^P_\beta\vert_\text{SM}} \left(R^{P}_{\alpha\beta}\right)^2
= 
\dfrac{\Gamma^P_\alpha\vert_\text{SM}}{\Gamma^P_\beta\vert_\text{SM}} \left( 1+2\eta_{\alpha \alpha}-2\eta_{\beta\beta} \right)\,,
\label{eq:univ}
\end{equation}
become thus powerful observables to indirectly probe for the existence of heavy fermion triplets. In the above equation, the phase space, chirality flip factors and SM loop corrections are encoded in $\Gamma^P_\alpha\vert_\text{SM}$, the SM expectation of the decay width of the parent particle $P$ involving a charged fermion of flavor $\alpha$. The decay rates containing the SM loop corrections from~\cite{Pich:2013lsa} have been used to derived the experimental constraints on $R^P_{\alpha \beta}$ shown in Table~\ref{Tab:obs:universality:values}.

\begin{table}[!ht]
\centering
\begin{tabular}{|c|c|c|}
\hline
Observable & SM prediction & Experimental value \\
\hline
\hline
$R^\pi_{\mu e}= \left(1+\left(\eta_{\mu\mu}-\eta_{ee}\right)\right)$ & 1 & $1.0042\pm0.0022$ \\
$R^\pi_{\tau \mu}= \left(1+\left(\eta_{\tau\tau}-\eta_{\mu\mu}\right)\right)$ & 1 & $0.9941\pm0.0059$\\
$R^W_{\mu e}= \left(1+\left(\eta_{\mu\mu}-\eta_{ee}\right)\right)$ & 1 & $0.992\pm0.020$\\
$R^W_{\tau \mu}= \left(1+\left(\eta_{\tau\tau}-\eta_{\mu\mu}\right)\right)$ & 1 & $1.071\pm0.025$\\
$R^K_{\mu e}= \left(1+\left(\eta_{\mu\mu}-\eta_{ee}\right)\right)$ & 1 & $0.9956\pm0.0040$\\
$R^K_{\tau \mu}= \left(1+\left(\eta_{\tau\tau}-\eta_{\mu\mu}\right)\right)$ & 1 & $0.978\pm0.014$\\
$R^l_{\mu e}= \left(1+\left(\eta_{\mu\mu}-\eta_{ee}\right)\right)$ & 1 & $1.0040\pm 0.0032$\\
$R^l_{\tau \mu}= \left(1+\left(\eta_{\tau\tau}-\eta_{\mu\mu}\right)\right)$ & 1 & $1.0029\pm0.0029$ \\
\hline
\end{tabular}
\caption{List of universality ratios considered for the global fit: in
  the first column, the leading order corrections on the $\eta$
  parameters; in the second, the SM prediction; in the third, the experimental values~\cite{Pich:2013lsa} used in the fit.}
\label{Tab:obs:universality:values}
\end{table}

\subsection{Unitarity of the CKM matrix}

Even though the Unitarity of the CKM quark mixing matrix is not directly affected in the Type-III Seesaw, the elements of the CKM matrix $V_{qq^\prime}$ are measured through processes which are modified by the new degrees of freedom. These modifications will happen not only in a direct way via Eq.~(\ref{eq:Leff}), but also in an indirect way through the determinations of $G_\mu$ and $s_\text{W}$ via Eq.~(\ref{eq:Gmu}) and Eq.~(\ref{eq:sw}), as discussed above. 

Starting from the Unitarity relation among the three elements $V_{uq^\prime}$ 
of the CKM matrix, the following relation between $\vert V_{ud}\vert$
and $\vert V_{us}\vert$ is obtained:
\begin{equation}
\vert V_{ud}\vert=\sqrt{1-\vert V_{us}\vert^{2}}\,,
\label{eq:udus}
\end{equation}
where $V_{ub} = \left(4.13\pm0.49\right)\times 10^{-3}$~\cite{Tanabashi:2018oca} has been neglected since $|V_{ub}|^2$ is much smaller than the present accuracy on $|V_{us}|^2$.

In the following, the dependence on the $\eta_{\alpha \beta}$ parameters of the different processes used to constrain the CKM elements $\vert V_{ud}\vert$ and $\vert V_{us}\vert$ will be discussed. These observables will be incorporated in the global fit as a function of $V_{us}$ via Eq.~(\ref{eq:udus}). Finally, $|V_{us}|$ will be treated as a nuisance parameter of the global fit, choosing its value in such a way that $\chi^2$ is minimized for each value of the involved $\eta_{\alpha \beta}$ parameters.   

Table~\ref{Tab:obs:CKM:values} summarizes the dependence on the $\eta_{\alpha \beta}$ parameters of the 9 observables constraining the CKM Unitarity used in the global fit and their experimental values. 

\subsubsection{Determination of $\left|V_{ud}\right|$ via Superallowed $\beta$ decays}

The best determination of $\left|V_{ud}\right|$ comes from Superallowed $0^+\to 0^+$ nuclear $\beta$ decays. It receives both a direct 
correction with $\left(1+\eta_{ee}\right)$ from the CC coupling to $e$ in Eq.~(\ref{eq:Leff}), and an indirect one from $G_F$ via Eq.~(\ref{eq:Gmu}), resulting in the following expression
\begin{equation}
\left|V_{ud}^{\beta}\right|=
\left(1-\eta_{\mu\mu}\right)\left|V_{ud}\right|\,.
\label{eq:betadec}
\end{equation}

Table~\ref{Tab:obs:CKM:values} shows the value of $\left|V_{ud}^{\beta}\right|$ based on the 20 different Superallowed $\beta$ transitions~\cite{Hardy:2018zsb}.

\begin{table}[!t]
\centering
\begin{tabular}{|c|c|c|}
\hline
Observable & SM prediction & Experimental value \\
\hline
\hline
$\left|V_{ud}^\beta\right| = \sqrt{1-|V_{us}|^2}(1-\eta_{\mu\mu})$ & $\sqrt{1-|V_{us}|^2}$ & $0.97420\pm0.00021$~\cite{Hardy:2018zsb}\\
\hline
$\left|V_{us}^{\tau\rightarrow K\nu}\right|= \left|V_{us}\right|\left(1-\eta_{ee}-\eta_{\mu\mu}+\eta_{\tau\tau}\right)$ & $\left|V_{us}\right|$ & $0.2186\pm0.0021$~\cite{Amhis:2016xyh} \\
$\left|V_{us}^{\tau\rightarrow K,\pi}\right|=\left|V_{us}\right|\left(1-\eta_{\mu\mu}\right)$ & $\left|V_{us}\right|$ & $0.2236\pm0.0018$~\cite{Amhis:2016xyh} \\
$\left|V_{us}^{K_{L}\rightarrow \pi e\overline{\nu}}\right|=\left|V_{us}\right|\left(1-\eta_{\mu\mu}\right)$ & $\left|V_{us}\right|$ & $0.2237\pm0.0011$ \cite{Antonelli:2010yf}\\
$\left|V_{us}^{K_{L}\rightarrow \pi \mu\overline{\nu}}\right|=\left|V_{us}\right|\left(1-\eta_{ee}\right)$ & $\left|V_{us}\right|$ & $0.2240\pm0.0011$ \cite{Antonelli:2010yf}\\
$\left|V_{us}^{K_{S}\rightarrow \pi e\overline{\nu}}\right|=\left|V_{us}\right|\left(1-\eta_{\mu\mu}\right)$ & $\left|V_{us}\right|$ & $0.2229\pm0.0016$ \cite{Antonelli:2010yf}\\
$\left|V_{us}^{K^{\pm}\rightarrow \pi e\overline{\nu}}\right|=\left|V_{us}\right|\left(1-\eta_{\mu\mu}\right)$ & $\left|V_{us}\right|$ & $0.2247\pm0.0012$ \cite{Antonelli:2010yf}\\
$\left|V_{us}^{K{\pm}\rightarrow \pi \mu\overline{\nu}}\right|=\left|V_{us}\right|\left(1-\eta_{ee}\right)$ & $\left|V_{us}\right|$ & $0.2245\pm0.0014$ \cite{Antonelli:2010yf}\\
$\left|V_{us}^{K,\pi \rightarrow \mu\nu}\right| =\left|V_{us}\right|\left(1-\eta_{\mu\mu}\right)$ & $\left|V_{us}\right|$ & $0.2315\pm0.0010$ \cite{Moulson:2014cra}\\
\hline
\end{tabular}
\caption{List of observables testing the Unitarity of the CKM matrix
  used as input for the global fit: in the first column,
  the leading order corrections on the $\eta$ parameters; in the second, the SM predictions; in the third, the experimental values~\cite{Hardy:2018zsb,Amhis:2016xyh,Antonelli:2010yf,Moulson:2014cra} used in the fit.}
\label{Tab:obs:CKM:values}
\end{table}

\subsubsection{Determination of $\left|V_{us}\right|$}

$\left|V_{us}\right|$ is presently determined via the measurement of $\tau$ decays into $K$ or $\pi$
and semileptonic and leptonic $K$ decays. In this work the values of the form factor $f_{+}(0)$ and decay constant $f_{K}/f_{\pi}$ involved in these processes have been taken from Ref.~\cite{Aoki:2016frl}. 

\begin{itemize}

\item{Via $\tau$ decays}

The $\tau\rightarrow K\nu$ decay rate is proportional to the CKM element $\left|V_{us}\right|$ and 
is directly corrected via the $\tau$ CC coupling and indirectly via $G_\mu$ 
\begin{equation}
\vert V_{us}^{\tau\rightarrow K\nu}\vert=
\left(1-\eta_{ee}-\eta_{\mu\mu}+\eta_{\tau\tau}\right)\vert V_{us}\vert\,.
\end{equation}
Notice that the present experimental value of $\left|V_{us}^{\tau\rightarrow K\nu}\right|$ given in Table~\ref{Tab:obs:CKM:values} is in tension with other determinations~\cite{Amhis:2016xyh}. \\

A complementary way to constrain $\vert V_{us}\vert$ is via the ratio $Br\left(\tau\rightarrow K\nu\right)/Br\left(\tau\rightarrow \pi\nu\right)$. In this case, the dependence on the $\eta_{\alpha\beta}$ parameters cancels out. Thus, this observable, sensitive to $\vert V_{us}\vert/\vert V_{ud}\vert$, remains unaffected by the presence of extra degrees of freedom. When combined with $\vert V_{ud}^\beta\vert$ from Eq.~(\ref{eq:betadec}), we obtain for $\left|V_{us}^{\tau\to K,\pi }\right|$

\begin{equation}
\left|V_{us}^{\tau\rightarrow K,\pi}\right|=\left(1-\eta_{\mu\mu}\right)\left|V_{us}\right|\,.
\label{eq:Vus_tau_decay}
\end{equation}

\item{Via $K$ decays}

In the decay rate $K\to \pi l_\alpha \overline{\nu}_\alpha$ (with $\alpha=\mu,e$), 
the direct $\eta_{\mu \mu}$ ($\eta_{ee}$) correction from the $W$ coupling to $\mu$ ($e$) cancels with the indirect $\eta_{\mu \mu}$ ($\eta_{ee}$) one introduced by $G_\mu$, leading to the following dependence on the new physics parameters
\begin{eqnarray}
\left|V_{us}^{K\rightarrow \pi e\overline{\nu}}\right|&=&
\left(1-\eta_{\mu\mu}\right)\left|V_{us}\right|\,,\\
\left|V_{us}^{K\rightarrow \pi \mu\overline{\nu}}\right|&=&
\left(1-\eta_{ee}\right)\left|V_{us}\right|\,.
\end{eqnarray}
The present determinations of $\left|V_{us}^{K\rightarrow \pi e\overline{\nu}}\right|$ and $\left|V_{us}^{K\rightarrow \pi \mu\overline{\nu}}\right|$ come from measurements of the different decays of $K_L$, $K_S$ and $K^\pm$ listed in Table \ref{Tab:obs:CKM:values}.
These observables have been included in the $\chi^2$ of the global fit taking into account the correlation matrix~\cite{Antonelli:2010yf} among them.\\

Finally, as in the $\left|V_{us}^{\tau\to K,\pi }\right|$ case discussed above, the ratio $Br\left(K \rightarrow \mu \nu \right)/Br\left(\pi \rightarrow \mu \nu \right)$ is sensitive to $\vert V_{us}\vert/\vert V_{ud}\vert$ and independent of $\eta$ . However, when the information of $\vert V_{ud}^\beta\vert$ from Eq.~(\ref{eq:betadec}) is introduced, an indirect dependence on $\eta_{\mu\mu}$ is induced
\begin{equation}
\left|V_{us}^{K,\pi \rightarrow \mu\nu }\right|=\left(1-\eta_{\mu\mu}\right)\left|V_{us}\right|\,.\\
\end{equation}

\end{itemize}

\subsection{LFV observables}
\label{subsec:LFV}

In the Type-III Seesaw LFV processes can occur already at
tree-level and they are driven by the off-diagonal $\vert \eta_{\alpha
  \beta}\vert$ parameters. The Lepton Flavour Conserving (LFC) processes discussed above constrain, on the other hand, the diagonal parameters $\vert \eta_{\alpha \alpha}\vert$. In principle, these are two separate set of bounds since they constrain a priori independent parameters. However, $\eta$ is a positive-definite matrix, and thus its off-diagonal elements $\eta_{\alpha \beta}$ are bounded by the diagonal ones via the Schwarz inequality given in Eq.~(\ref{eq:schwarz}). Thus, the off-diagonal elements of $\eta$ are bounded both directly via the LFV processes that will be discussed in this section, and indirectly 
via LFC processes. Furthermore, in the 3$\Sigma$-SS and 2$\Sigma$-SS cases, the Schwarz inequality is saturated to $|\eta_{\alpha \beta}| = \sqrt{\eta_{\alpha \alpha} \eta_{\beta \beta}}$ (see Eq.~(\ref{eq:etaandthetas})) and thus the LFV observables will be 
included together with the LFC ones in the global fit.\\

In the following, the most relevant LFV processes are described. Notice that since these observables are already proportional to $\vert \eta_{\alpha \beta}\vert^2$, the additional dependence on $\eta$ from $G_\mu$ and $s_\text{W}$ is subleading and, therefore, it will be neglected in the remainder of this section. Table~\ref{Tab:LFV:values} summarizes the present experimental bounds and future sensitivities to $\vert\eta_{\alpha \beta}\vert$ associated to each LFV process. 

\begin{table}[!t]
\centering
\begin{tabular}{|c|c|c|}
\hline
Observable & Present bound on $\vert\eta_{\alpha \beta}\vert$ & Future sensitivity on $\vert\eta_{\alpha \beta}\vert$ \\
\hline
\hline
$\boldsymbol{\mu\rightarrow e}$ \textbf{(Ti)} & $\boldsymbol{\vert \eta_{\mu e}\vert <3.0\cdot 10^{-7}}$~\cite{Dohmen:1993mp} & $\boldsymbol{\vert \eta_{\mu e}\vert <1.4\cdot 10^{-10}}$~\cite{Barlow:2011zza} \\
\hline
$\mu\to eee$ & $\vert \eta_{\mu e}\vert < 8.7\cdot 10^{-7}$~\cite{Tanabashi:2018oca} & $\vert \eta_{\mu e}\vert < 1.1\cdot 10^{-8}$~\cite{Blondel:2013ia}\\
$\tau\to eee$ & $\vert \eta_{\tau e}\vert < 3.4 \cdot 10^{-4}$~\cite{Tanabashi:2018oca} & $\vert \eta_{\tau e}\vert < 2.9 \cdot 10^{-5}$~\cite{OLeary:2010hau}\\
$\tau\to \mu\mu\mu$ & $\vert \eta_{\tau \mu}\vert < 3.0\cdot 10^{-4}$~\cite{Tanabashi:2018oca} & $\vert \eta_{\tau \mu}\vert < 2.9\cdot 10^{-5}$~\cite{OLeary:2010hau} \\
$\boldsymbol{\tau\to e\mu\mu}$ & $\boldsymbol{\vert \eta_{\tau e}\vert < 3.0 \cdot 10^{-4}}$~\cite{Tanabashi:2018oca} & $\boldsymbol{\vert \eta_{\tau e}\vert < 2.6 \cdot 10^{-5}}$~\cite{OLeary:2010hau} \\
$\boldsymbol{\tau\to \mu ee}$ & $\boldsymbol{\vert \eta_{\tau \mu}\vert < 2.5 \cdot 10^{-4}}$~\cite{Tanabashi:2018oca} & $\boldsymbol{\vert \eta_{\tau \mu}\vert < 2.6 \cdot 10^{-5}}$~\cite{OLeary:2010hau} \\
\hline
$Z\to \mu e$ & $\vert \eta_{\mu e}\vert < 8.5 \cdot 10^{-4}$~\cite{Tanabashi:2018oca} & \textemdash \\
$Z\to \tau e$ & $\vert \eta_{\tau e}\vert < 3.1\cdot 10^{-3}$~\cite{Tanabashi:2018oca} & \textemdash \\
$Z\to \tau \mu$ & $\vert \eta_{\tau \mu}\vert < 3.4\cdot 10^{-3}$~\cite{Tanabashi:2018oca} & \textemdash \\
\hline
$h\to \mu e$ & $\vert \eta_{\mu e}\vert < 0.54$~\cite{Tanabashi:2018oca} & \textemdash\\
$h\to \tau e$ & $\vert \eta_{\tau e}\vert < 0.14$~\cite{Tanabashi:2018oca} & \textemdash\\
$h\to \tau \mu$ & $\vert \eta_{\tau \mu}\vert < 0.20$~\cite{Tanabashi:2018oca} & \textemdash \\
\hline
$\mu\rightarrow e \gamma$ & $\vert \eta_{\mu e}\vert < 1.1\cdot 10^{-5}$~\cite{Tanabashi:2018oca} & $\vert \eta_{\mu e}\vert < 5.3\cdot 10^{-6}$~\cite{Baldini:2018nnn}\\
$\tau\rightarrow e \gamma$ & $\vert \eta_{\tau e}\vert < 7.2\cdot 10^{-3}$~\cite{Tanabashi:2018oca} & $\vert \eta_{\tau e}\vert < 1.2\cdot 10^{-3}$~\cite{Bona:2007qt}\\
$\tau\rightarrow \mu \gamma$ & $\vert \eta_{\tau \mu}\vert < 8.4 \cdot 10^{-3}~$\cite{Tanabashi:2018oca} & $\vert \eta_{\tau \mu}\vert < 1.2 \cdot 10^{-3}$~\cite{Bona:2007qt}\\
\hline
\end{tabular}
\caption{ Summary of the present constraints and expected future sensitivities for the most relevant LFV observables considered. The corresponding present and future bounds on $\vert\eta_{\alpha\beta}\vert$ have been computed assuming no correlations among the elements of the matrix $\eta$ and are shown at $2\sigma$. The dominant limits and future sensitivities are highlighted in bold face. 
}
\label{Tab:LFV:values}
\end{table}

\subsubsection{$\mu\rightarrow e$ conversion in nuclei}

The branching ratio for $\mu-e$ conversion in nuclei with atomic
number $Z$ and mass number $A\lesssim 100$ normalized to the total
nuclear muon capture rate $\Gamma_\text{capt}$ is given
by~\cite{Bernabeu:1993ta}
\begin{equation}
R^{\mu\to e}_{^{A}_{Z}X}= \dfrac{2\,G_\mu^2\alpha^3 m_\mu^5 Z_\text{eff}^4}{\pi^2 \Gamma_\text{capt}}\vert F(q)\vert^2 \dfrac{Q_\text{W}^2}{Z} \vert \eta_{\mu e}\vert^2\,,
\label{eq:mu_e_conversion}
\end{equation}
where $Z_\text{eff}$ is the effective atomic number due to the
screening effect, $F(q)$ is the nuclear form factor as measured from
electron scattering, and
\begin{equation}
Q_\text{W}=\left(A+Z\right) \left(1-\dfrac{8}{3}s^2_\text
{W}\right)+\left(2A-Z\right) \left(-1+\dfrac{4}{3}s^2_\text
{W}\right)\,.
\end{equation}
$\Gamma_\text{capt}$ is also measured experimentally with good precision and therefore, information on $\eta_{\mu e}$ can be extracted from Eq.~(\ref{eq:mu_e_conversion}) in a nuclear-model independent fashion. The strongest experimental bound on this LFV transition is stated by $\mu$ to $e$ conversion 
in $^{48}_{22}$Ti, measured by the experiment SINDRUM II~\cite{Dohmen:1993mp}. In this case, $Z_\text{eff}\simeq 17.6$ and $F\left(q\simeq-m_\mu\right)\simeq0.54$~\cite{Bernabeu:1993ta}, $\Gamma_\text{capt}\simeq \left(2.590 \pm 0.012\right) \cdot 10^6\text{ s}^{-1}$~\cite{Suzuki:1987jf}, and Eq.~(\ref{eq:mu_e_conversion}) thus reads
\begin{equation}
R^{\mu\to e}_{^{48}_{22}\text{Ti}}\simeq 58.88 \vert \eta_{\mu e}\vert^2 \, .
\end{equation}
The corresponding bound on $\vert\eta_{\mu e}\vert$ 
is shown in Table~\ref{Tab:LFV:values}. This is the strongest present
bound on $\vert\eta_{\mu e}\vert$. The future
PRISM/PRIME~\cite{Barlow:2011zza} sensitivity to this parameter is
expected to be improved by three orders of magnitude.

\subsubsection{LFV 3-body lepton decays}

The LFV decay rate of a lepton with flavor $\alpha$ into three leptons with flavor $\beta\neq \alpha$ is given by 
\begin{equation}
\Gamma \left(l_\alpha \to l_\beta l_\beta \bar{l}_\beta\right)=
\dfrac{G_\mu^2 s^4_\text{W} m_{\alpha}^5}{6\pi^3}\vert\eta_{\alpha
  \beta}\vert^2 \, ,
\end{equation}
while the $\tau$ decay rate into two leptons with flavor $\beta\neq\tau$ and one with 
flavor $\alpha\neq\beta,\tau$ reads
\begin{equation}
\Gamma\left(\tau \to l_\alpha \bar{l}_\beta l_\beta\right) = \dfrac{G_\mu^2m_\tau^5\left(1-4 s^2_\text{W}+8 s^4_\text{W}\right)}{48\pi^3} \vert \eta_{\tau \alpha}\vert^2.
\end{equation}
The corresponding bounds are listed in Table~\ref{Tab:LFV:values}. The $\tau$ decays into three leptons set the present and near-future most constraining bounds on $\vert\eta_{\tau e}\vert$ and $\vert\eta_{\tau \mu}\vert$. 

\subsubsection{LFV $Z$ and $h$ decays}

The decay rate of $Z$ and Higgs fields into two leptons of different flavors is given by
\begin{equation}
\Gamma \left(Z \to l_\alpha^\pm l_\beta^\mp\right) = \dfrac{4M_Z^3 G_\mu}{3 \sqrt{2}\pi} \vert \eta_{\alpha \beta} \vert^2 \,,
\label{eq:LFV_Z_decay}
\end{equation}
and
\begin{equation}
\Gamma \left( h\to l_\alpha^\pm l_\beta^\mp\right) = \dfrac{9 G_\mu M_h}{2 \sqrt{2} \pi} \left( m_{\alpha}^2+m_{\beta}^2\right) \vert \eta_{\alpha \beta} \vert^2 \,,
\label{eq:LFV_h_decay}
\end{equation}
respectively. Although these processes are not competitive with respect to LFV lepton decays and $\mu\rightarrow e$ conversion in nuclei, we list the bounds on $\vert\eta_{\alpha\beta}\vert$ they would lead to in Table~\ref{Tab:LFV:values}.

\subsubsection{Radiative decays}

Finally, radiative decays of the type $l_\alpha\to l_\beta\gamma$ would be induced at loop-level~\cite{Abada:2007ux,Abada:2008ea}
\begin{equation}
\dfrac{\Gamma\left(l_\alpha\to l_\beta\gamma\right)}{\Gamma\left(l_\alpha\to\l_\beta \nu_\alpha \nu_\beta\right)}\simeq \dfrac{3\alpha}{8\pi}\left(\dfrac{13}{3}-6.56\right)^2\vert\eta_{\alpha\beta}\vert^2\,.
\end{equation}
The present bound on $|\eta_{\mu e}|$ from MEG~\cite{TheMEG:2016wtm} together with the bounds from
$\tau\to e \gamma$ and $\tau\to \mu \gamma$ are listed in
Table~\ref{Tab:LFV:values}. Even though in other extensions of the
SM $\mu\to e \gamma$ is the dominant LFV channel, in the Type-III
Seesaw other LFV transitions, as $\mu$ to $e$ conversion in nuclei, can already occur at tree level and therefore set more stringent bounds than those stemming from radiative decays.

\subsection{Summary}
\label{summary}

Summarizing, the following set of 43 observables will be used to derive the most updated global constraints on the Type-III Seesaw parameters:
\begin{itemize}
\item the mass of the $W$ boson: $M_W$;
\item the invisible width of the $Z$: $\Gamma_\text{inv}$;
\item 6 ratios of $Z$ fermionic decays: $R_e$, $R_\mu$, $R_\tau$, $R_c$, $R_b$ and $\sigma^0_\text{had}$;
\item 6 $Z$ asymmetry parameters: $A_e$, $A_\mu$, $A_\tau$, $A_b$, $A_c$ and $A_s$;
\item 8 ratios of weak decays constraining EW universality: $R^\pi_{\mu e}$, $R^\pi_{\tau \mu}$, $R^W_{\mu e}$, $R^W_{\tau \mu}$, $R^K_{\mu e}$, $R^K_{\tau \mu}$, $R^l_{\mu e}$ and $R^l_{\tau \mu}$;
\item 9 weak decays constraining the CKM Unitarity;
\item 12 LFV processes: $\mu$ to $e$ conversion in Ti, 5 rare lepton decays, 3 radiative decays and 3 $Z$ LFV decays.
\end{itemize}
The expressions for these observables in the Type-III Seesaw model, the SM predictions and the experimental values are given in Tables~\ref{Tab:obs:MW:values}-\ref{Tab:LFV:values}.

\section{Results and discussion}
\label{s:results}

With all the observables introduced in the previous section, a $\chi^2$ function has been built under the assumption of gaussianity so as to test the bounds that the experimental constraints 
can set globally on the Type-III Seesaw parameters. In order to achieve an efficient exploration of the parameter space of the three scenarios introduced in Section~\ref{s:parametrization}, particularly for the G-SS and 3$\Sigma$-SS scenarios which are characterized by a relatively large number of parameters, a Markov chain Monte Carlo (MCMC) method has been employed. 

For each scenario, $\mathcal{O}(10^7)$ distinct samples have been generated through 20 chains running simultaneously, achieving a convergence for all the free parameters better than $R - 1 < 0.0008$~\cite{Gelman:1992zz}. These scans are sufficiently well-sampled to allow a frequentist analysis by profiling the $\chi^2$ function of the different subsets of parameters, obtaining the 1D and 2D contours for the preferred regions. The processing of the chains has been performed with the MonteCUBES~\cite{Blennow:2009pk} user interface. 

As discussed in Section~\ref{s:parametrization}, the $\theta_\alpha$ parameters are not independent in the 3$\Sigma$-SS and 2$\Sigma$-SS, and the information from neutrino oscillations needs to be considered so as to reproduce the correct neutrino masses and mixings. Thus, $\theta_{12}$, $\theta_{13}$, $\theta_{23}$, $\Delta m_\text{sol}^2$, and $\Delta m_\text{atm}^2$, have been fixed to their best fit values listed in Table~\ref{tab:osc_params}, while the Dirac phase $\delta$, the Majorana phases $\alpha_1 $ and $\alpha_2$, and the lightest neutrino mass are free parameters in the scan\footnote{In the 2$\Sigma$-SS, the lightest neutrino is exactly massless and therefore $\alpha_1$ is unphysical.}, with a constraint on the sum of the light neutrino masses (from Planck) $\Sigma m_i < 0.12$ eV at 95\% CL \cite{Aghanim:2018eyx}.

Even though present oscillation data show a preference for some values of $\delta$, at the $2\sigma$ level about half of the parameter space is still allowed and there is some tension between the values favoured by T2K and NO$\nu$A analyses. Thus, we allow $\delta$ to vary freely in the fit. Moreover, in the NH case, both for the 3$\Sigma$-SS and 2$\Sigma$-SS scenarios, we have verified that if instead of fixing the mass splittings and mixing angles to their best fit values we introduce them as free parameters with their corresponding priors from Ref.~\cite{Esteban:2018azc}, the change in the results is negligible. Finally, even though neutrino oscillation data presently disfavor IH at more than $2\sigma$, in order to illustrate the impact of the mass hierarchy in our analysis, we present our results for both IH and NH.

We will show our individual constraints on the $d=6$ effective operator coefficients $\eta_{\alpha\beta}$, but also the 2D allowed regions projected in the $\sqrt{2 \eta_{\alpha\alpha}}-\sqrt{2 \eta_{\beta\beta}}$ planes for the three cases under study. In the 3$\Sigma$-SS and 2$\Sigma$-SS scenarios $\sqrt{2\eta_{\alpha\alpha}}=\theta_\alpha$
is the mixing of the active neutrino $\nu_\alpha$ with the neutral component of the heavy Dirac fermion triplet (see Eq.~(\ref{eq:etaandthetas})). Therefore, our results in the G-SS projected on the $\sqrt{2 \eta_{\alpha\alpha}}-\sqrt{2 \eta_{\beta\beta}}$ can be easily compared with the corresponding bounds on the mixing for the the 3$\Sigma$-SS and 2$\Sigma$-SS scenarios. For completeness, the individual bounds on $\sqrt{2|\eta_{\alpha\beta}|}$ (the mixing $|\theta_{\alpha}|$ in the 3$\Sigma$-SS and 2$\Sigma$-SS cases) are also reported in Appendix~\ref{s:Appendix-B}.

\subsection{General scenario (G-SS)}

\begin{figure}
\centering
\includegraphics[width=0.45\textwidth]{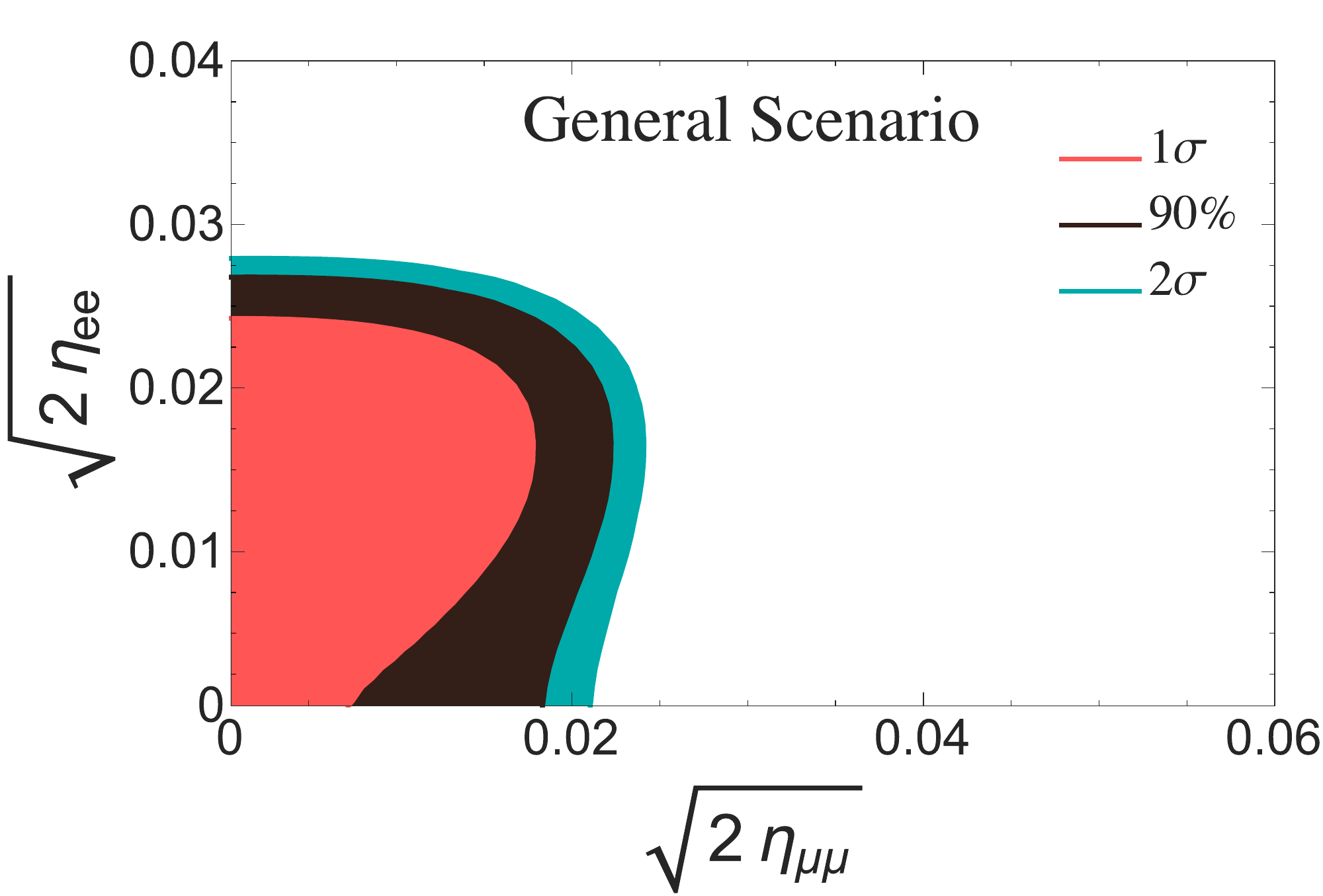}
\includegraphics[width=0.45\textwidth]{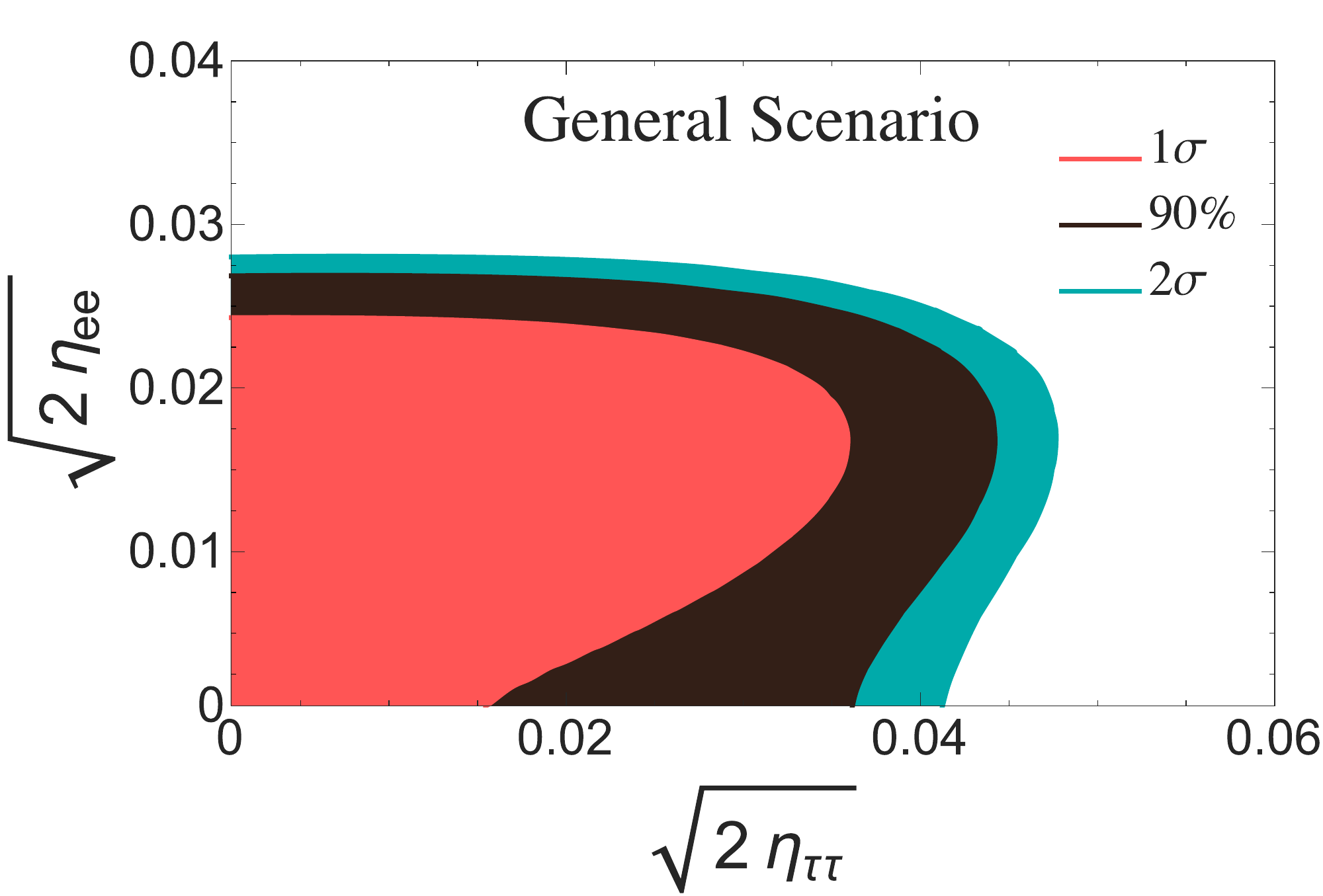}
\caption{ Frequentist confidence intervals at $1 \sigma$, $90\%$ and $2 \sigma$ on the parameter space of the G-SS.}
\label{fig:contours-GSS}
\end{figure}

In this scenario all the elements of the $d=6$ operator $\eta_{\alpha\beta}$ are independent free parameters in the fit. In Fig.~\ref{fig:contours-GSS} we present the 2 dof frequentist allowed regions of the parameter space 
at $1\sigma$, 90\%, and $2\sigma$ in red, black and cyan, respectively, projected in the $\sqrt{2 \eta_{ee}}-\sqrt{2 \eta_{\mu\mu}}$ (left panel) and $\sqrt{2 \eta_{ee}}-\sqrt{2 \eta_{\tau\tau}}$ (right panel) planes. The individual constraints on each $|\eta_{\alpha\beta}|$ at $2\sigma$ after profiling over all other parameters are summarized in Table~\ref{tab:bounds_eta}. With the information on the diagonal elements, and due to the fact that $\eta$ is a positive-definite Hermitian matrix,  
bounds on the off-diagonal elements can be derived via Eq.~(\ref{eq:schwarz}). These values, collected in the first column of Table~\ref{tab:bounds_eta}, are derived from data sets independent from the LFV processes discussed in Sec.~\ref{subsec:LFV} and thus we will refer to them as LFC. 
In the second column (LFV) of Table~\ref{tab:bounds_eta} we show the present constraints on the off-diagonal parameters directly derived from the set of LFV observables considered in Sec.~\ref{subsec:LFV}.
Regarding the bounds on the diagonal parameters, both $|\eta_{e e}|$
and $|\eta_{\mu\mu}|$ are $\mathcal{O}\left(10^{-4}\right)$ while the
constraint on $|\eta_{\tau\tau}|$ is $\sim 3-4$ times weaker. The
constraints from the LFC and LFV independent sets of data are remarkably
similar in magnitude for the $\eta_{\tau e}$ and $\eta_{\tau\mu}$
elements, $\mathcal{O}\left(10^{-4}\right)$, being the LFV ones
slightly more constraining. For the $\eta_{\mu e}$ element, however,
the extremely stringent constraint from $\mu$ to $e$ conversion in nuclei allows to set an $\mathcal{O}\left(10^{-7}\right)$ upper bound, three orders of magnitude stronger than the one derived from the LFC data set.

\begin{table}[!t]
\centering
\begin{tabular}{|c||c|c||c|c||c|c|}
\hline
{\multirow{2}{*}{}} & \multicolumn{2}{c||}{G-SS} & \multicolumn{2}{c||}{$3\Sigma$-SS} & \multicolumn{2}{c|}{$2\Sigma$-SS} \\
\cline{2-7}
 & LFC & LFV & NH & IH & NH & IH \\ 
\hline
\hline
$\eta_{ee}$ & $\boldsymbol{<3.2\cdot 10^{-4}}$ & \textemdash & $\boldsymbol{<3.1\cdot 10^{-4}}$ & $<3.2\cdot 10^{-4}$ & $\boldsymbol{<2.3\cdot 10^{-7}}$ & $<1.4\cdot 10^{-5}$ \\
$\eta_{\mu\mu}$ & $\boldsymbol{<2.1\cdot 10^{-4}}$ & \textemdash & $\boldsymbol{<1.4\cdot 10^{-4}}$ & $<1.1\cdot 10^{-4}$ & $\boldsymbol{<3.8\cdot 10^{-6}}$ & $<1.1\cdot 10^{-6}$ \\
$\eta_{\tau\tau}$ & $\boldsymbol{<8.5\cdot 10^{-4}}$ & \textemdash & $\boldsymbol{<6.5\cdot 10^{-4}}$ & $<3.9\cdot 10^{-4}$ & $\boldsymbol{<6.1\cdot 10^{-6}}$ & $<1.4\cdot 10^{-6}$\\
\hline
$\eta_{\mu e}$ & $<2.0\cdot 10^{-4}$ & $\boldsymbol{<3.0\cdot 10^{-7}}$ & $\boldsymbol{<3.0\cdot 10^{-7}}$ & $<3.0\cdot 10^{-7}$ & $\boldsymbol{<3.0\cdot 10^{-7}}$ & $<3.0\cdot 10^{-7}$\\
$\eta_{\tau e}$ & $<4.1\cdot 10^{-4}$ & $\boldsymbol{<2.7\cdot 10^{-4}}$ & $\boldsymbol{<2.5\cdot 10^{-4}}$ & $<2.3\cdot 10^{-4}$ & $\boldsymbol{<5.4\cdot 10^{-7}}$ & $<3.6\cdot 10^{-6}$\\
$\eta_{\tau\mu}$ & $<2.8\cdot 10^{-4}$ & $\boldsymbol{<2.2\cdot 10^{-4}}$ & $\boldsymbol{<1.4\cdot 10^{-4}}$ & $<1.3\cdot 10^{-4}$ & $\boldsymbol{<3.6\cdot 10^{-6}}$ & $<1.2\cdot 10^{-6}$\\
\hline
\end{tabular}
\caption{The $2\sigma$ constraints on the coefficient of the $d=6$ operator $\eta$ are shown. For the G-SS, the off-diagonal entries are bounded in two independent ways: indirectly from LFC observables via Eq.~(\ref{eq:schwarz}) and directly from LFV processes. The bold face highlights the most constraining G-SS bounds. For the $3\Sigma$-SS and $2\Sigma$-SS, the constraints are shown separately for normal hierarchy (NH) and inverted hierarchy (IH). The constraints for the NH scenarios are highlighted in bold face since NH provides a better fit to present neutrino oscillation data, while IH is disfavored at more than $2\sigma$~\cite{Esteban:2018azc}. The corresponding bounds on the mixing $\theta_{\alpha}$ between $\nu_{\alpha}$ and the heavy triplets are shown in Appendix~\ref{s:Appendix-B}.}
\label{tab:bounds_eta}
\end{table}

\subsection{Three and two triplets Seesaw scenarios (3$\Sigma$-SS and 2$\Sigma$-SS)}

When the number of fermion triplets is $\leq 3$, Eq.~(\ref{eq:schwarz})
is saturated to an equality
$\vert\eta_{\alpha\beta}\vert=\sqrt{\eta_{\alpha\alpha}\eta_{\beta\beta}}$, and thus the LFV processes, which a priori constrain only the off-diagonal elements of $\eta$,  will also contribute to the bounds on the diagonal elements.

\begin{figure}[!t]
\centering
\includegraphics[width=0.42\textwidth]{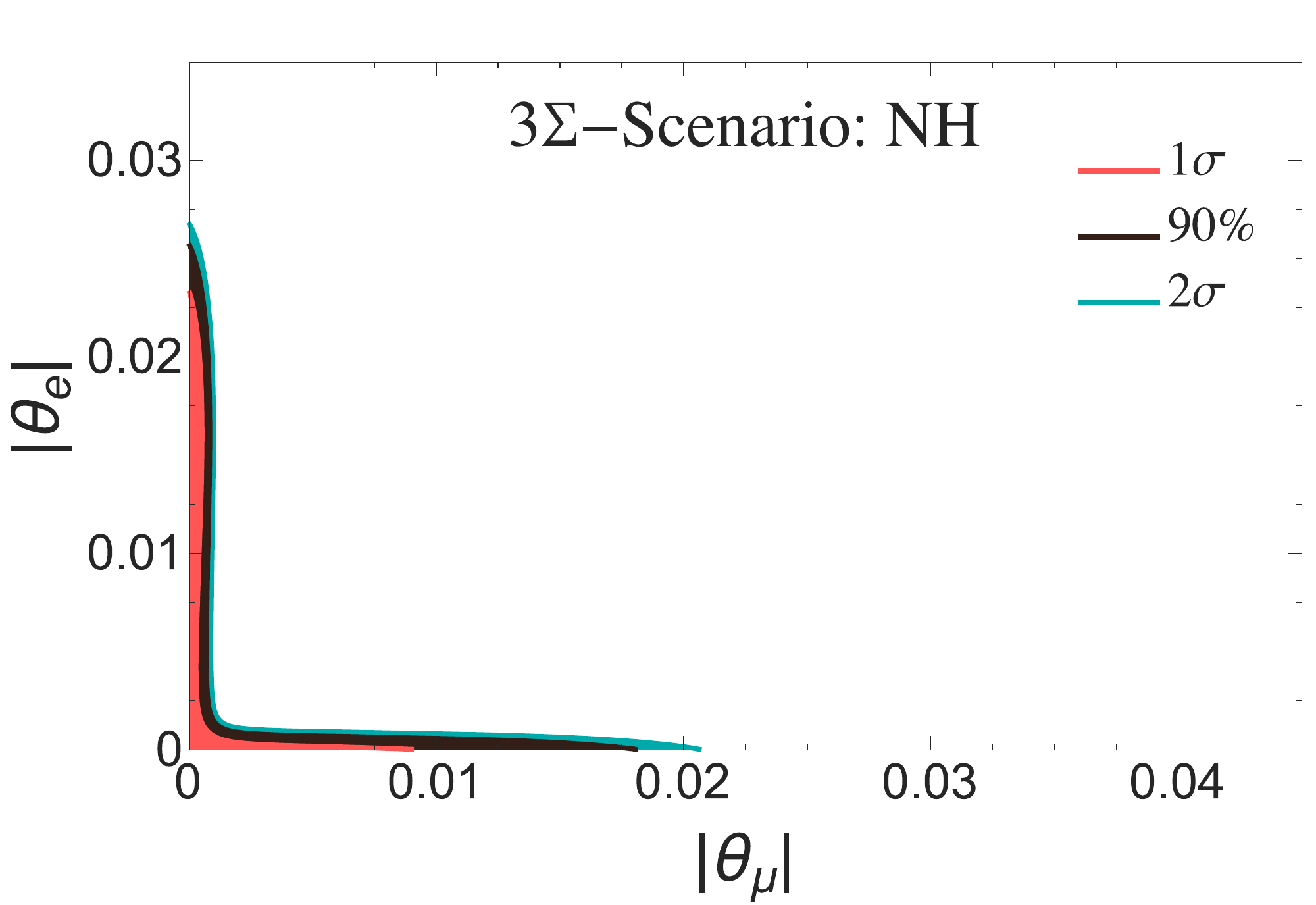}
\includegraphics[width=0.42\textwidth]{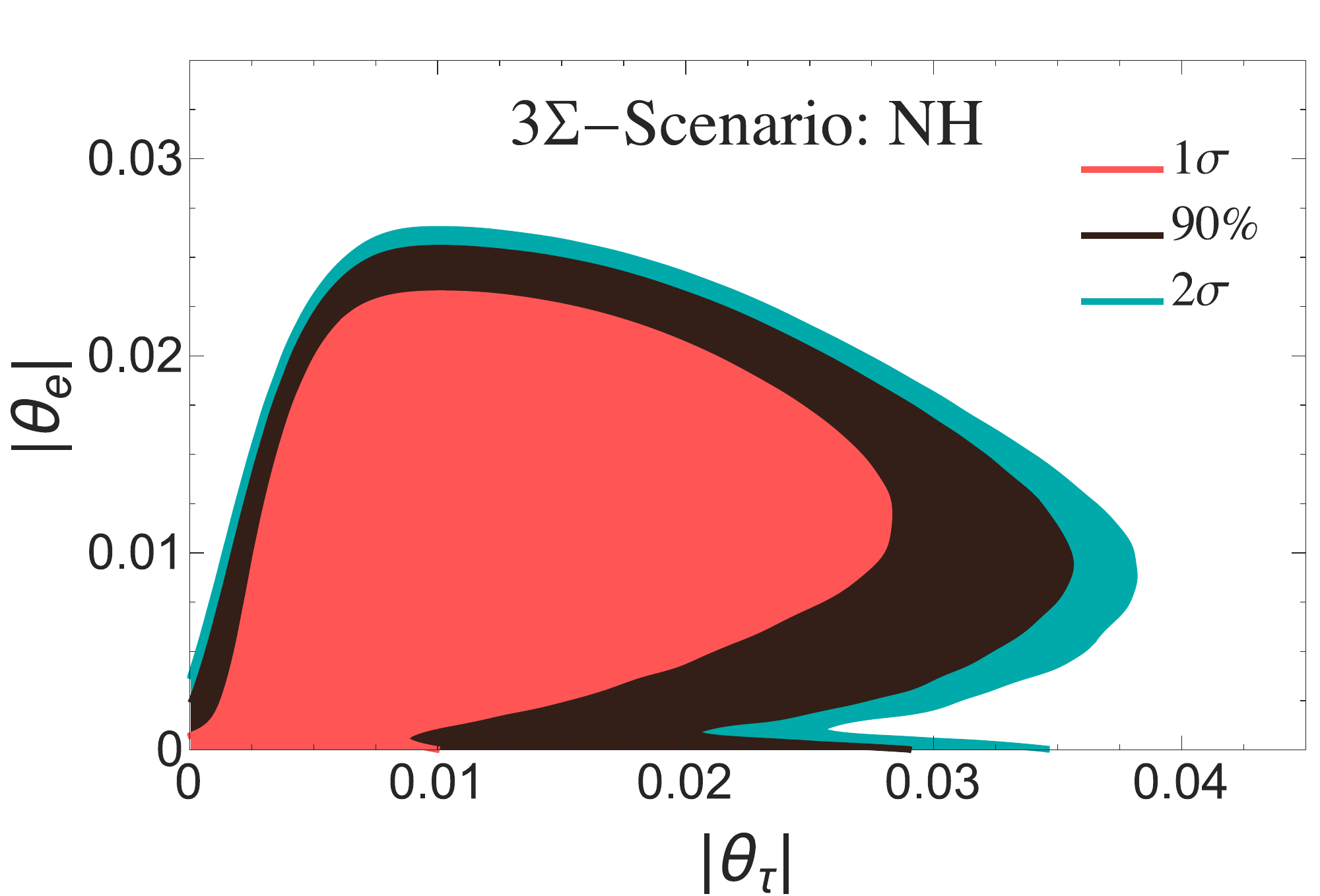}
\includegraphics[width=0.42\textwidth]{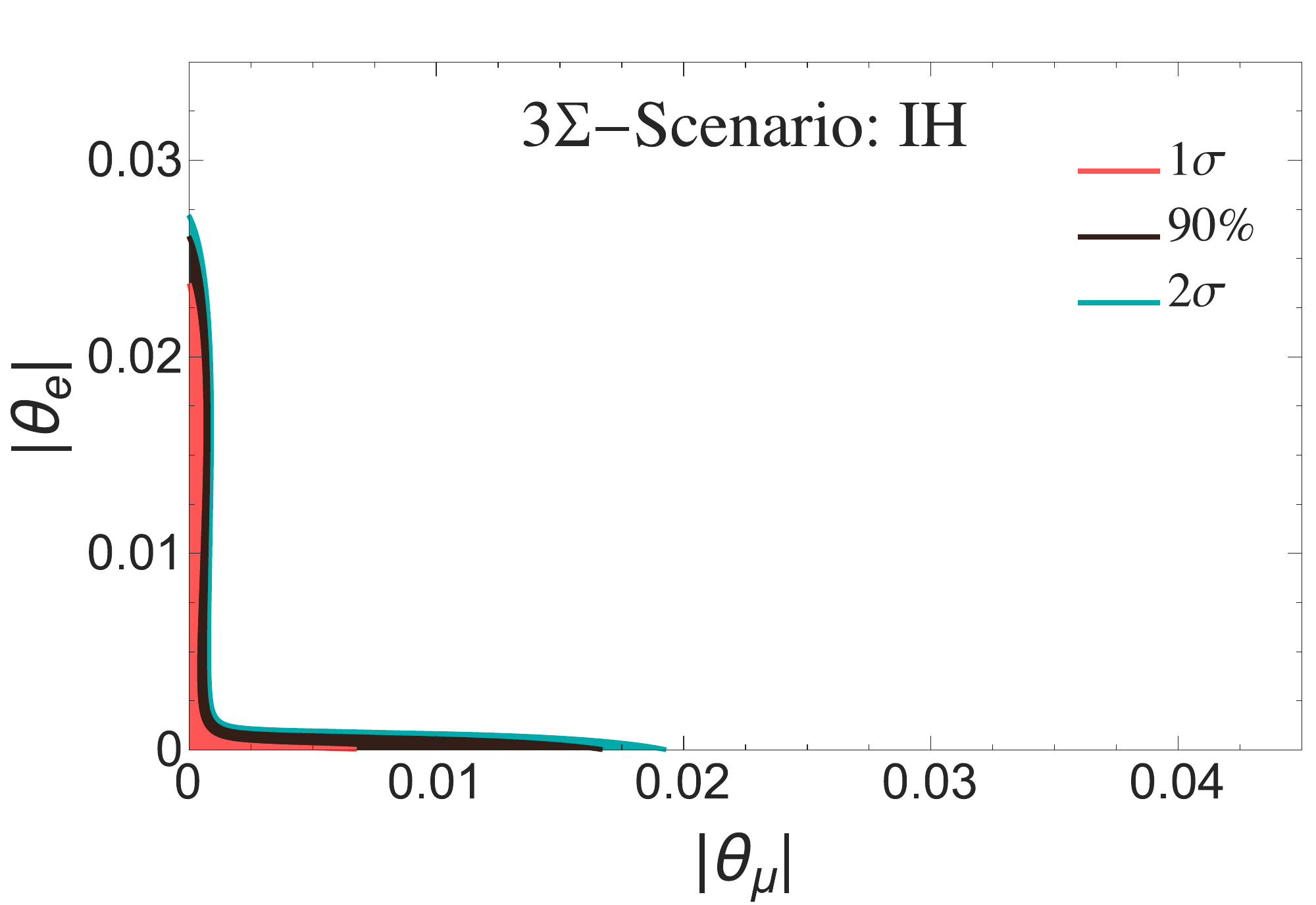}
\includegraphics[width=0.42\textwidth]{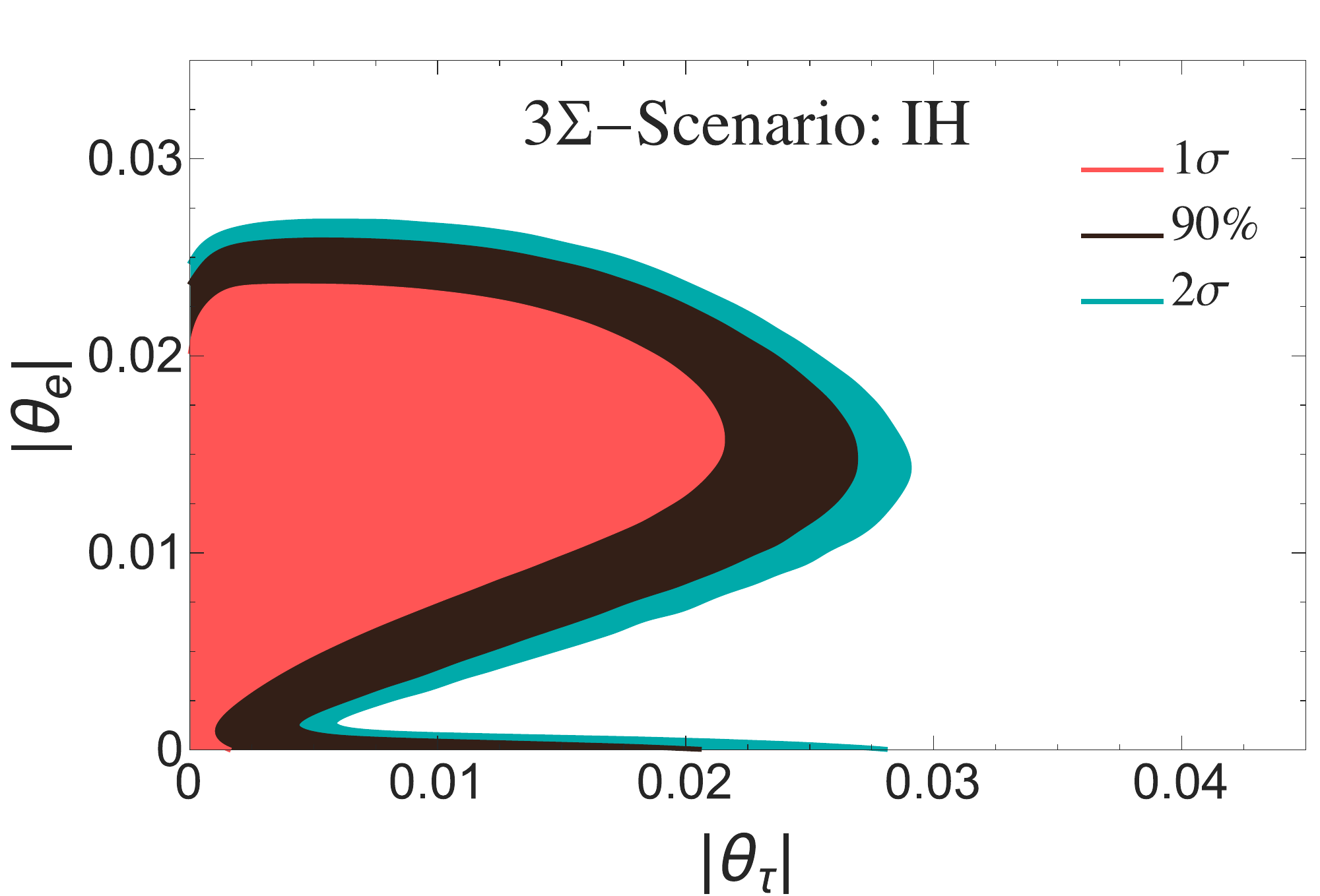}
\caption{Frequentist confidence intervals at $1 \sigma$, $90\%$ CL and $2 \sigma$ on the parameter space of the $3\Sigma$-SS for normal hierarchy (upper panels) and inverted hierarchy (lower panels).
}
\label{fig:contours-3TSS}
\end{figure}

\begin{figure}[!h]
\centering
\includegraphics[width=0.42\textwidth]{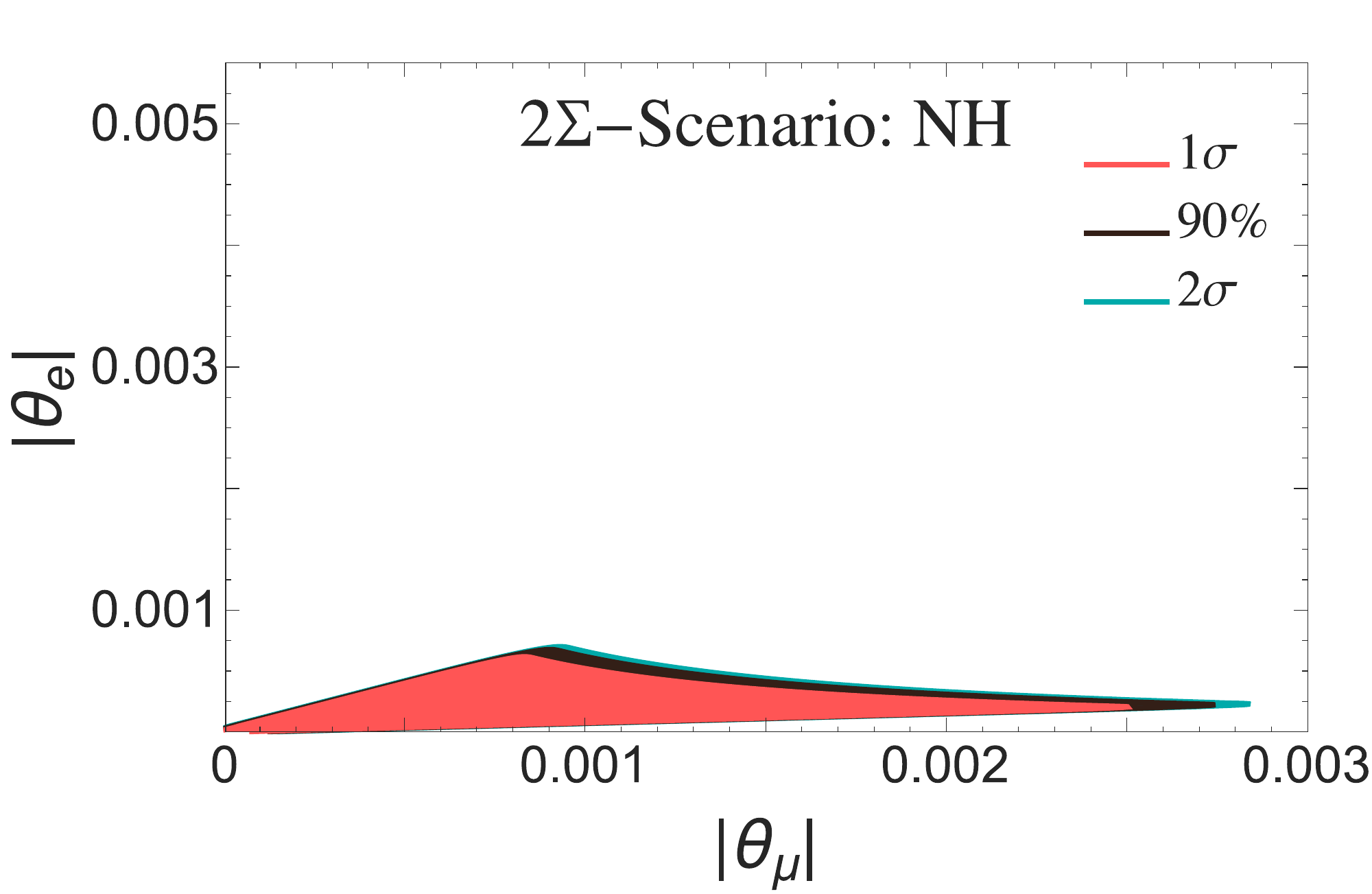}
\includegraphics[width=0.42\textwidth]{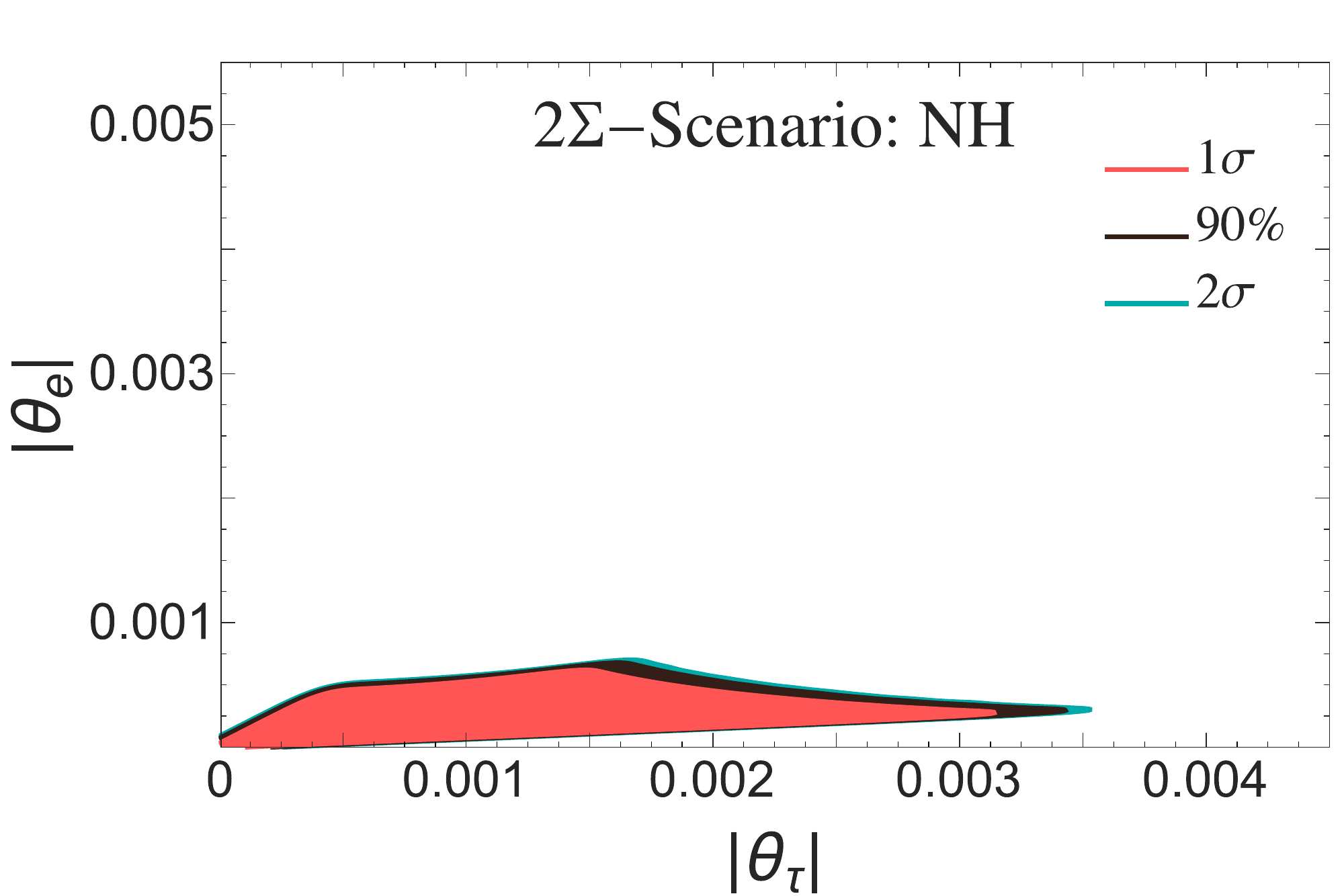}
\includegraphics[width=0.42\textwidth]{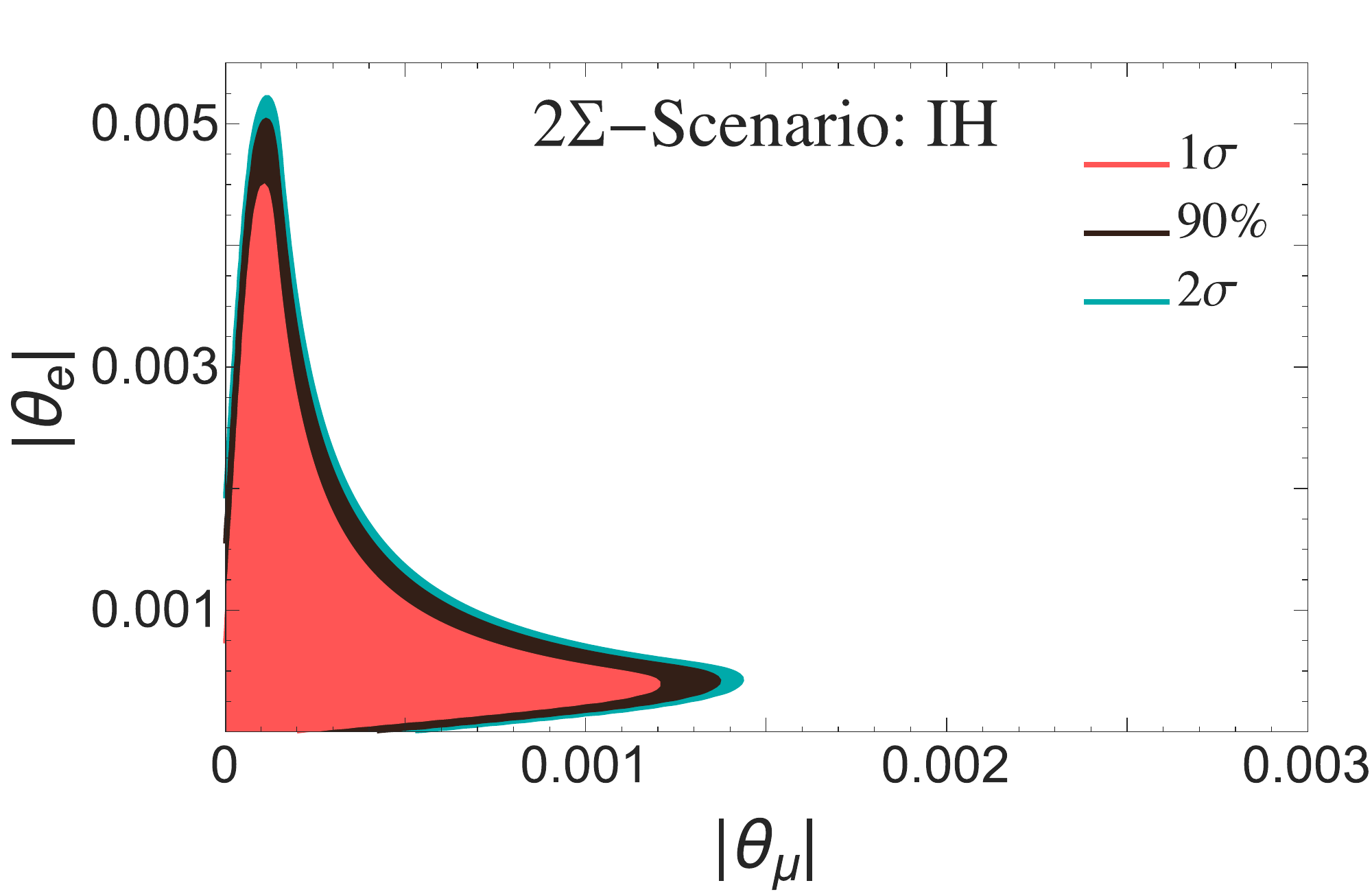}
\includegraphics[width=0.42\textwidth]{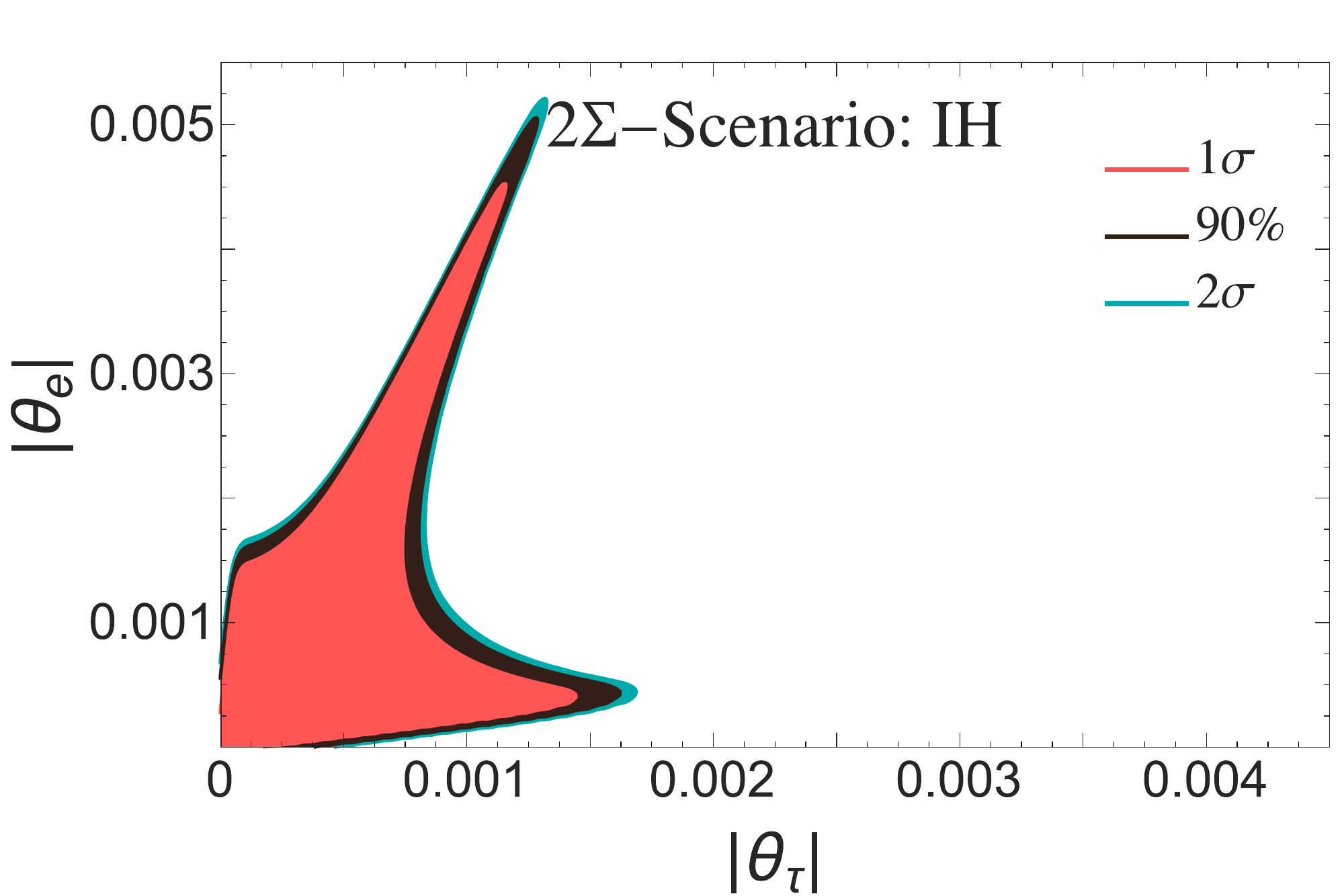}
\caption{Frequentist confidence intervals at $1 \sigma$, $90\%$ CL and $2 \sigma$ on the parameter space of the $2\Sigma$-SS for normal hierarchy (upper panels) and inverted hierarchy (lower panels).
}
\label{fig:contours-2TSS}
\end{figure}

In Figure~\ref{fig:contours-3TSS} (Figure~\ref{fig:contours-2TSS}) we present the 2 dof frequentist contours on the mixings $\vert\theta_\alpha\vert$ at $1\sigma$, 90\% CL, and $2\sigma$ in red, black and cyan, respectively for the 3$\Sigma$-SS (2$\Sigma$-SS) scenario. The left panels show the allowed regions in the plane $\vert \theta_e\vert - \vert \theta_\mu\vert$ while the right panels show the allowed regions in the plane $\vert \theta_e\vert - \vert \theta_\tau\vert$, for normal (top panels) and inverted (bottom panels) neutrino mass hierarchy.

The pronounced hyperbolic shape of the contours on the $\vert
\theta_e\vert - \vert \theta_\mu\vert$ plane of Figure~\ref{fig:contours-3TSS} are driven by the fact that in this scenario the product of both mixings is directly bounded by $\mu$ to
$e$ conversion in Ti nuclei. The allowed parameter space is thus dramatically reduced with respect to the G-SS scenario, even if the bounds on the individual parameters are similar. On the other hand the correlation shown in the right panels of the same Figure is determined by the constraints due to the generation of the light neutrino masses and mixing reflected in Eq.~(\ref{eq:hellishRelation}). To be precise, in the 3$\Sigma$-SS scenario, $\theta_\tau$ is determined by $\theta_e$, $\theta_\mu$ and the light neutrino free parameters. 
As can be observed in Figure~\ref{fig:contours-2TSS}, these features are even more pronounced in the minimal scenario with 2 fermion triplets 2$\Sigma$-SS
since the three mixings are directly proportional to $\theta_e$ with a proportionality constant which depends only on the light neutrino free parameters (see Eqs.~(\ref{eq:thetamu}) and~(\ref{eq:thetatau})). The overall scale of the three mixings is thus controlled by the most constraining observable, namely $\mu$ to $e$ conversion in nuclei, which sets quite stringent individual bounds on all $\theta_\alpha$. The particular correlations observed arise because in this minimal model the flavour structure is completely determined by the light neutrino parameters.


The constraints on the corresponding $\eta_{\alpha\beta}$ elements in both scenarios, and for both hierarchies, are summarized in Table~\ref{tab:bounds_eta}. The individual bounds on the 3$\Sigma$-SS case are pretty similar to the G-SS scenario, however they are considerably 
stronger in the 2$\Sigma$-SS, ranging from $\mathcal{O}(10^{-5})$ to $\mathcal{O}(10^{-7})$. Finally, notice that in the 3$\Sigma$-SS case the results for both light neutrino hierarchies are similar. However, in the 2$\Sigma$-SS scenario, the hierarchy has a strong impact in the results since it is a very relevant input regarding the constrained flavor structure of this minimal model (see Eqs.~(\ref{eq:thetamu}) and~(\ref{eq:thetatau})).

\section{Present LHC bounds on the (2$\Sigma$-SS and 3$\Sigma$-SS) new physics scale}
\label{s:LHC}

In contrast to the Type-I Seesaw, the Type-III  has better prospects to detect the neutrino mass mediator, since the heavy fermion triplets are charged under the SM gauge group. The minimal 2$\Sigma$-SS and next to minimal 3$\Sigma$-SS symmetry protected scenarios are particularly interesting since the constrained Yukawa flavor structure offers the possibility of testing this neutrino mass generation mechanism. The potential of the LHC regarding this possibility has been already analyzed in \cite{Eboli:2011ia} and, more recently, in~\cite{Agostinho:2017biv} in the context of the 2$\Sigma$-SS model. 

Here we will simply recast the present most constraining CMS bounds on the new physics scale of the 2$\Sigma$-SS and 3$\Sigma$-SS models. In Ref.~\cite{Sirunyan:2017qkz} bounds on the fermion triplet masses were derived in a simplified model with just one triplet as a function of the individual branching fractions 
\be
B_\alpha=\frac{|(Y_{\Sigma})_\alpha|^2}{\sum_{\beta}|(Y_{\Sigma})_\beta|^2}\, ,
\ee
as suggested in Ref.~\cite{Aguilar-Saavedra:2013twa}.
The analysis includes $13\,\rm{TeV}$ LHC data with $35.9\,\rm{fb}^{-1}$ collected by the CMS experiment in 2016. The results can be interpreted as conservative constraints on the minimal 2$\Sigma$-SS and next to minimal 3$\Sigma$-SS scenarios with branching fractions
\be
B_\alpha=\frac{|\theta_\alpha|^2}{\sum_{\beta}|\theta_\beta|^2}\,,
\label{eq:BR_theta}
\ee
where $\theta_{\alpha}$ should satisfy Eqs.~(\ref{eq:thetamu}) and
(\ref{eq:thetatau}) in the 2$\Sigma$-SS, and
Eq.~(\ref{eq:hellishRelation}) in the 3$\Sigma$-SS. In contrast to the
simplified model of Ref.~\cite{Sirunyan:2017qkz}, the 2$\Sigma$-SS has
two quasi-degenerate triplets with the same branching fractions
instead of one. Therefore, a fully dedicated analysis could yield somewhat stronger constraints than the ones summarized here, but it is beyond the scope of the present work. The 3$\Sigma$-SS has also two quasi-degenerate triplets with the same branching fractions and includes, in addition, another triplet which, however, has no impact in the CMS analysis since its Yukawa couplings are extremely small and it would thus not decay
inside the detector. In summary, the bounds from CMS fully apply to both cases.

\begin{figure}[!t]
\centering
\includegraphics[width=0.6\textwidth]{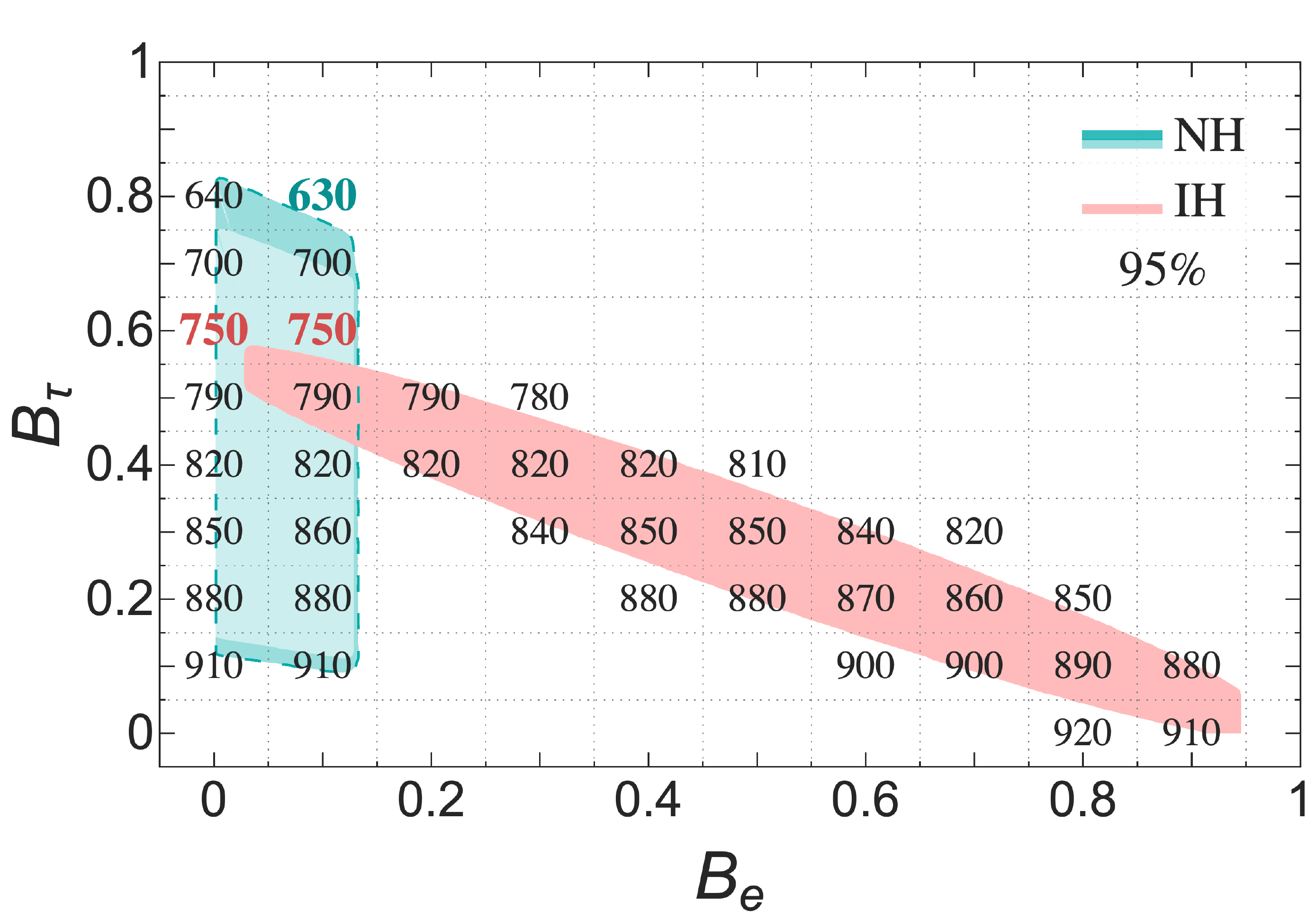}
\caption{$95\%$ CL lower bounds on heavy fermion triplet masses $M_\Sigma$ (GeV) in the 2$\Sigma$-SS scenario for each combination of branching fractions $B_e$ and $B_\tau$ to the individual lepton flavors $e$ and $\tau$ extracted from Ref.~\cite{Sirunyan:2017qkz}. The cyan (pink) solid region corresponds to the theoretically allowed parameter space given by Eqs.~(\ref{eq:thetamu}) and (\ref{eq:thetatau}) for NH (IH) when the neutrino mixing parameters are fixed to their best fit values given in Table~\ref{tab:osc_params}. The darker dashed contour shows the effect of including the corresponding priors from~\cite{Esteban:2018azc}. The highlighted bins in cyan (pink) indicates the general LHC lower bound on the new physics scale of this minimal model for NH (IH).}
\label{CMSbounds}
\end{figure}

In Fig.~\ref{CMSbounds} we show the lower bounds on the heavy fermion triplet masses $M_\Sigma$ derived by the CMS collaboration in Ref.~\cite{Sirunyan:2017qkz} and applied to the 2$\Sigma$-SS scenario. In this minimal scenario, the overall scale $\theta_e$ cancels when computing the flavor ratios $B_{\alpha}$ (see Eq.~(\ref{eq:BR_theta})) using Eqs.~(\ref{eq:thetamu}) and (\ref{eq:thetatau}). Thus, the colored regions correspond to the allowed part of the parameter space compatible with the neutrino mixings as determined in oscillation experiments. The solid contour in cyan (pink) corresponds to the allowed region when the mixing parameters are set to their best fit values from Table~\ref{tab:osc_params}, and when a NH (IH) is assumed. For the NH case, we have also studied the impact of present uncertainties in the mixing parameters when computing the allowed parameter space. In particular, the darker contour in Fig.~\ref{CMSbounds} shows the allowed region at $95\%$CL when we introduce the mixing angles and mass splittings as free parameters with their corresponding priors from~\cite{Esteban:2018azc}. The dominant contribution comes from the degeneracy on the octant of $\theta_{23}$ with larger values of $B_\tau$ allowed for $\theta_{23}$ in the lower octant. The most conservative lower bounds on the new physics scale of this minimal model are indicated by the colored numbers in bold face in Fig.~\ref{CMSbounds} and given by
\be
M_\Sigma\gtrsim \,630\; (750)\; \rm{GeV}\,,
\nonumber
\ee
for NH (IH). 

These same correlations from Eqs.~(\ref{eq:thetamu}) and (\ref{eq:thetatau}) imply that, in case of a discovery in searches such as those by the CMS collaboration, the measurement of the different $B_\alpha$ allows a direct test of this  scenario and the ability to rule out its connection to the origin of neutrino masses.

Conversely, the 3$\Sigma$-SS scenario has more freedom, given our present lack of knowledge on the absolute neutrino mass scale, the mass hierarchy and the Majorana phases. Hence, the
full area below $B_e+B_\tau=1$ in Fig.~\ref{CMSbounds} is allowed for
both NH and IH and therefore, for the 3$\Sigma$-SS case, the present
lower bound on the new physics scale is relaxed to the generic bound
obtained in Ref.~\cite{Sirunyan:2017qkz}:
$M_\Sigma\gtrsim390$~GeV. A direct test of this scenario would also require additional information on the absolute neutrino mass scale and the Majorana phases. On the other hand,
if the mass of the decoupled fermion triplet is at the LHC reach, a
potential characteristic signal of this model would be the decay into
a long lived heavy charged component of the triplet and missing
energy: $pp \rightarrow W^{\pm} \rightarrow \Sigma^0\,\Sigma^{\pm}$. 

Notice that, since the heavy triplets are produced via gauge interactions, the global bounds derived in Section~\ref{s:results} do not play any role in the production mechanism but only in the decay of the triplets via mixing with the SM particles. Since the constraints from flavor and electroweak precision data obtained here are not strong enough to exclude the prompt decay of the triplets, they are not relevant for LHC searches in this respect. As discussed above, the constraints from the light neutrino mass and mixing generation in the 2$\Sigma$-SS and 3$\Sigma$-SS scenarios are the most relevant ones in this context.

\section{Conclusions}
\label{s:conclusions}

In this work we have performed a global fit to present lepton flavor
and electroweak precision data to constrain the Type-III Seesaw as the
mechanism to generate the light neutrino masses. We have presented our
constraints in terms of the mixing among the neutral component of the heavy fermion triplets
and the SM neutrinos or, equivalently, the coefficient of the effective
dimension 6 operator generated upon integrating out the heavy
fermions. We have analyzed a completely general scenario in which an
arbitrary number of heavy fermion triplets is added to the SM field
content (G-SS), and the 3$\Sigma$-SS and 2$\Sigma$-SS scenarios in
which three and two fermion triplets are considered respectively. In
order for these constraints to be meaningful, the $d=6$ operator
cannot be suppressed neither by the heavy new physics scale nor by the
small Yukawa couplings, which are naively required to explain the
smallness of neutrino masses. This is possible, and technically natural, if the smallness of neutrino masses is explained with an approximate lepton number (LN) symmetry. These symmetry-protected scenarios are particularly interesting since the smallness of neutrino masses is suppressed not only by the new physics scale but also by the small parameters breaking LN. As a result, new physics scales close to the EW scale can be realized leading to a potentially testable phenomenology. 

Once the heavy fermions are integrated out from the low energy spectrum, all the new physics effects are encoded in the coefficient of the $d=6$ effective operator, $\eta$. The results of our global fit are summarized in Table~\ref{tab:bounds_eta}. At the $2\sigma$ level, the upper bounds on $\eta_{\alpha\beta}$ are $\mathcal{O}(10^{-4})$ for both the G-SS and 3$\Sigma$-SS, with the notable exception of $\eta_{\mu e}$ at $\mathcal{O}(10^{-7})$. As expected, LFV observables play a very relevant role in all the scenarios since FCNC are already present at tree level. In particular, $\mu$ to $e$ conversion in Ti sets the most constraining present bound: $\eta_{\mu e}<3.0\cdot10^{-7}$. This is indeed a very promising channel since the future PRISM/PRIME experiment is expected to improve the sensitivity by three orders of magnitude. Interestingly, in the G-SS scenario, the direct bounds from LFV processes on the remaining off-diagonal $\eta$ parameters are comparable to the indirect LFC bounds derived from the Schwarz inequality given in Eq.~(\ref{eq:schwarz}). The individual bounds are rather similar for the G-SS and 3$\Sigma$-SS scenarios. However, when the results are projected in two dimensions a drastic reduction of the 3$\Sigma$-SS allowed parameter space with respect to the G-SS can be observed. This is mainly due to the Schwarz inequality saturating to an equality when only three or less triplets are considered so that the stringent constraints from $\mu$ to $e$ conversion propagate to other elements. Moreover, the constraint stemming from the light neutrino mass generation leads to additional flavor correlations in the 3$\Sigma$-SS. The minimal 2$\Sigma$-SS scenario is even more constrained since in this case only the overall scale of the $d=6$ effective operator coefficient $\eta_{\alpha\beta}$ is free, while its flavor structure is completely determined by the light neutrino masses and mixings. This overall scale is then bounded by the most stringent observable, namely $\mu$ to $e$ conversion in nuclei. Thus, the bounds on this scenario are much stronger ranging from $\mathcal{O}(10^{-6})$ to $\mathcal{O}(10^{-7})$, with the exception of $\eta_{ee}$ for IH which is of the order of $\mathcal{O}(10^{-5})$. 

Although a previous analysis of the Type-III Seesaw global bounds applying to the G-SS was performed more than ten years ago in Ref.~\cite{Abada:2007ux}, the bounds derived in this work represent the most updated set of constraints for this scenario. The upper bounds on $\eta_{ee}$ and $\eta_{\mu\mu}$ have improved by one order of magnitude, while the constraints on $\eta_{\tau\tau}$ and the off-diagonal elements of $\eta$ have improved by a factor 3 approximately with respect to Ref.~\cite{Abada:2007ux}. To our knowledge, there are no similar previous analysis of the 3$\Sigma$-SS and 2$\Sigma$-SS cases, with the exception of Ref.~\cite{Kamenik:2009cb} in which the case with two triplets without explicitly requiring approximate LN conservation was also analyzed. Our constraints obtained for the 2$\Sigma$-SS case apply also to the scenario analyzed in Ref.~\cite{Kamenik:2009cb} since approximate LN conservation is required to saturate the constraints from EW precision and LFV observables.
 
The indirect searches analyzed in this work are strongly complementary to collider searches since only the latter can be directly sensitive to the new physics scale of the model. This is particularly relevant in the symmetry-protected scenarios in which the new physics scale can be at the reach of LHC. Furthermore, in contrast to the Type-I Seesaw, in which the right handed neutrinos are singlets of the SM gauge group, in the Type-III Seesaw the fermion triplets are charged under $\text{SU}\left(2\right)$, and thus its production is not necessarily penalized by the mixing.  In case of a discovery at colliders, the exciting prospect of directly testing the origin of neutrino masses presents itself. Indeed, in the minimal 2$\Sigma$-SS scenario the requirement to reproduce the observed masses and mixings implies correlations among the triplet branching ratios that are directly testable. We have exploited these correlations to recast the present limits on the new physics scale $M_\Sigma$. The most constraining bounds are from CMS and lead to the lower bound $M_\Sigma\gtrsim \,630\; (750)\; \rm{GeV}$ for NH (IH). On the other hand, in the 3$\Sigma$-SS case, there is at present enough freedom given our lack of knowledge on the neutrino mass hierarchy, absolute mass scale and Majorana phases to avoid additional constraints so that the lower bound is the one set by CMS in a general scenario, $M_\Sigma\gtrsim390$~GeV. However, with future information on these presently unknown parameters, this scenario would also imply non-trivial testable correlations.

\section*{Acknowledgments}

JLP would like to thank Miha Nemev\v{s}ek and Richard Ruiz for useful discussions regarding the collider phenomenology of the Type-III Seesaw. JHG warmly thanks Jordi Salvado for useful discussion.
This work made extensive use of the HPC-Hydra cluster at IFT.
This work is supported in part by the European Union's Horizon 2020 research and innovation programme under the Marie Sklodowska-Curie grant agreements 674896-Elusives and 690575-InvisiblesPlus. 
EFM acknowledges support from the ``Spanish Agencia Estatal de Investigaci\'on'' (AEI) and the EU ``Fondo Europeo de Desarrollo Regional'' (FEDER) through the project FPA2016-78645-P; and the Spanish MINECO through the Centro de Excelencia Severo Ochoa Program under grant SEV-2016-0597. JHG acknowledges support by the grant K125105 of the National Research, Development and Innovation Fund in Hungary. JLP acknowledges support by the ``Generalitat Valenciana" (Spain) through the ``plan GenT" program (CIDEGENT/2018/019).

\begin{appendix}

\section{Bounds on the mixing of heavy fermion triplets}
\label{s:Appendix-B}

\begin{figure}[!t]
\centering
\includegraphics[width=0.32\textwidth]{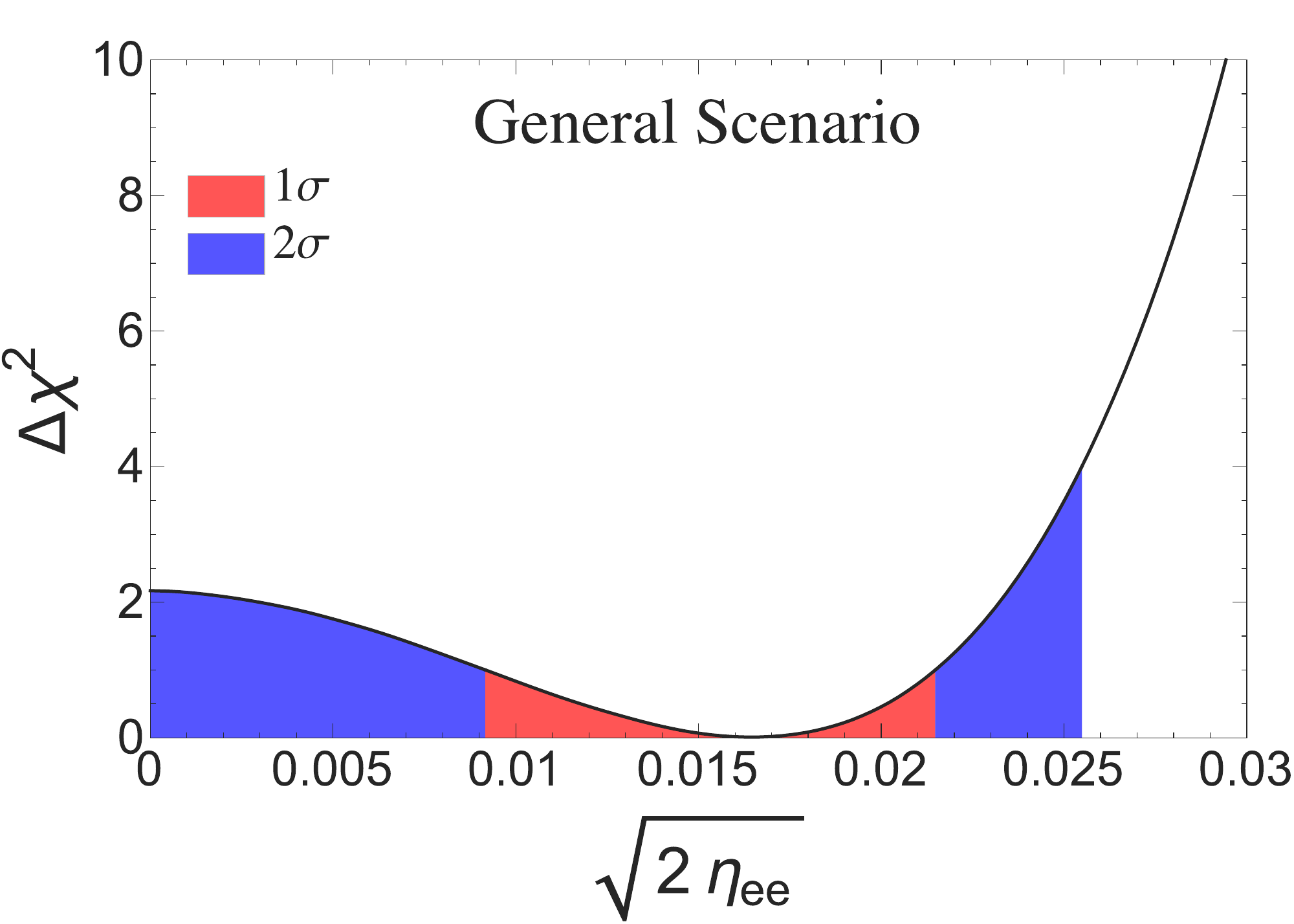}
\includegraphics[width=0.32\textwidth]{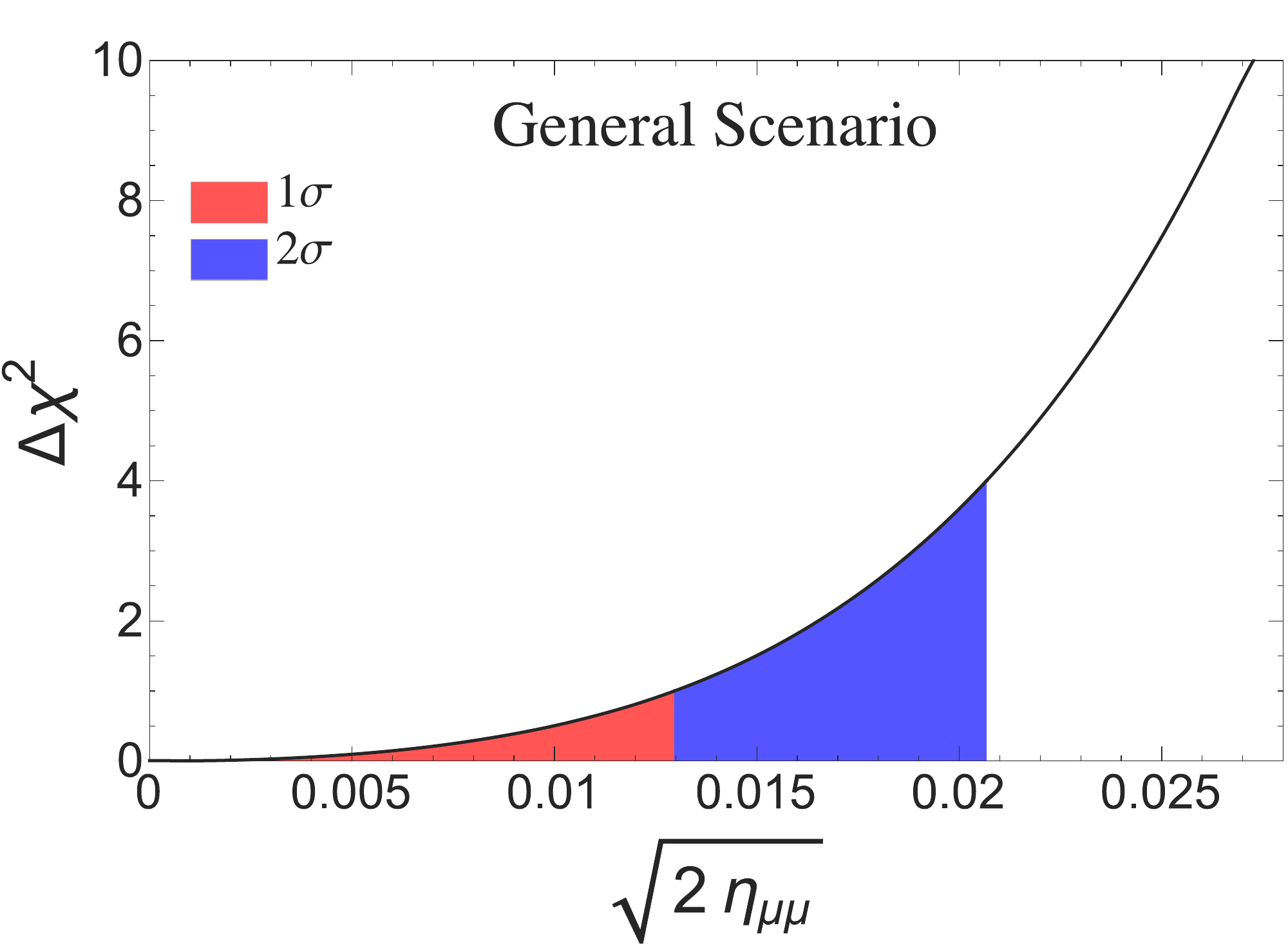}
\includegraphics[width=0.32\textwidth]{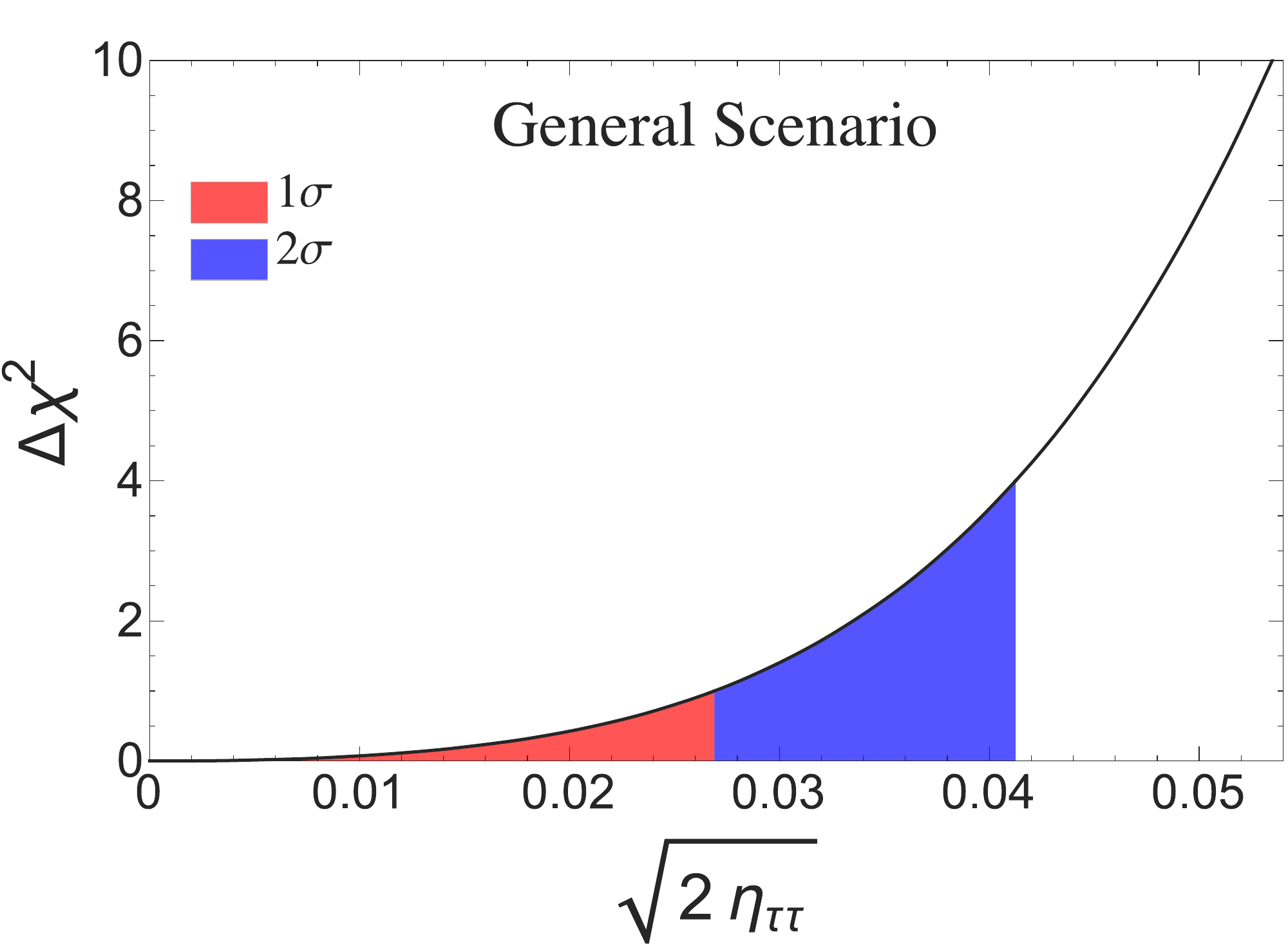} \\
\includegraphics[width=0.32\textwidth]{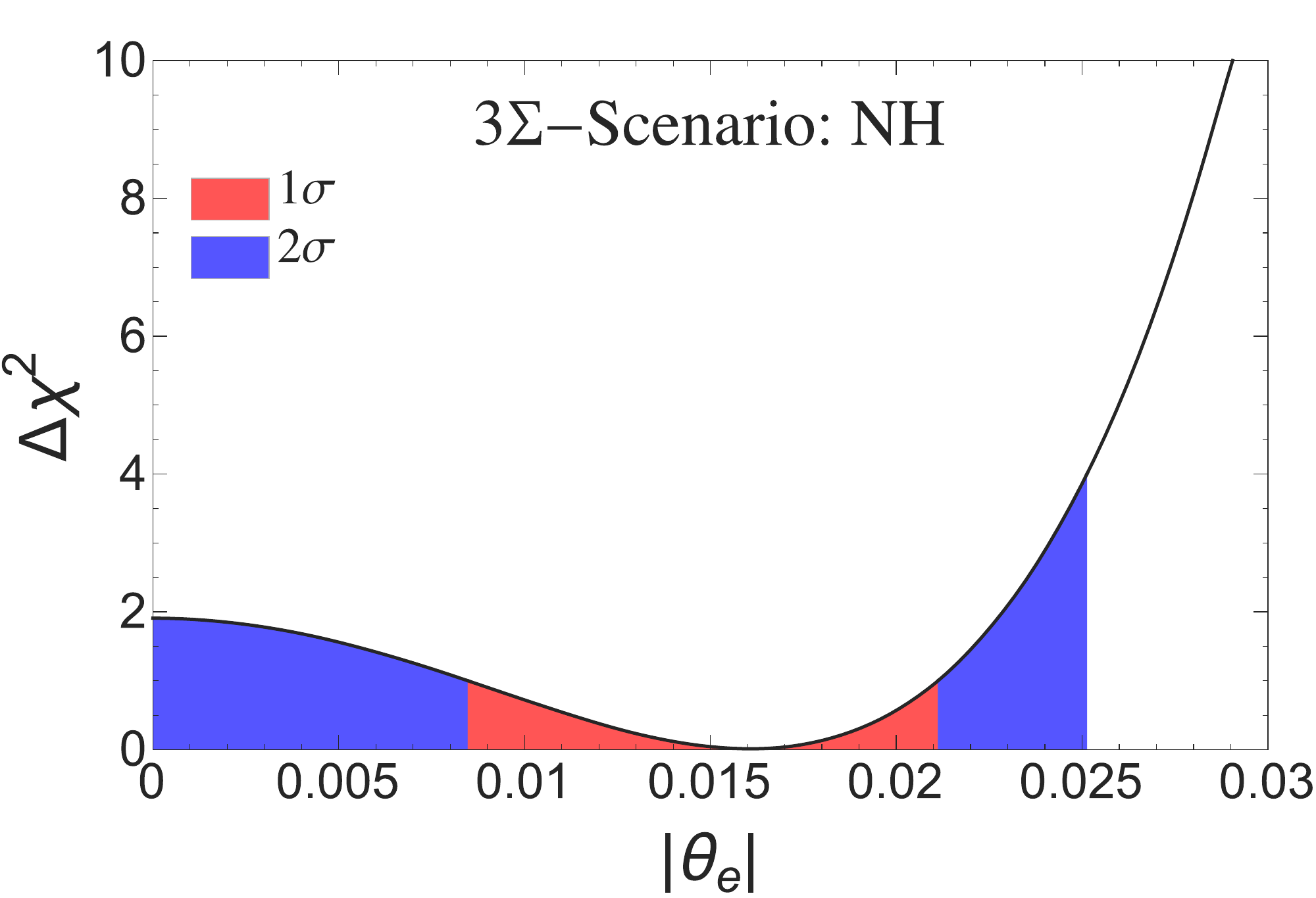}
\includegraphics[width=0.32\textwidth]{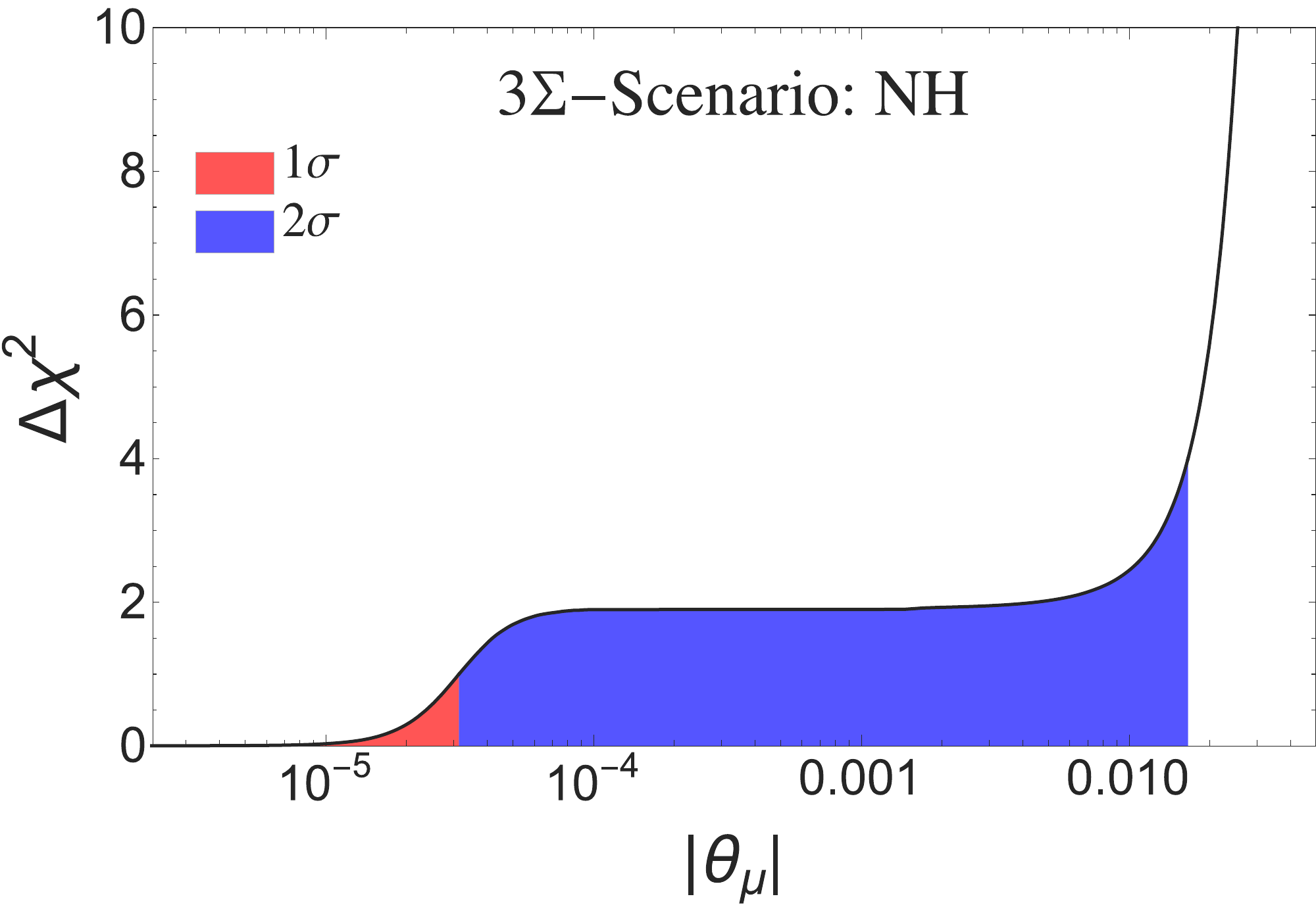}
\includegraphics[width=0.32\textwidth]{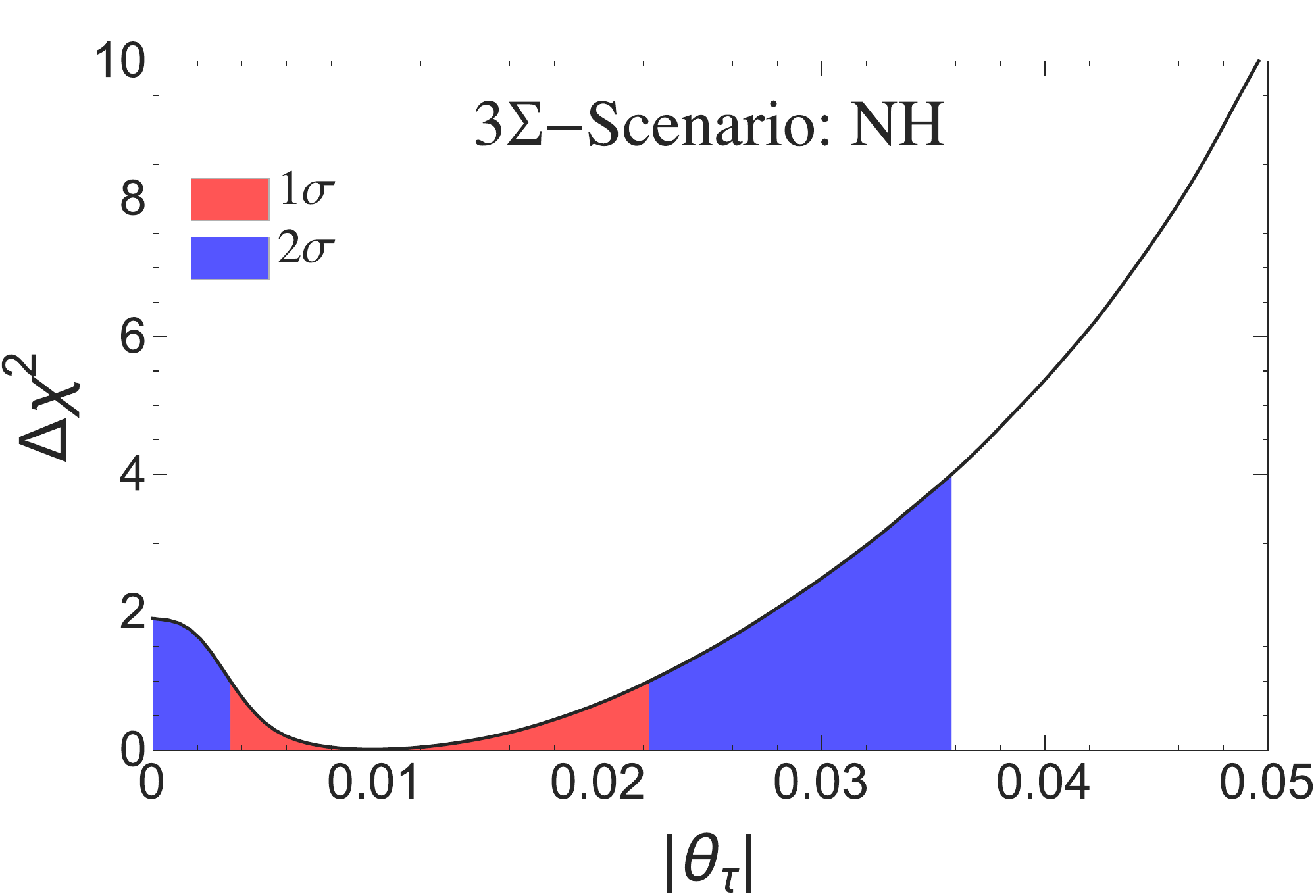} \\
\includegraphics[width=0.32\textwidth]{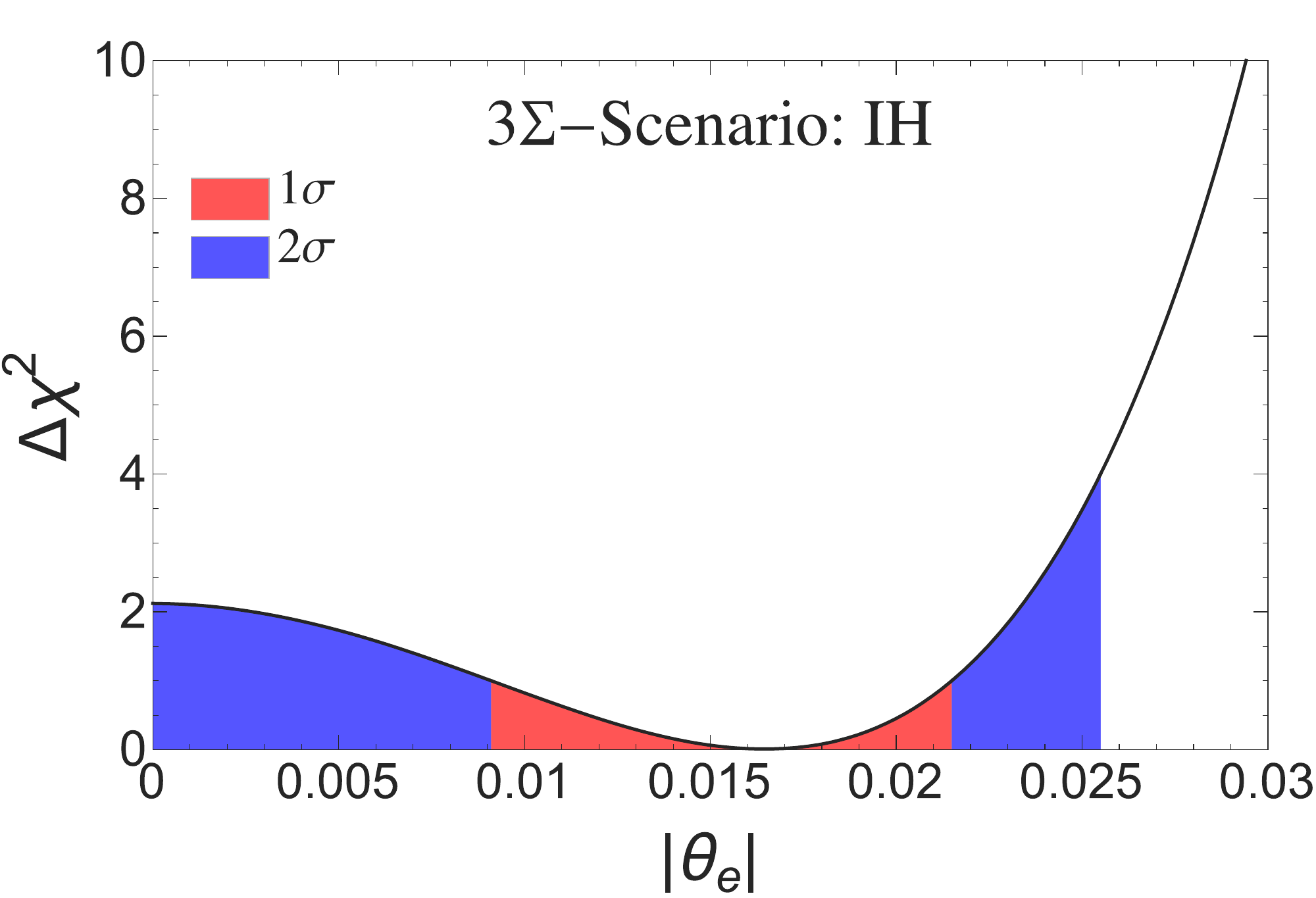}
\includegraphics[width=0.32\textwidth]{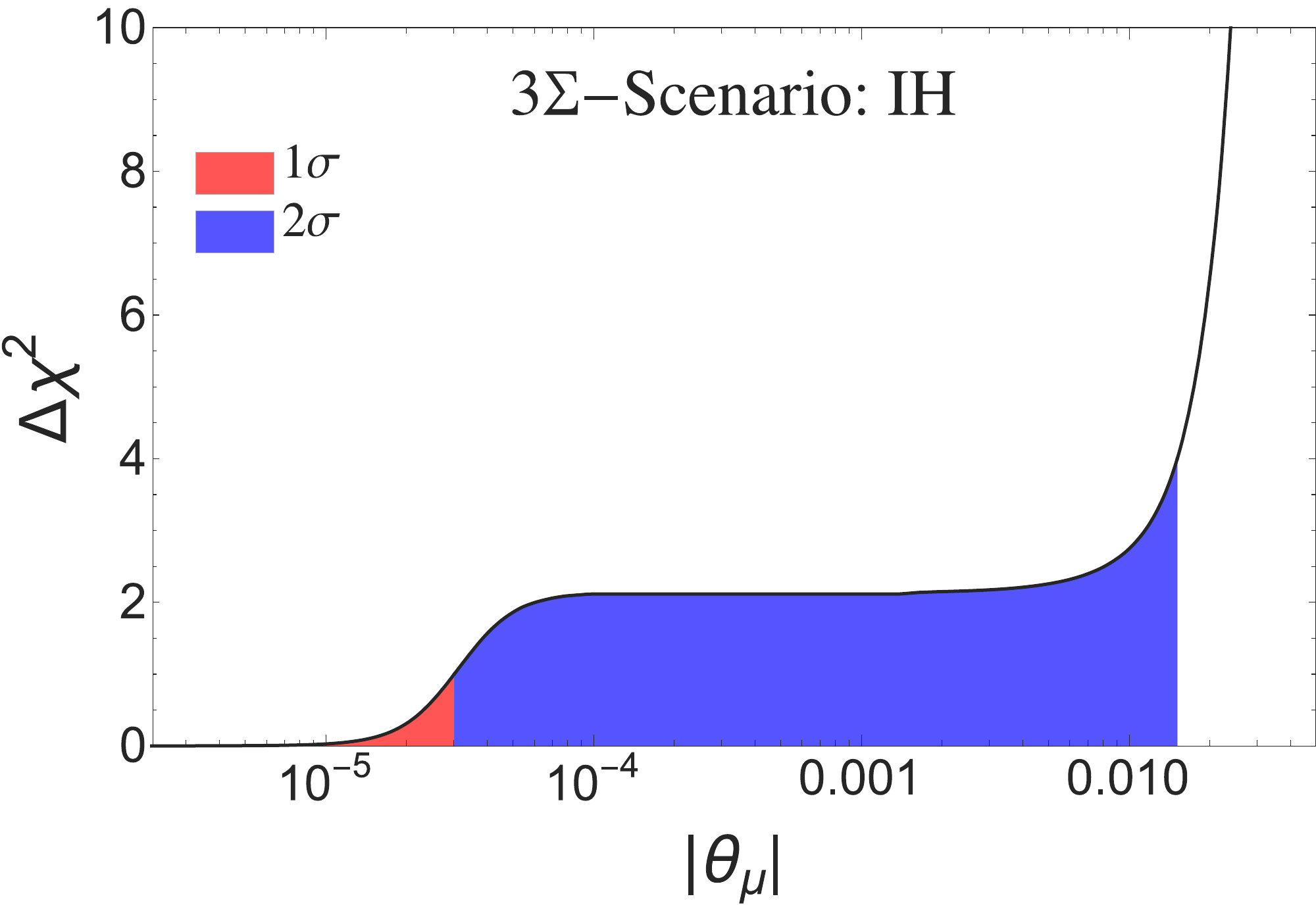}
\includegraphics[width=0.32\textwidth]{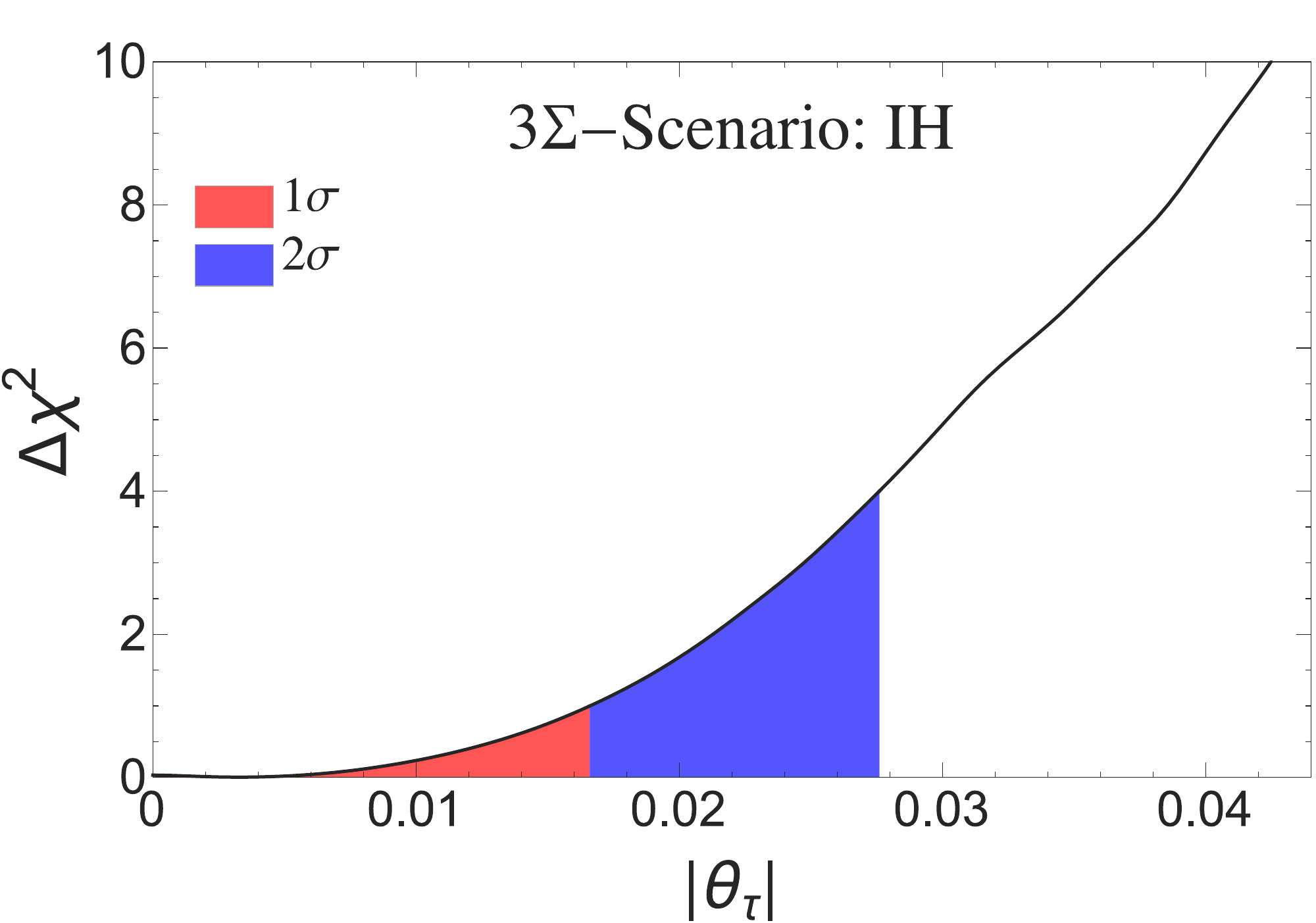} \\
\includegraphics[width=0.32\textwidth]{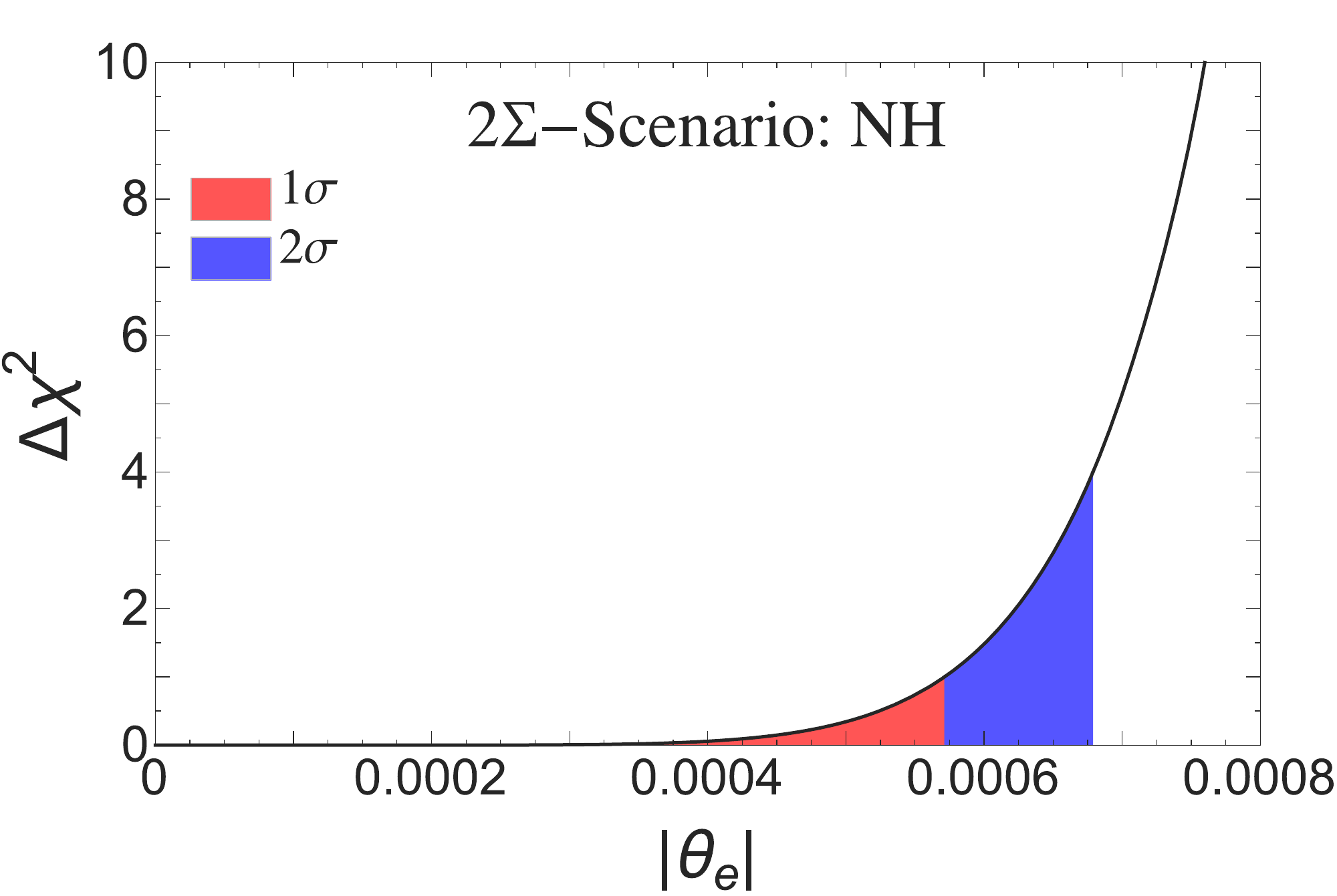}
\includegraphics[width=0.32\textwidth]{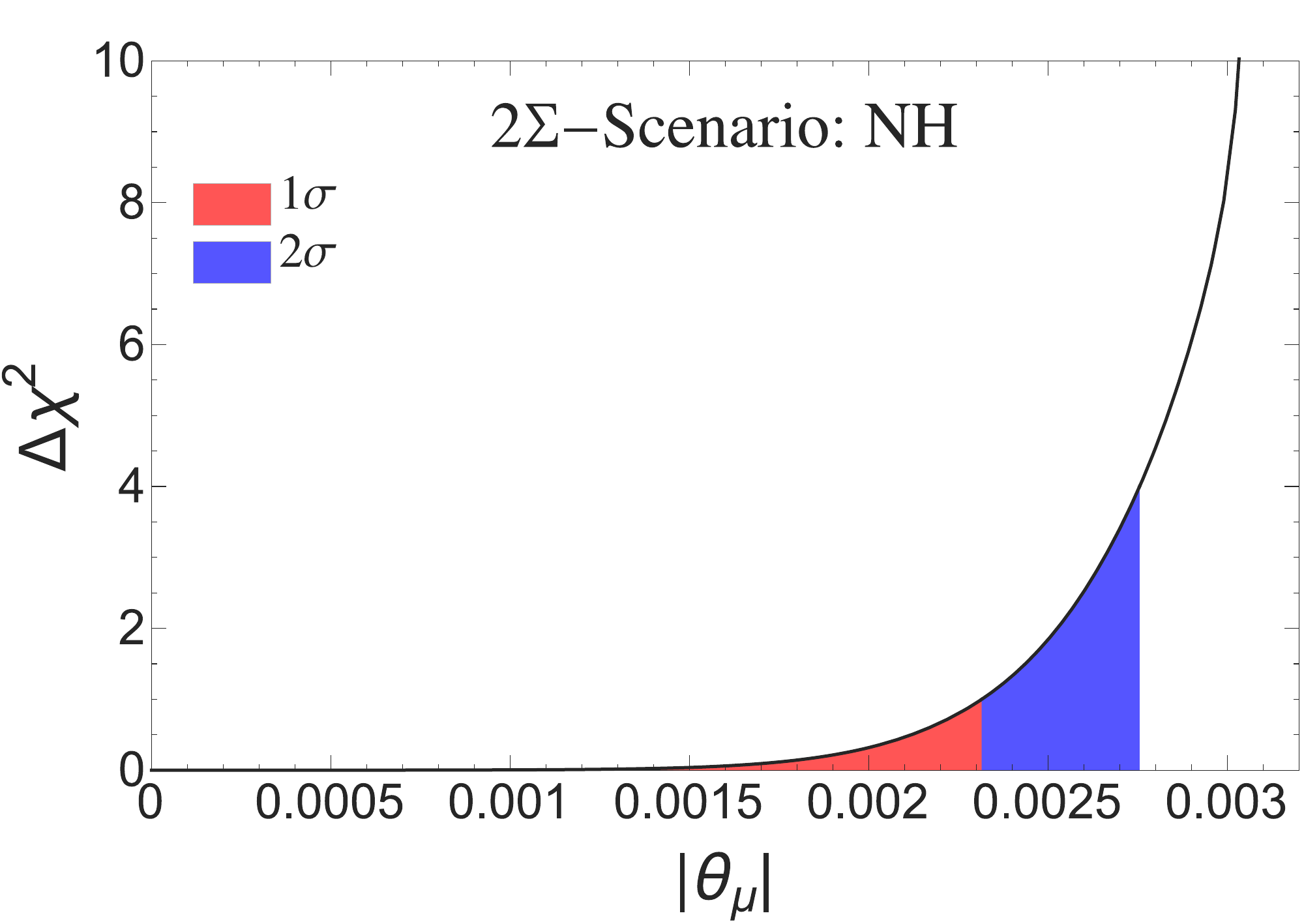}
\includegraphics[width=0.32\textwidth]{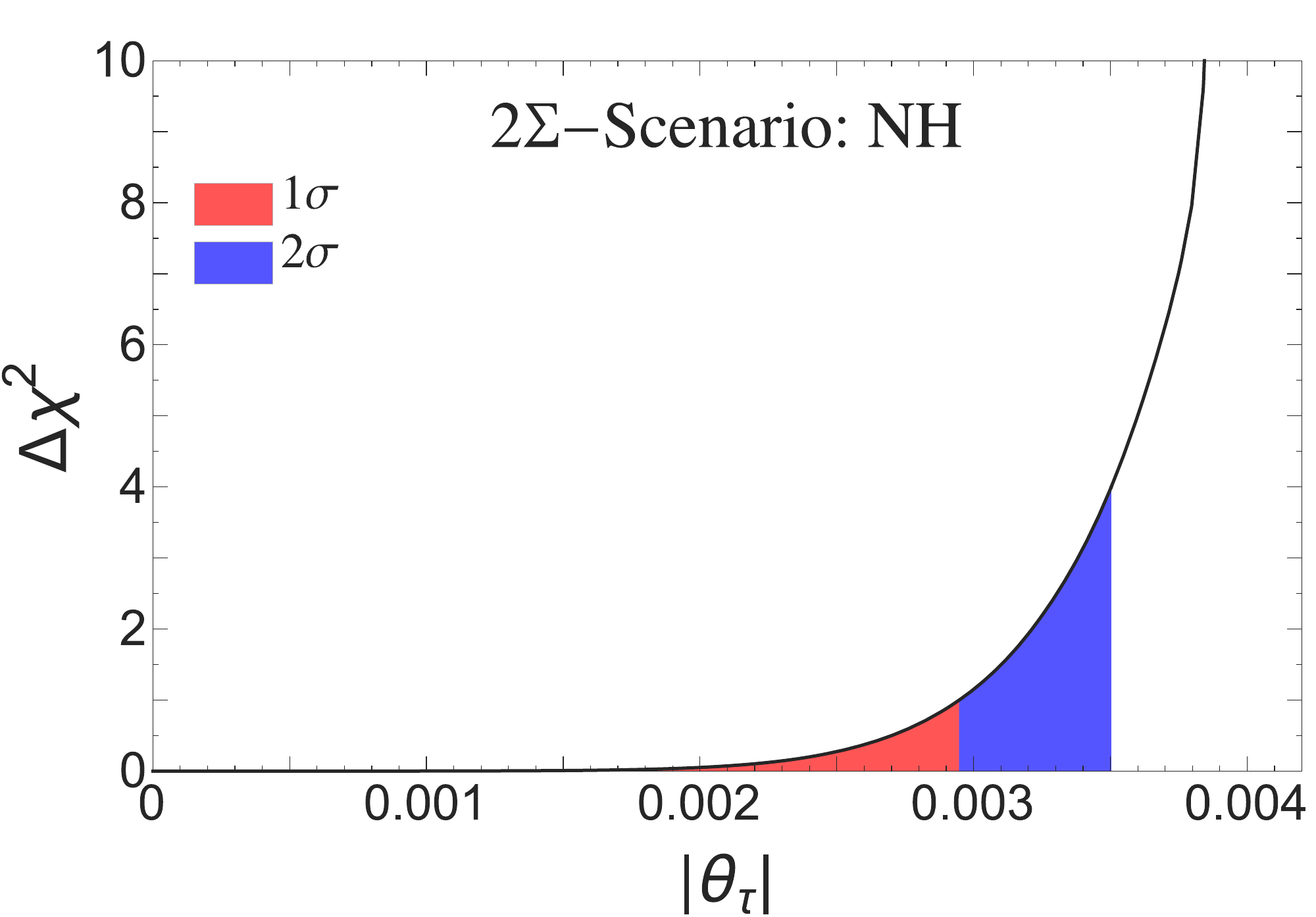} \\
\includegraphics[width=0.32\textwidth]{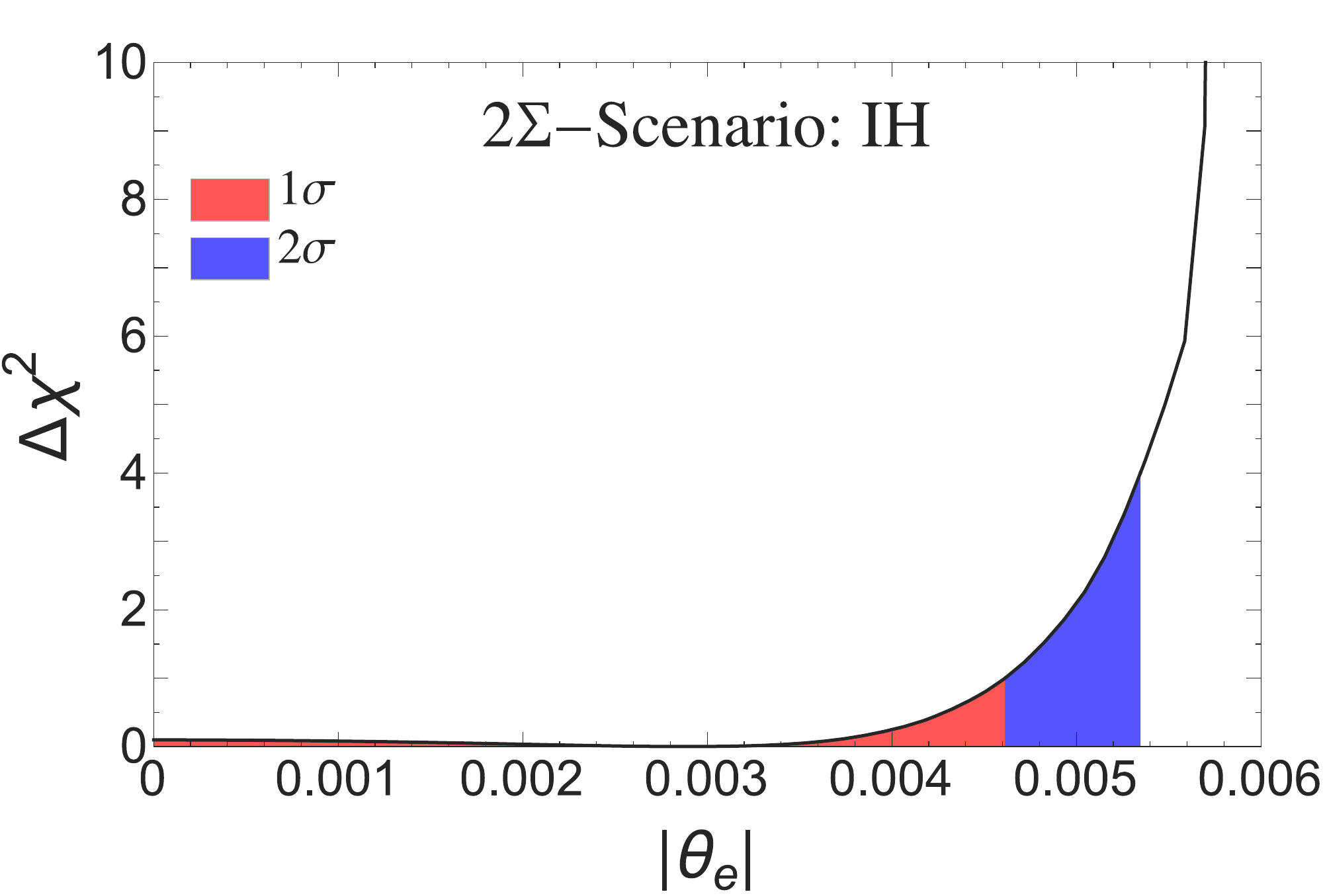}
\includegraphics[width=0.32\textwidth]{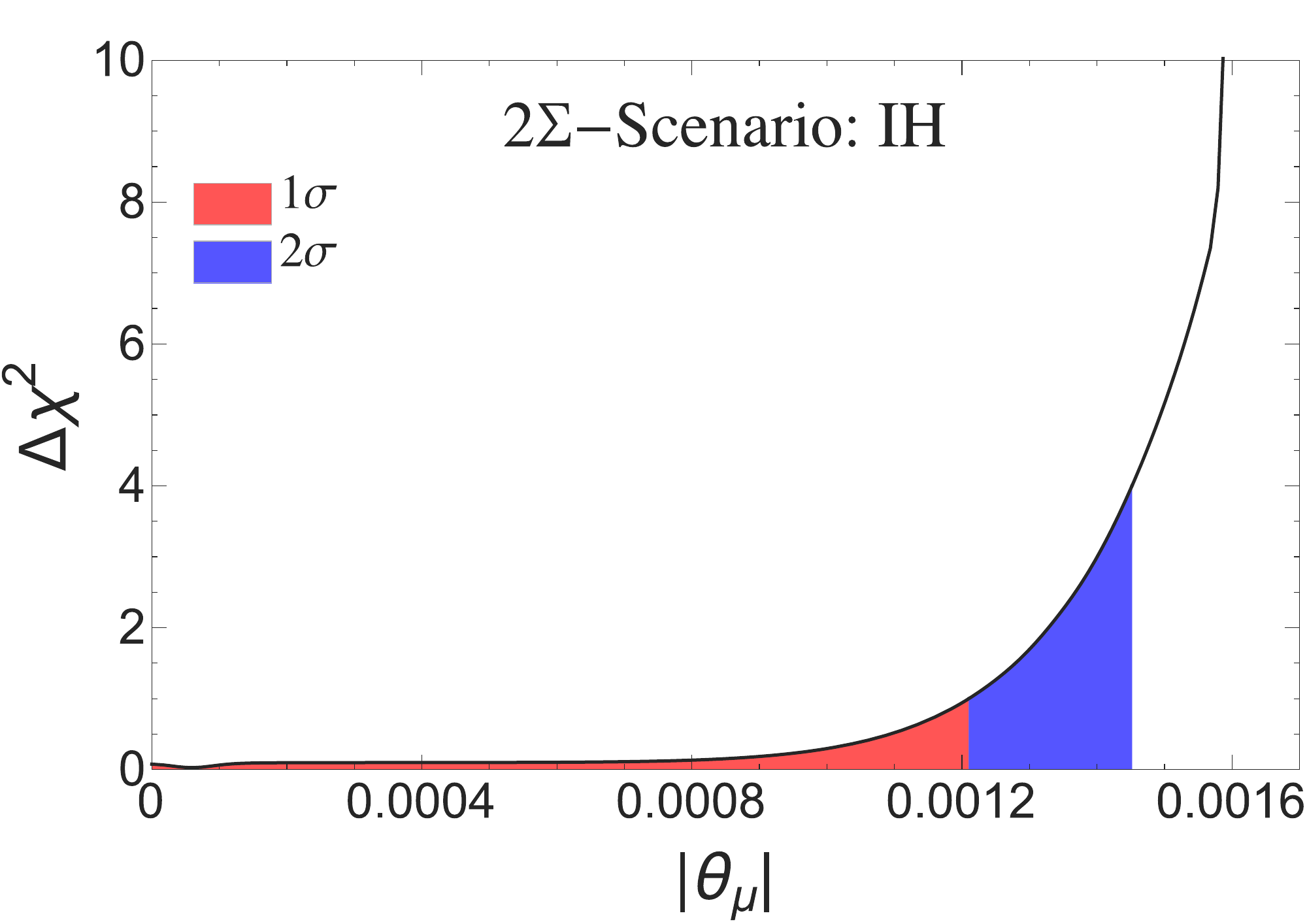}
\includegraphics[width=0.32\textwidth]{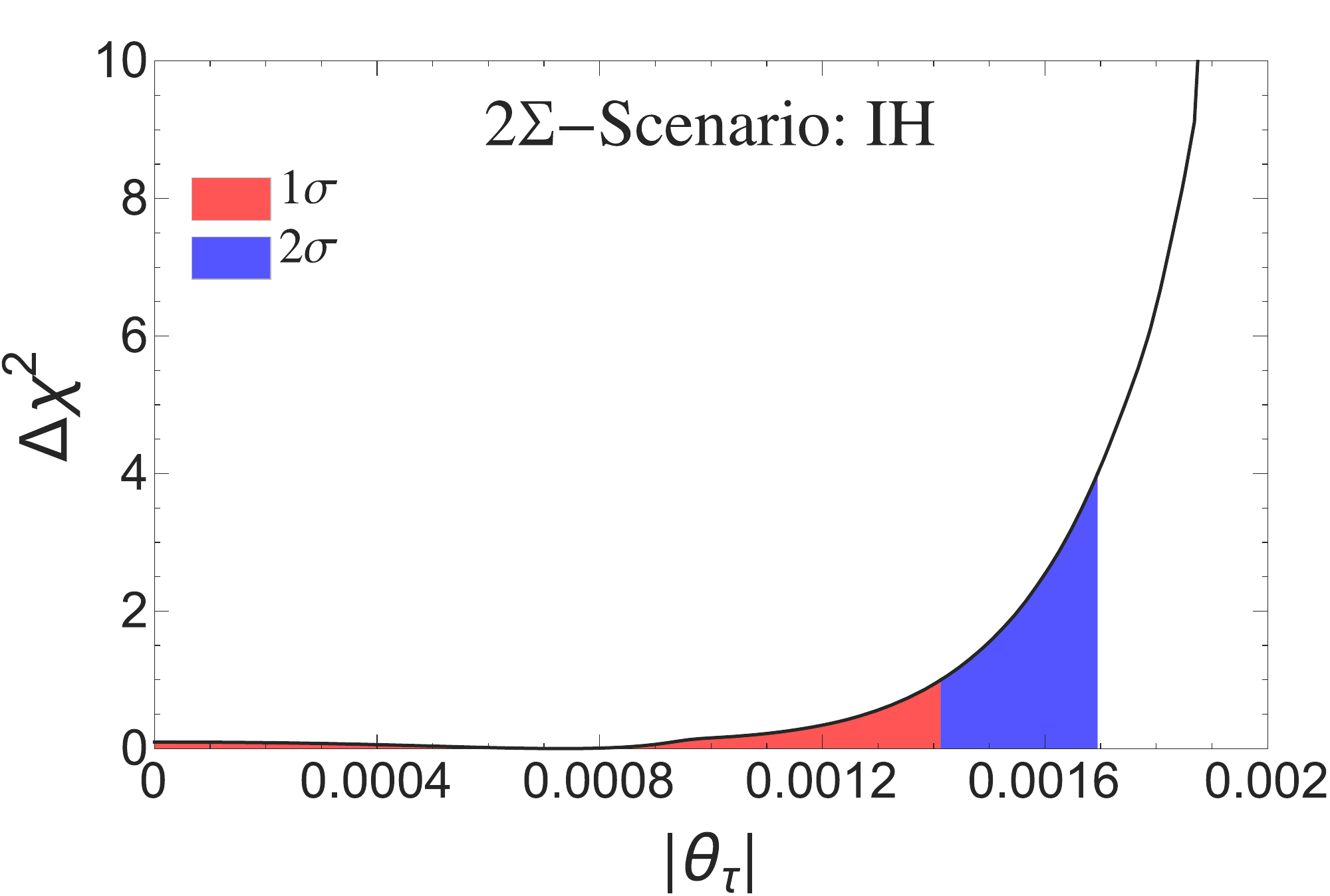}
\caption{$\Delta \chi^2$ profile minimized over all fit parameters but one single $\theta_\alpha$ (or $\sqrt{2 \eta_{\alpha \alpha}}$ for G-SS). In the upper panels the G-SS fit results are plotted, while the middle and lower panels show the results of the $3\Sigma$-SS and the $2\Sigma$-SS, for NH and IH respectively.}
\label{fig:chi2_diag}
\end{figure}

\begin{figure}[t!]
\centering
\includegraphics[width=0.32\textwidth]{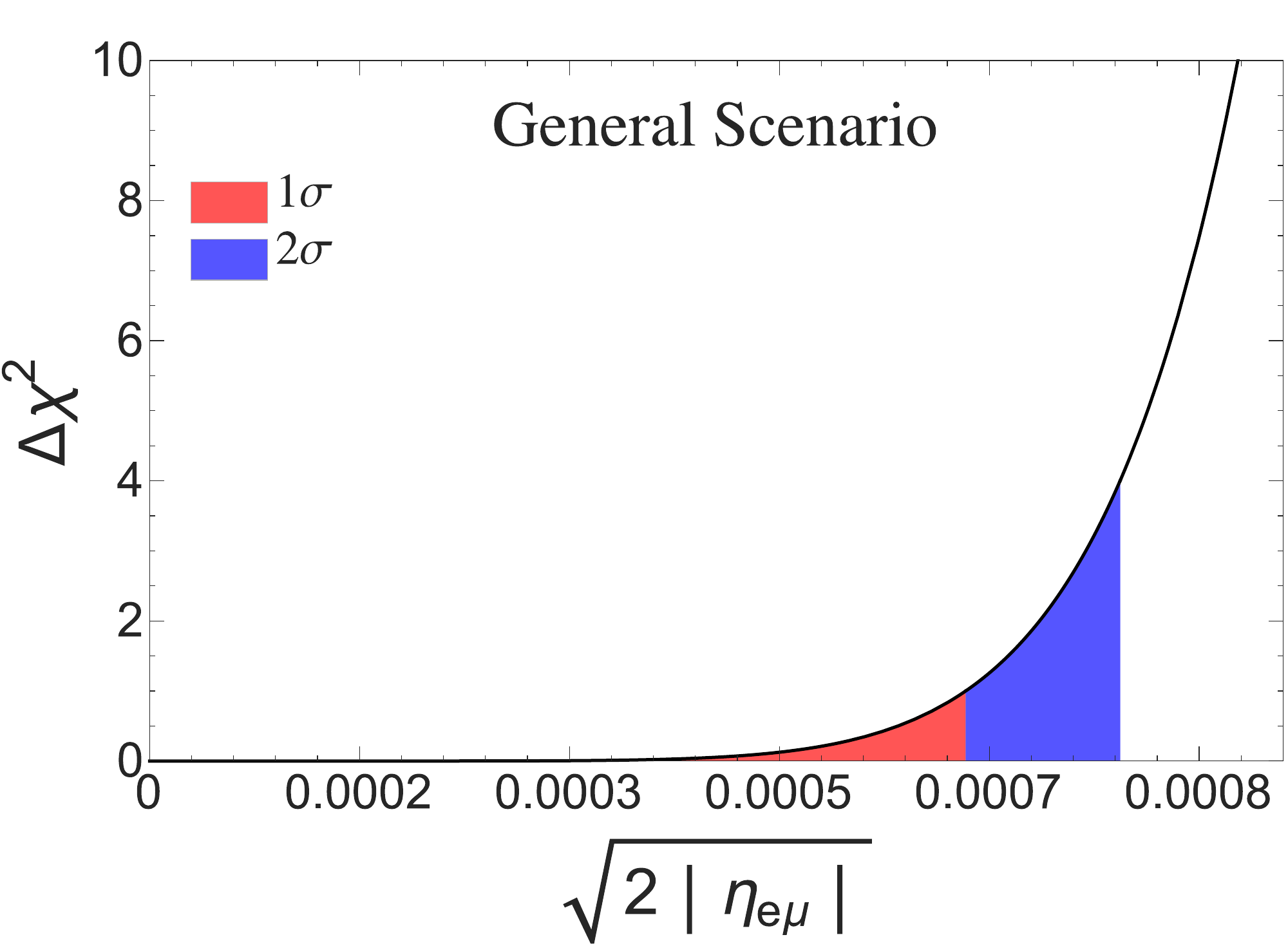}
\includegraphics[width=0.32\textwidth]{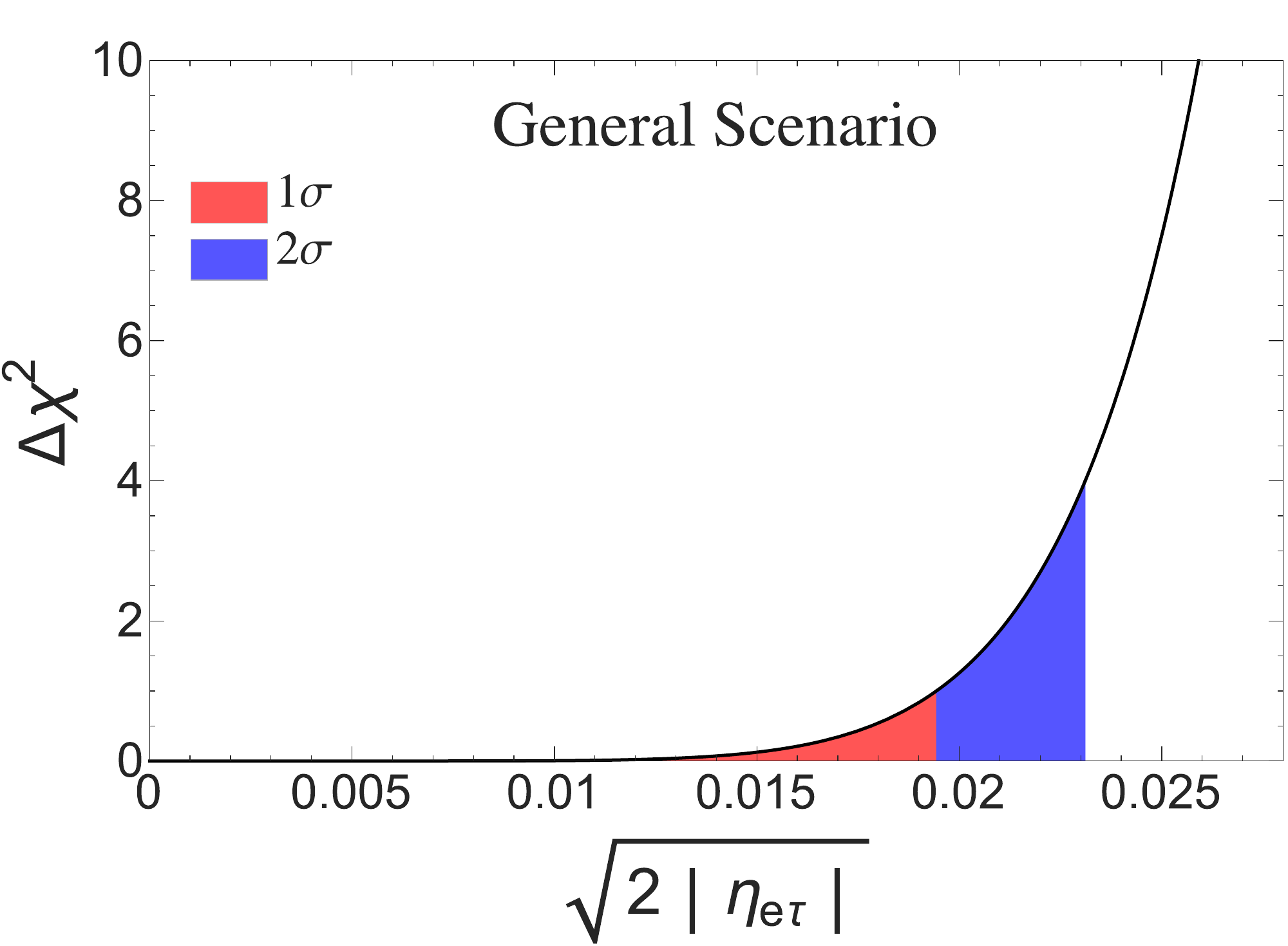}
\includegraphics[width=0.32\textwidth]{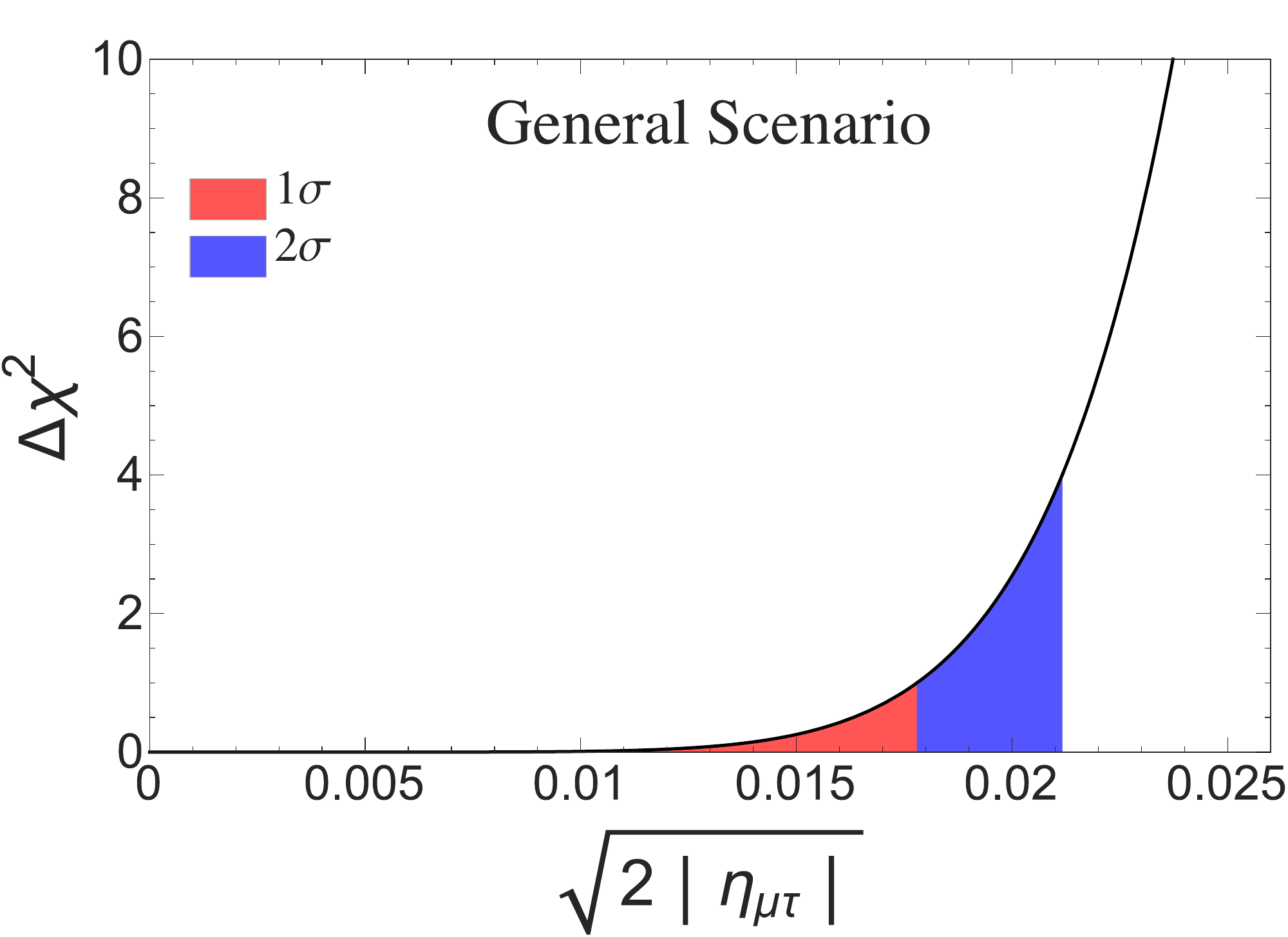} \\
\includegraphics[width=0.32\textwidth]{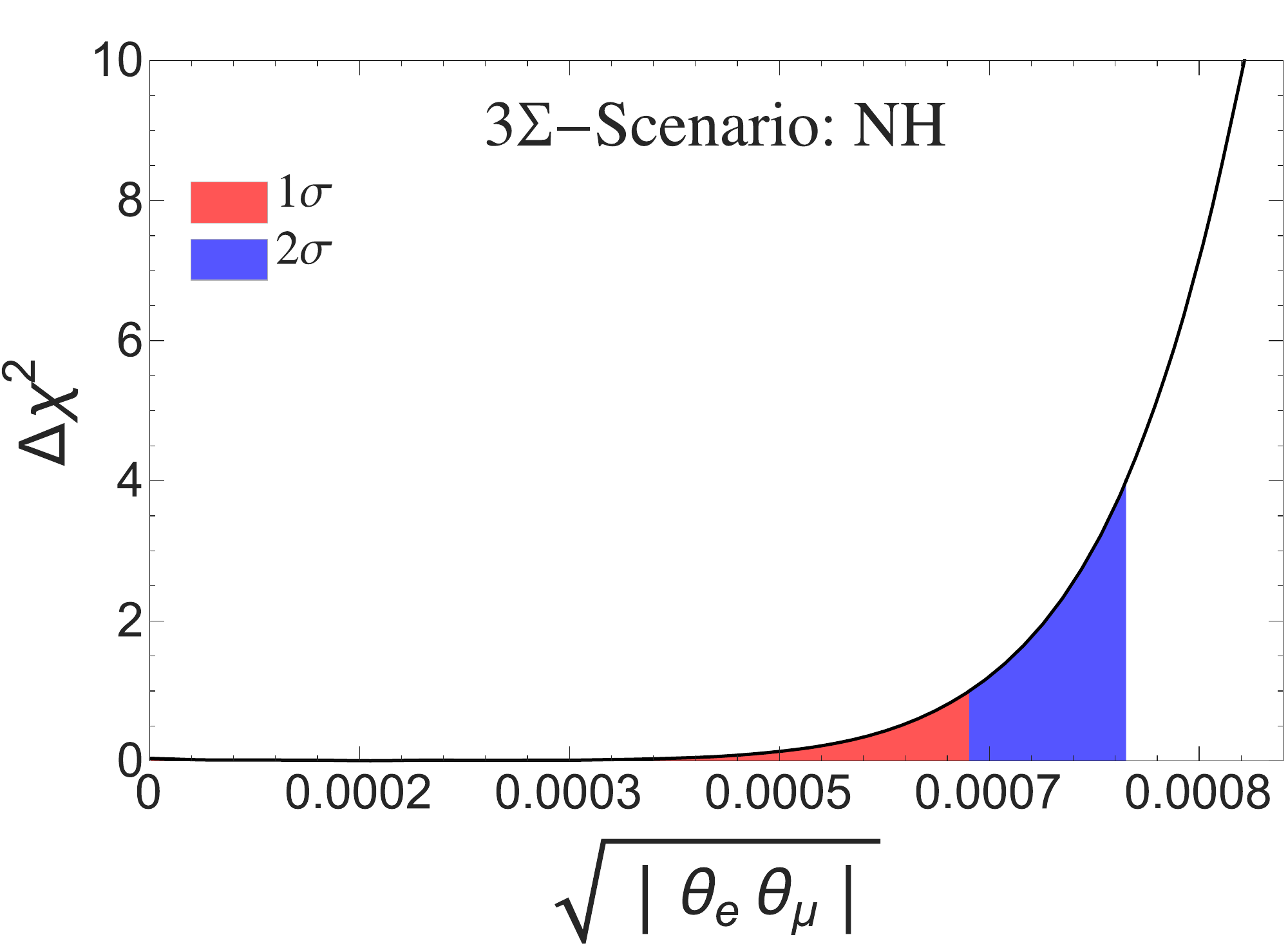}
\includegraphics[width=0.32\textwidth]{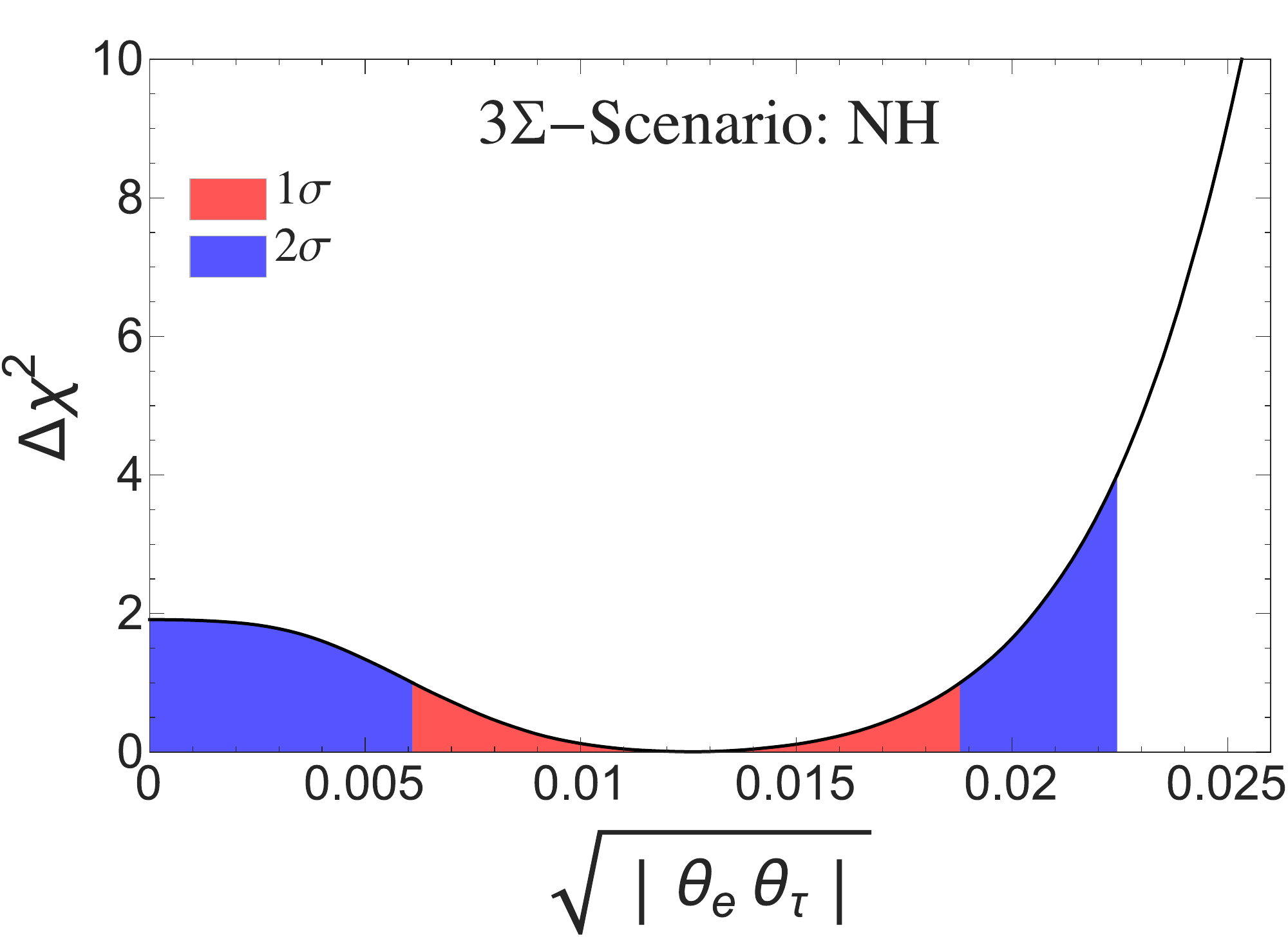}
\includegraphics[width=0.32\textwidth]{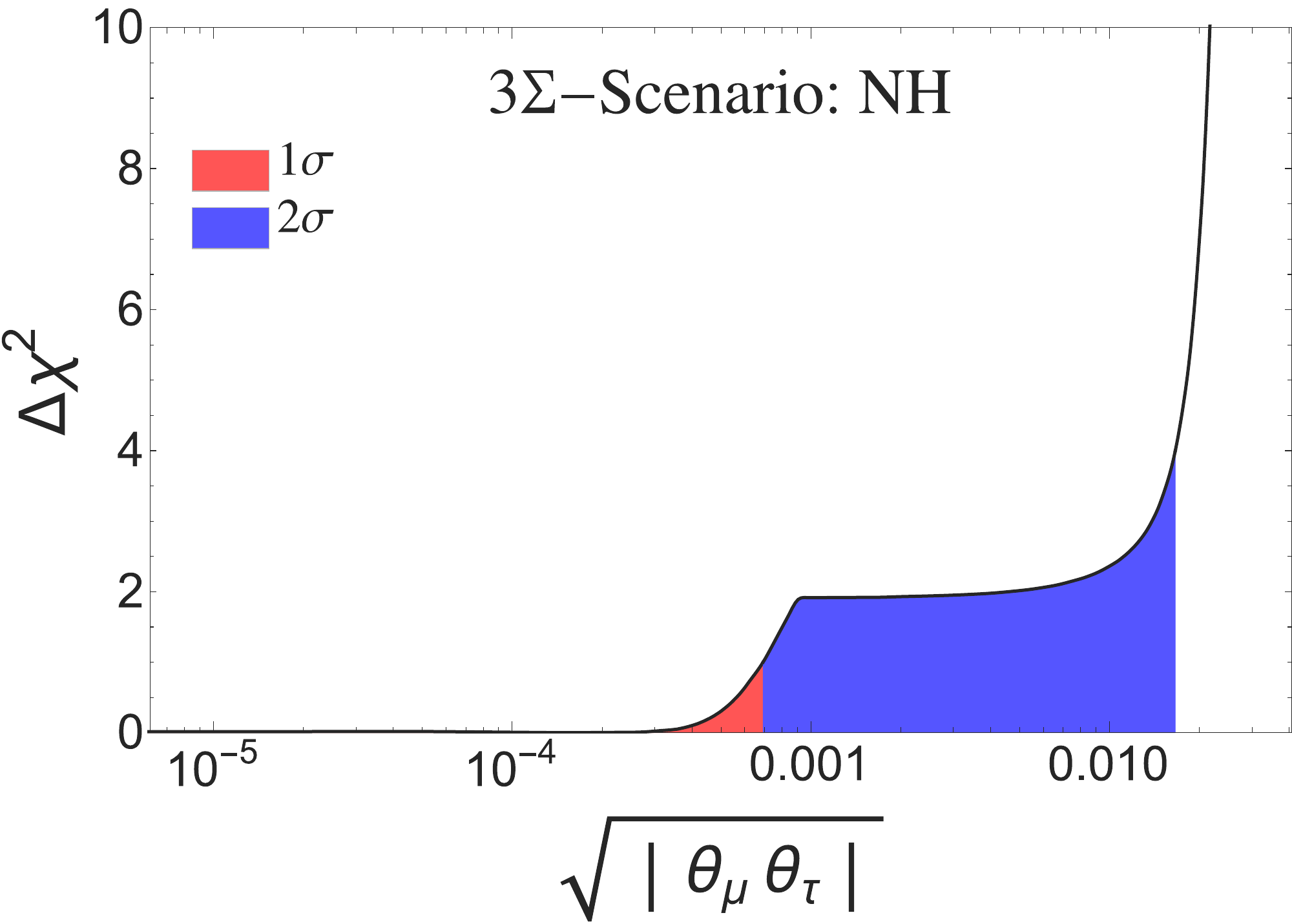} \\
\includegraphics[width=0.32\textwidth]{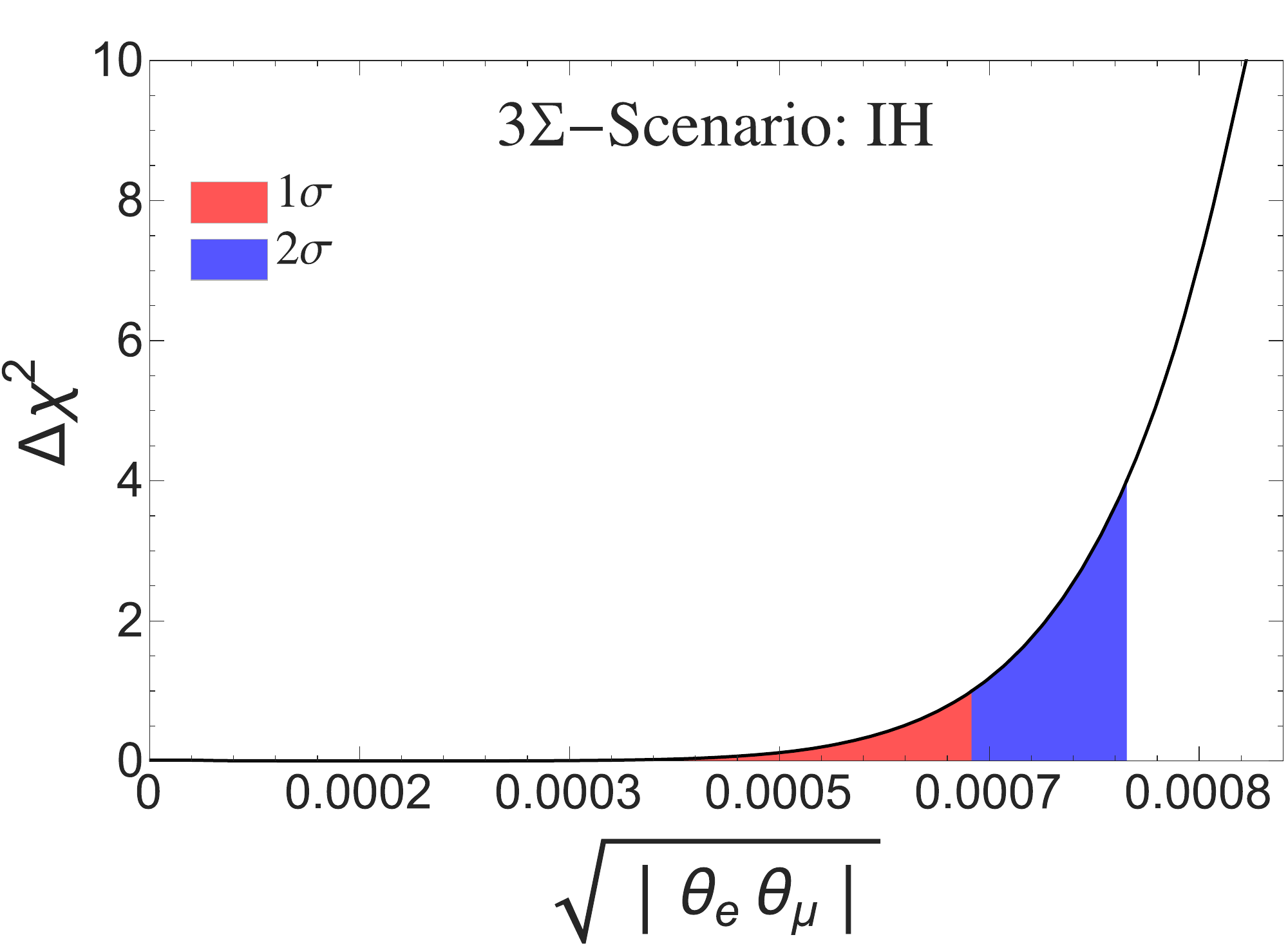}
\includegraphics[width=0.32\textwidth]{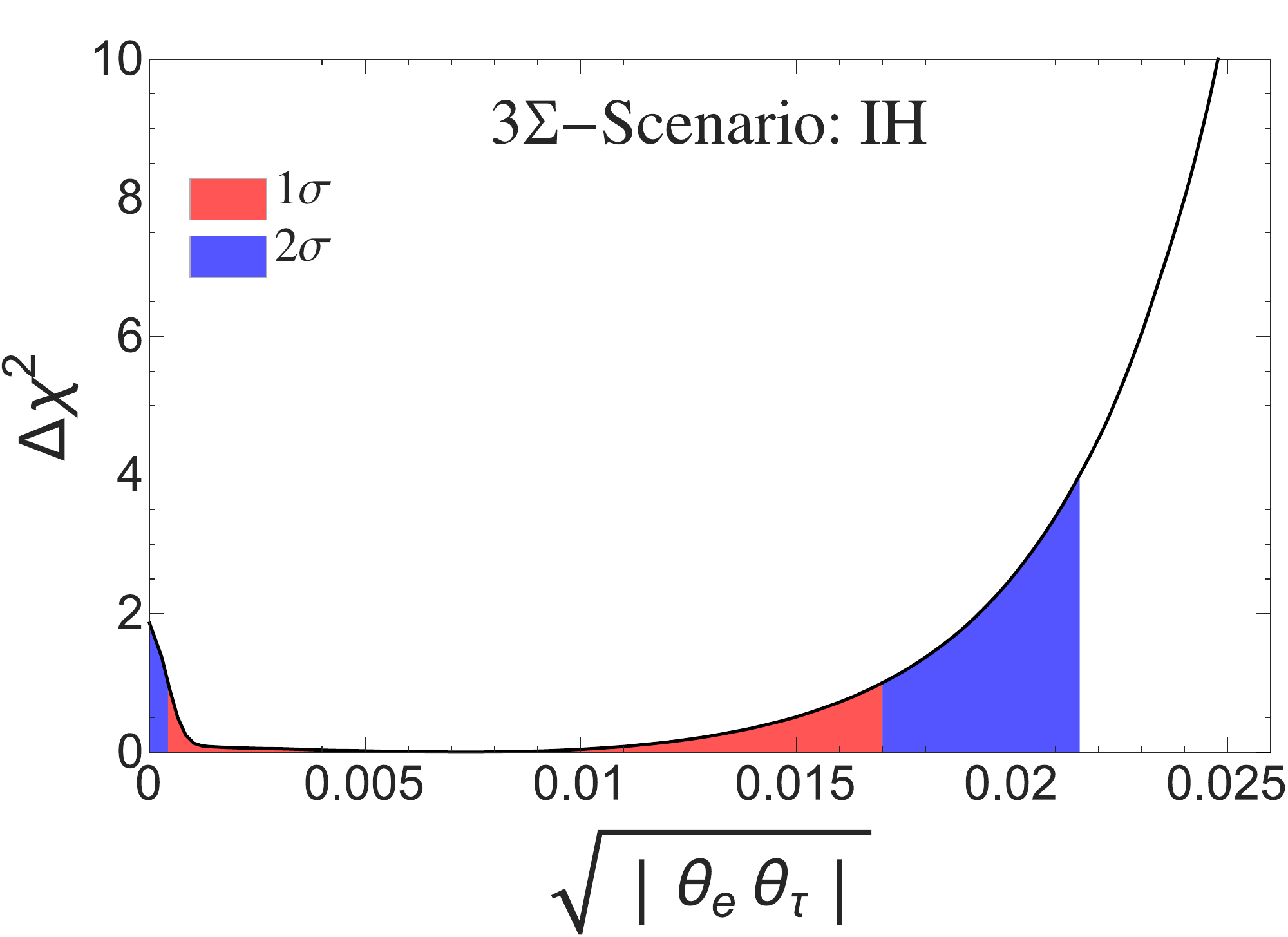}
\includegraphics[width=0.32\textwidth]{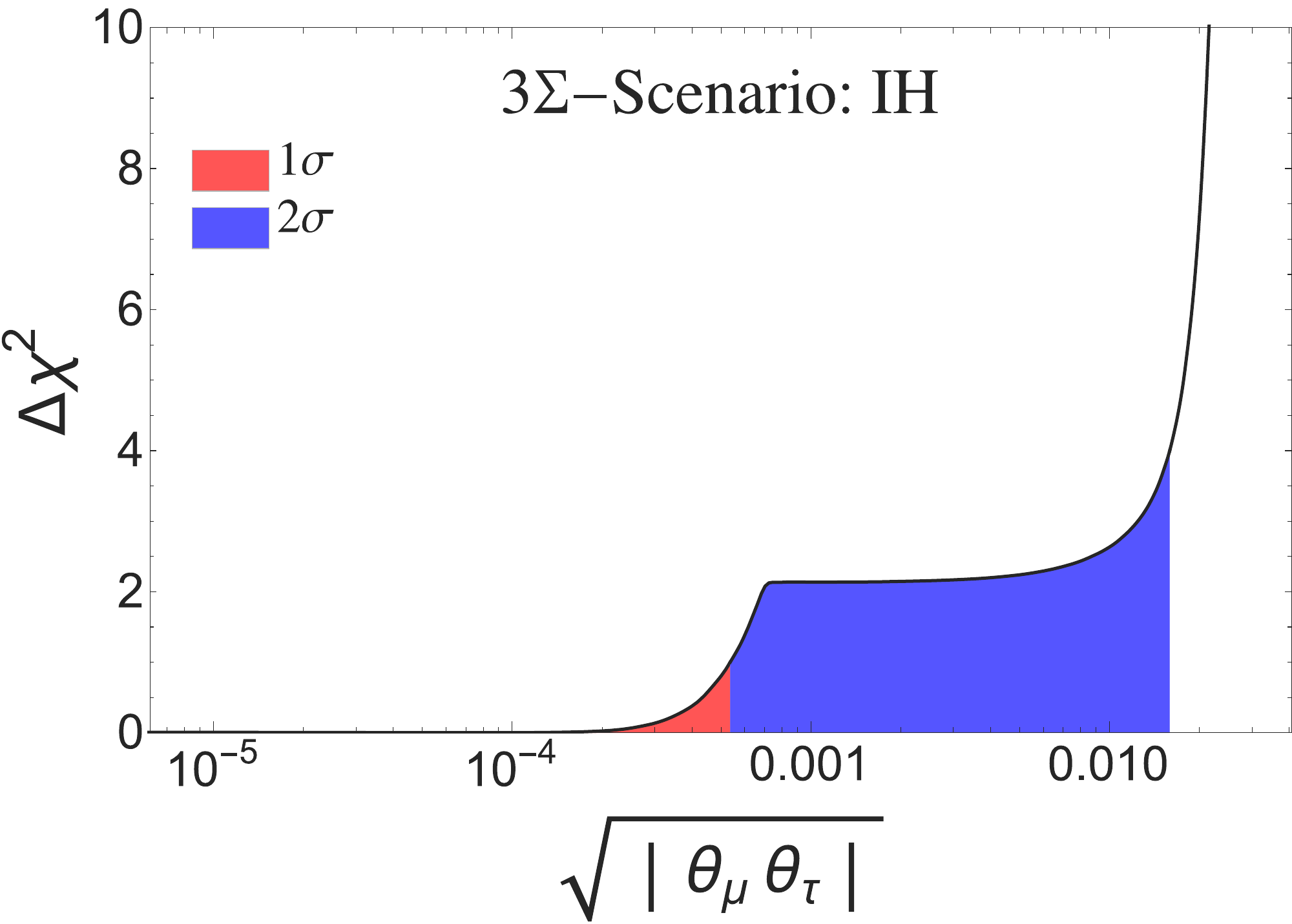} \\
\includegraphics[width=0.32\textwidth]{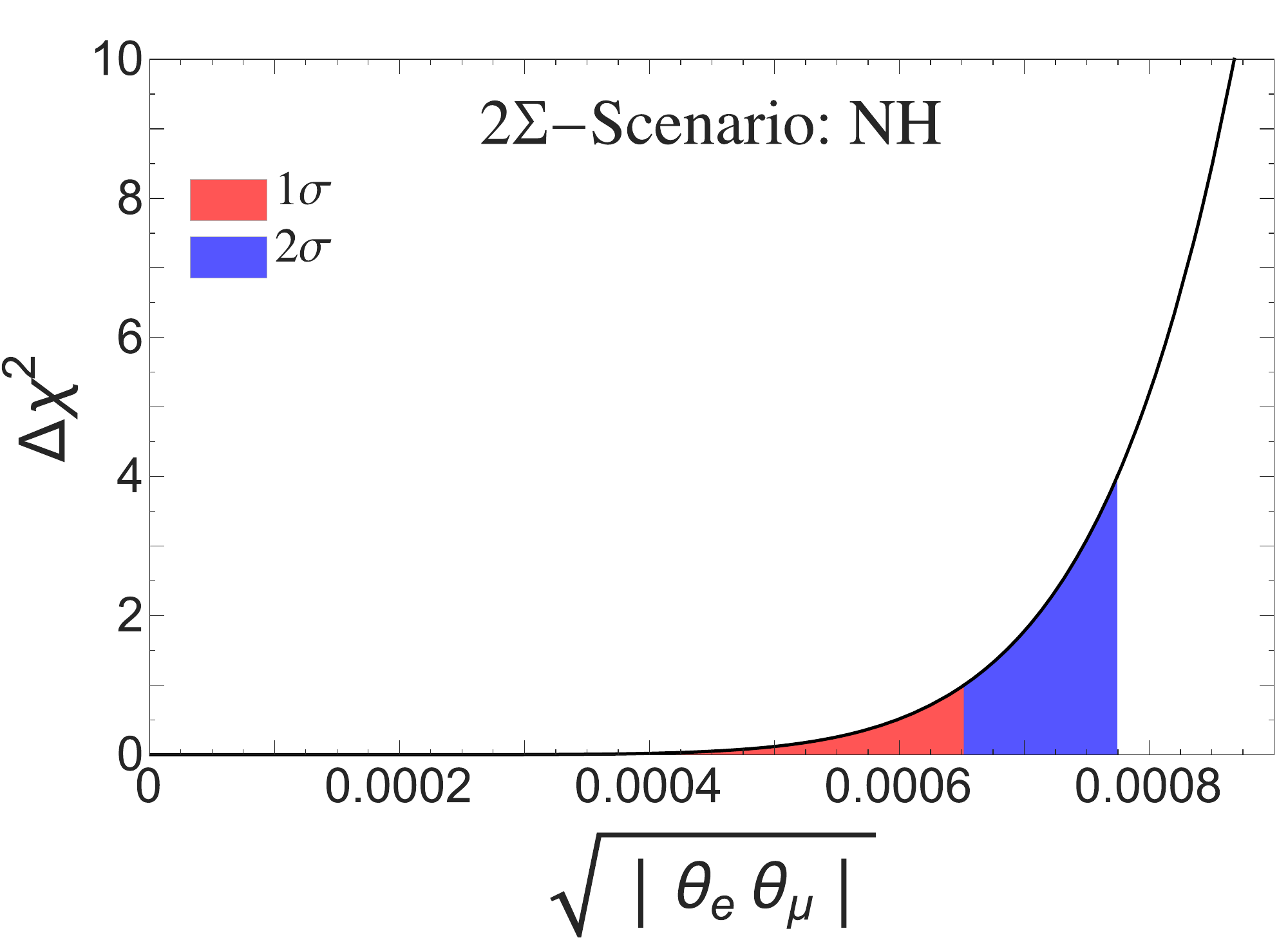}
\includegraphics[width=0.32\textwidth]{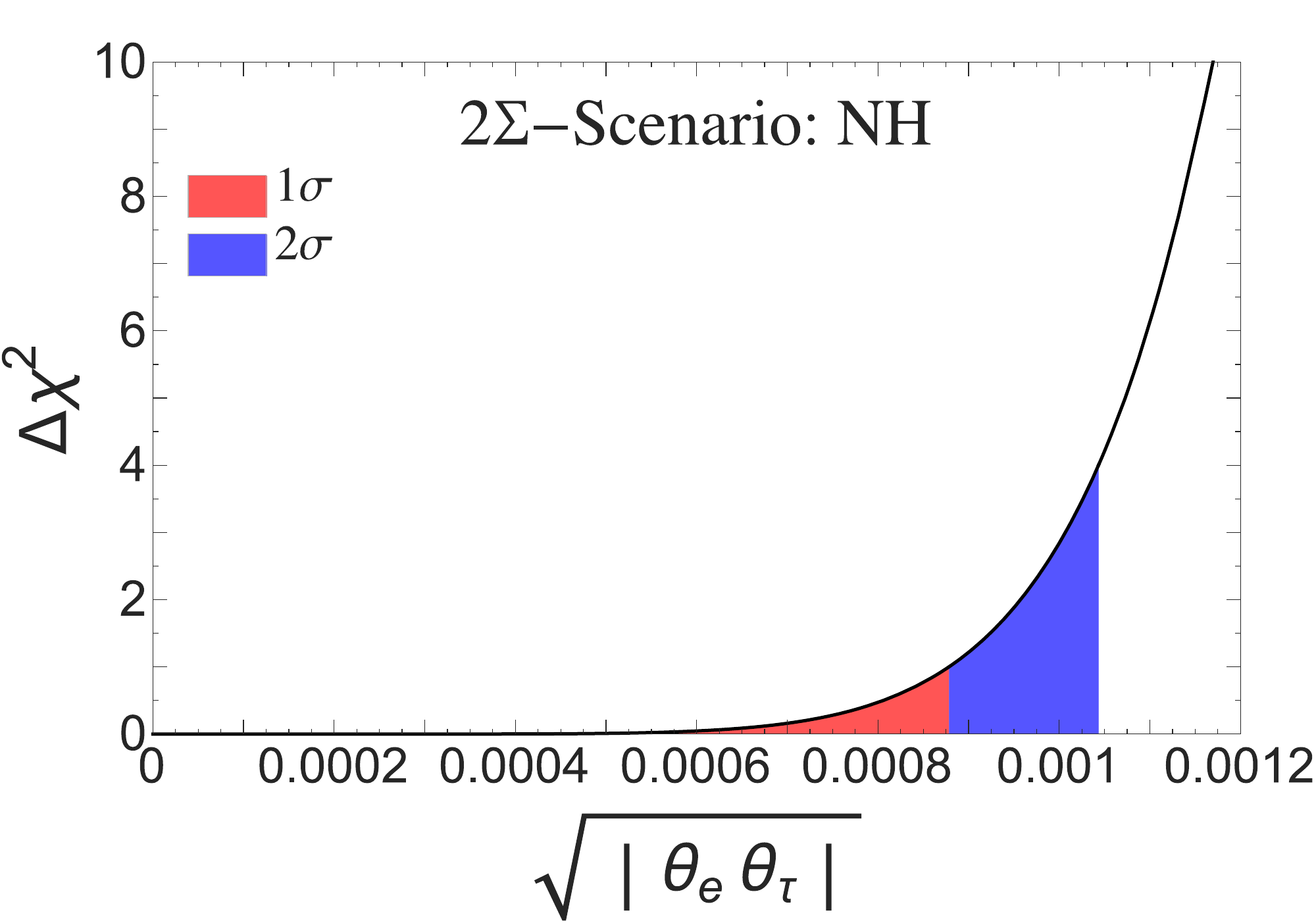}
\includegraphics[width=0.32\textwidth]{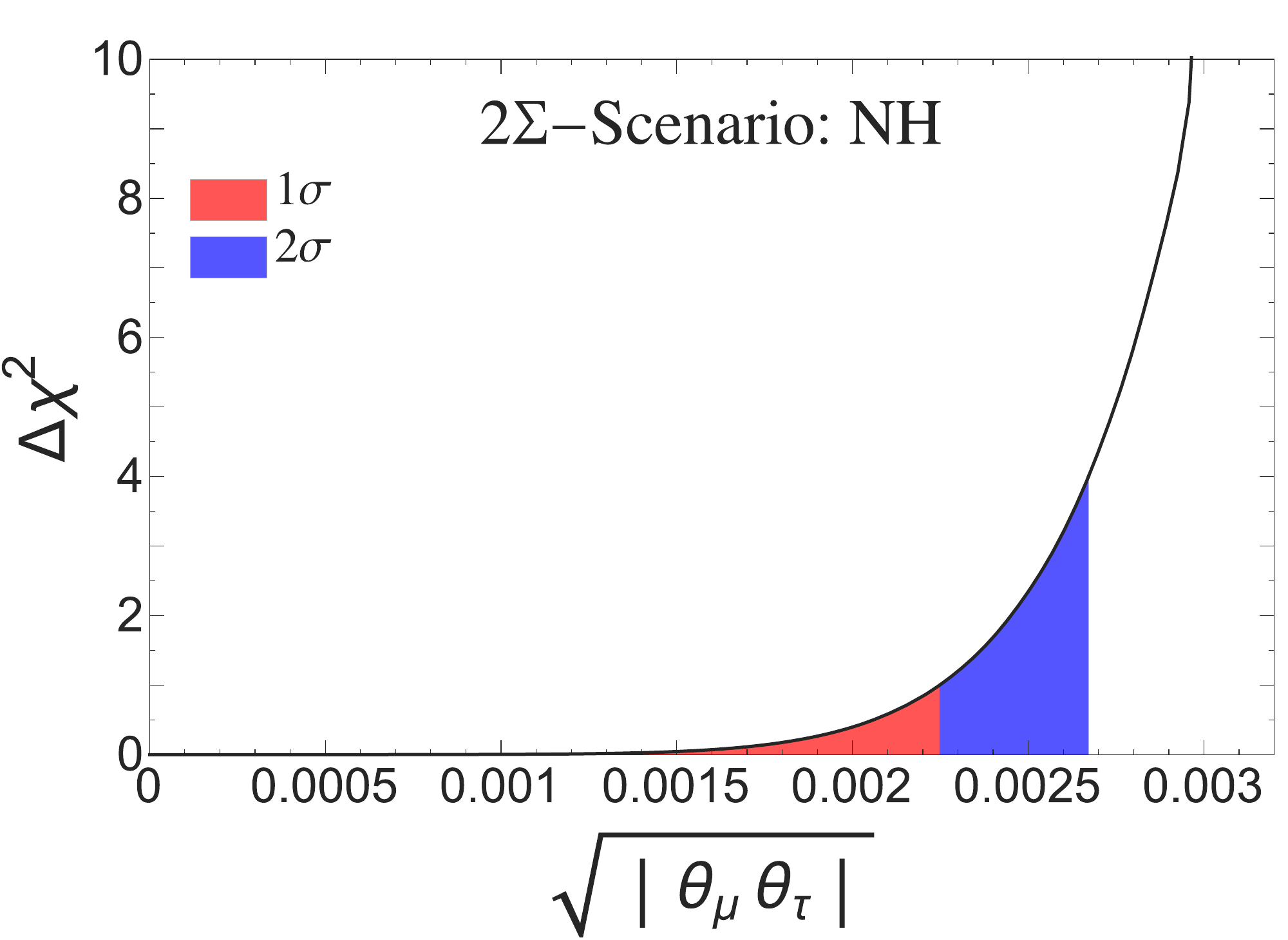} \\
\includegraphics[width=0.32\textwidth]{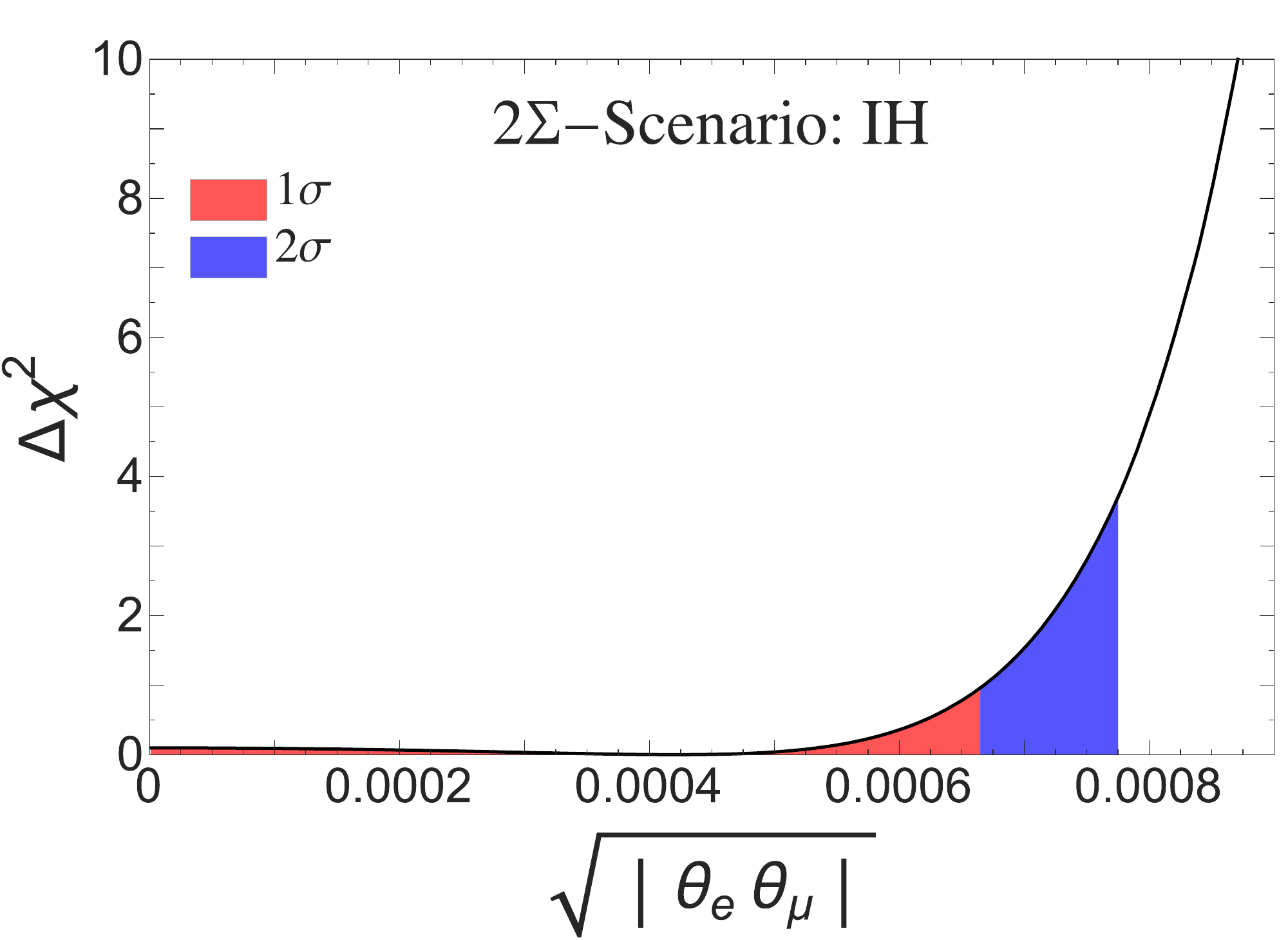}
\includegraphics[width=0.32\textwidth]{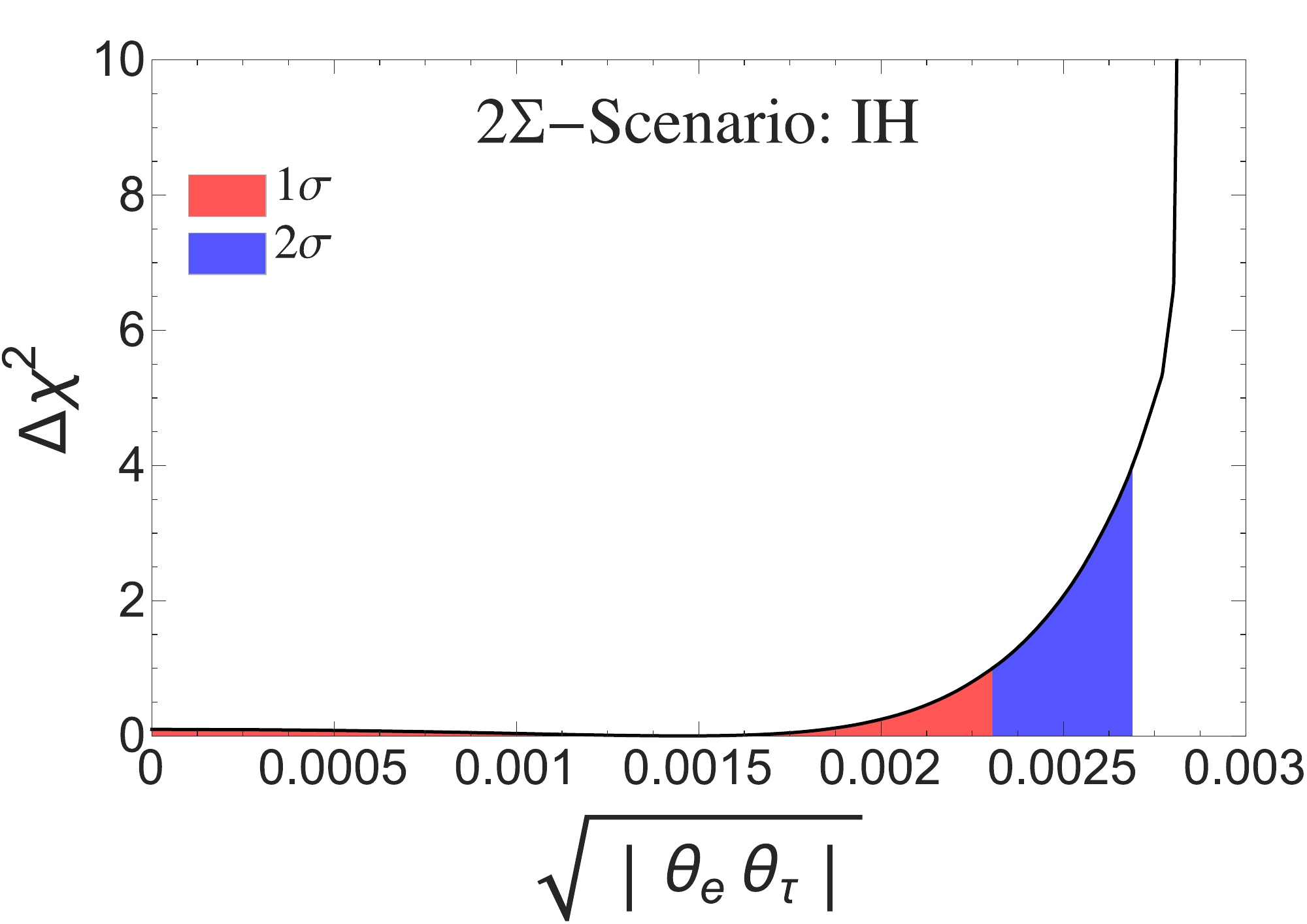}
\includegraphics[width=0.32\textwidth]{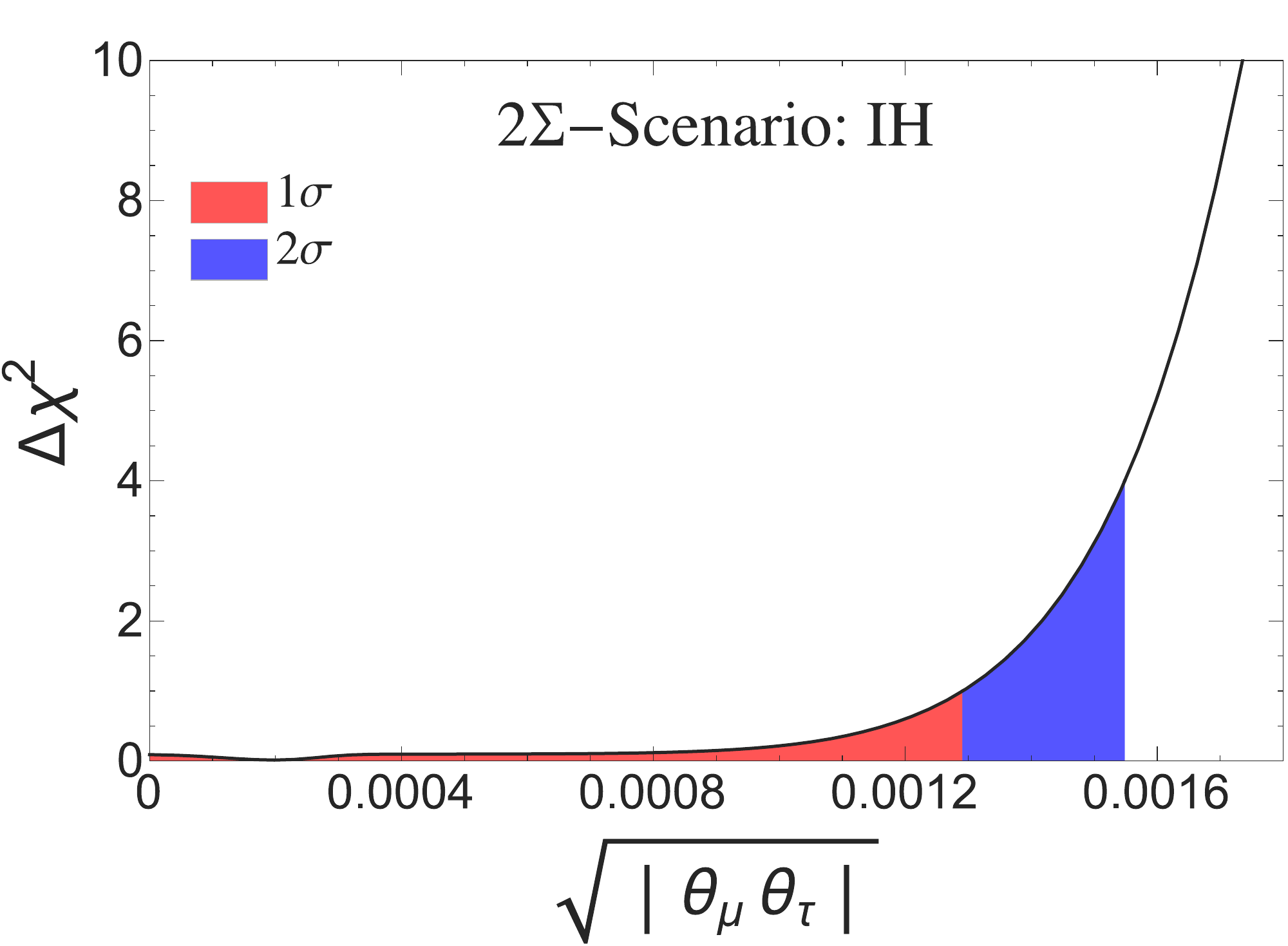}
\caption{Constraints on the off-diagonal entries of $\eta_{\alpha \beta}$ ( or $|\theta_\alpha \theta_\beta|$ for $3\Sigma$-SS and $2\Sigma$-SS). In the upper panels the G-SS fit results are plotted, while the middle and lower panels show the results of the $3\Sigma$-SS and the $2\Sigma$-SS, for NH and IH respectively.\\}
\label{fig:chi2_off}
\end{figure}

For completeness we present here the results of our analysis in terms of $\sqrt{2|\eta_{\alpha\beta}|}$ for the G-SS scenario, $\theta_{\alpha}$ and $|\theta_\alpha \theta_\beta|$ for the 3$\Sigma$-SS and 2$\Sigma$-SS cases. Figure~\ref{fig:chi2_diag} and Figure~\ref{fig:chi2_off} show the 1 dof projections of the $\chi^2$, and Table~\ref{tab:bounds} summarizes the constraints on the mixing at $1\sigma$ and $2\sigma$.

\begin{table}[h!]
\centering
\begin{tabular}{|c|c||c|c||c|c||c|c|}
\hline
 \multicolumn{2}{|c||}{\multirow{2}{*}{}} & \multicolumn{2}{c||}{G-SS} & \multicolumn{2}{c||}{$3\Sigma$-SS} & \multicolumn{2}{c|}{$2\Sigma$-SS} \\
\cline{3-8}
 \multicolumn{2}{|c||}{}  & LFC & LFV & NH & IH & NH & IH \\ 
\hline
\hline
 \multirow{2}{*}{$\sqrt{2\eta_{ee}}$, $|\theta_{e}|$} & $1\sigma$ & $\boldsymbol{0.016^{+0.005}_{-0.008}}$ & \textemdash & $\boldsymbol{0.016^{+0.005}_{-0.008}}$ & $0.016^{+0.005}_{-0.007}$ & $\boldsymbol{<5.7 \cdot 10^{-4}}$ & $<4.6 \cdot 10^{-3}$\\
\cline{2-8}
& $2\sigma$ & $\boldsymbol{<0.025}$ & \textemdash & $\boldsymbol{<0.025}$ & $< 0.025$ & $\boldsymbol{<6.8 \cdot 10^{-4}}$ & $<5.3 \cdot 10^{-3}$\\
\hline
 \multirow{2}{*}{$\sqrt{2\eta_{\mu \mu}}$, $|\theta_{\mu}|$} & $1\sigma$ & $\boldsymbol{<0.013}$ & \textemdash & $\boldsymbol{<3.1\cdot 10^{-5}}$ & $< 3.0 \cdot 10^{-5}$ & $\boldsymbol{< 2.3 \cdot 10^{-3}}$ & $<1.2 \cdot 10^{-3}$\\
\cline{2-8}
& $2\sigma$ & $\boldsymbol{<0.021}$ & \textemdash & $\boldsymbol{<0.017}$ & $< 0.015$ & $\boldsymbol{< 2.8\cdot 10^{-3}}$ & $<1.5 \cdot 10^{-3}$\\
\hline
 \multirow{2}{*}{$\sqrt{2\eta_{\tau \tau}}$, $|\theta_{\tau}|$} & $1\sigma$ & $\boldsymbol{<0.027}$ & \textemdash & $\boldsymbol{0.010^{+0.012}_{-0.006}}$ & $<0.017$ & $\boldsymbol{<2.9 \cdot 10^{-3}}$ & $< 1.4 \cdot 10^{-3}$\\
\cline{2-8}
& $2\sigma$ & $\boldsymbol{<0.041}$ & \textemdash & $\boldsymbol{<0.036}$ & $< 0.028$ & $\boldsymbol{<3.5 \cdot 10^{-3}}$ & $<1.7 \cdot 10^{-3}$\\
\hline
 \multirow{2}{*}{$\sqrt{2\eta_{e \mu}}$, $\sqrt{|\theta_{e} \theta_{\mu}|}$} & $1\sigma$ & $<0.015$ & $\boldsymbol{<6.5\cdot 10^{-4}}$ & $\boldsymbol{<6.5\cdot 10^{-4}}$ & $<6.5\cdot 10^{-4}$ & $\boldsymbol{<6.5 \cdot 10^{-4}}$ & $<6.5 \cdot 10^{-4}$\\
\cline{2-8}
& $2\sigma$ & $<0.020$ & $\boldsymbol{<7.7\cdot 10^{-4}}$ & $\boldsymbol{<7.7\cdot 10^{-4}}$ & $<7.7\cdot 10^{-4}$ & $\boldsymbol{<7.7\cdot 10^{-4}}$ & $<7.7\cdot 10^{-4}$\\
\hline
 \multirow{2}{*}{$\sqrt{2\eta_{e \tau}}$, $\sqrt{|\theta_{e} \theta_{\tau}|}$} & $1\sigma$ & $< 0.022$ & $\boldsymbol{<0.019}$ & $\boldsymbol{0.012^{+0.006}_{-0.006}}$ & $0.0074^{+0.0096}_{-0.0070}$ & $\boldsymbol{<8.8\cdot 10^{-4}}$ & $<2.3\cdot 10^{-3}$\\
\cline{2-8}
& $2\sigma$ & $<0.029$ & $\boldsymbol{<0.023}$ & $\boldsymbol{<0.022}$ & $< 0.022$ & $\boldsymbol{<1.0\cdot 10^{-3}}$ & $<2.7\cdot 10^{-3}$\\
\hline 
 \multirow{2}{*}{$\sqrt{2\eta_{\mu \tau}}$, $\sqrt{|\theta_{\mu} \theta_{\tau}|}$} & $1\sigma$ & $\boldsymbol{<0.015}$ & $<0.018$ & $\boldsymbol{<6.9\cdot 10^{-4}}$ & $< 5.4 \cdot 10^{-4}$ & $\boldsymbol{<2.2\cdot 10^{-3}}$ & $<1.3\cdot 10^{-3}$\\
\cline{2-8}
& $2\sigma$ & $<0.024$ & $\boldsymbol{<0.021}$ & $\boldsymbol{<0.017}$ & $< 0.016$ & $\boldsymbol{<2.7\cdot 10^{-3}}$ & $<1.5\cdot 10^{-3}$\\
\hline
\end{tabular}
\caption{Comparison of all $1\sigma$ and $2\sigma$ constraints on the heavy fermion triplets mixing. For the G-SS the bounds are expressed for $\sqrt{2 \eta_{\alpha \beta}}$. For the off-diagonal entries the indirect bounds from the LFC observables via the Schwartz inequality Eq.~(\ref{eq:schwarz}) are compared with the direct LFV bounds. The most stringent constraint in the G-SS is highlighted in bold face. In the $3\Sigma$-SS and $2\Sigma$-SS the bounds on $\theta_{\alpha}$ for normal (NH) and inverted hierarchy (IH) are shown. The bounds for the NH have been highlighted in bold face as an overall constraint on each scenario since this hierarchy is preferred by present neutrino oscillation data at more than $2\sigma$.}
\label{tab:bounds}
\end{table}

\end{appendix}

\clearpage

\providecommand{\href}[2]{#2}\begingroup\raggedright\endgroup

\end{document}